\documentclass[PAPER, cernpreprint, atlasdraft=false, UKenglish, texmf, orcidlogo]{atlasdoc}
\usepackage{atlaspackage}
\usepackage{atlasbiblatex}

\usepackage{atlasphysics}

\usepackage{makecell}
\usepackage{multirow}

\graphicspath{{logos/}{figures/}}

\usepackage{ANA-STDM-2021-04-PAPER-defs}

\AtlasTitle{Measurements of the electroweak production of a \mbox{$W$ boson}
in association with two jets at $\sqrt{s}=13\,\mathrm{TeV}$ with the ATLAS
detector}

\AtlasAbstract{%
Measurements of the electroweak production of a $W$ boson in association with two jets at high dijet invariant mass are performed using a dataset from $140\,\mathrm{fb}^{-1}$ of $\sqrt{s}=13\,\mathrm{TeV}$ proton--proton collisions recorded by the ATLAS detector at the LHC. The measurements are made in a fiducial region where the $W$ boson's production via vector-boson fusion is enhanced. The fiducial region requires at least two jets, an electron or a muon, and a neutrino, assuming a leptonic $W$ decay. In addition to the integrated production cross-section, differential cross-sections unfolded to particle level are determined as functions of observables which characterise angular properties of the two highest-energy jets, the kinematics of the $W$ boson, and the transverse momenta of the highest-energy jet and the charged lepton. Results are interpreted in the context of an effective field theory, and the limits placed on anomalous triple-gauge-boson couplings are among the most stringent to date for linear terms of the extended Lagrangian.
}

\AtlasRefCode{STDM-2021-04}

\PreprintIdNumber{CERN-EP-2026-182}

\AtlasJournal{Eur.\ Phys.\ J.\ C}
\hypersetup{pdftitle={ATLAS document},pdfauthor={The ATLAS Collaboration}}

\begin{document}

\maketitle

\tableofcontents

\section{Introduction}
%
%

Electroweak production of a $W$ boson in association with two
jets (\ewk \Wjj) in proton--proton ($pp$) collisions includes diagrams which feature
vector-boson fusion (VBF) at a triple-gauge-boson $WW\gamma$ or
$WWZ$ vertex.
Measurement of \ewk \Wjj is of interest for examining the non-Abelian nature
of the electroweak interactions, and for setting limits on anomalous
weak-boson self-interactions.

\begin{figure}[t]
\centering
\includegraphics[width=0.3\textwidth]{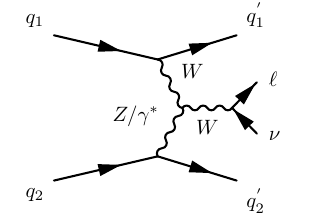}
\includegraphics[width=0.3\textwidth]{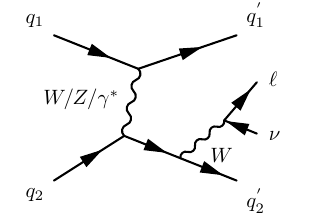}
\includegraphics[width=0.3\textwidth]{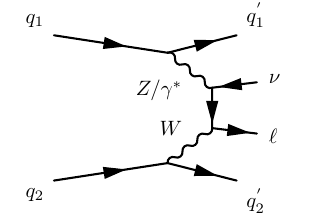}
\caption{Subset of Feynman diagrams for the \ewk \Wjj production at Born level constituting the
measured signal.}
\label{fig:ew_diag}
\end{figure}

Production of a $W$ boson via vector-boson fusion (VBF-$W$) cannot be measured directly
because it has the same final state as other \ewk \Wjj processes, and they interfere with each other.
Instead, this paper focuses on a kinematic
region where the contribution from VBF diagrams is enhanced.
A representative set of dominant Feynman diagrams for \ewk \Wjj production exemplifying the signal under study are shown in
Figure~\ref{fig:ew_diag} at Born level, $\mathcal{O}(\alpha_{\text{EW}}^{4})$, where the leftmost diagram depicts the VBF production mode.
In addition, the middle and rightmost diagrams each illustrate
the $W$ boson being radiated by one of the quarks involved.
The complete set of \ewk \Wjj diagrams includes $WW$ and $WZ$ production,
where one of the bosons decays hadronically and the other leptonically.
These contributions are suppressed by the selection criteria and explicitly excluded from the signal definition.

\begin{figure}[t]
\centering
\includegraphics[width=0.3\textwidth]{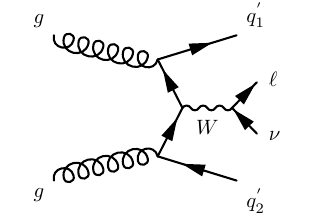}
\includegraphics[width=0.3\textwidth]{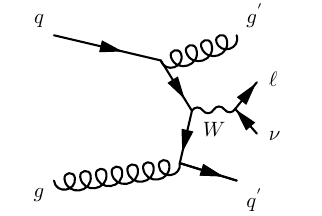}
\includegraphics[width=0.3\textwidth]{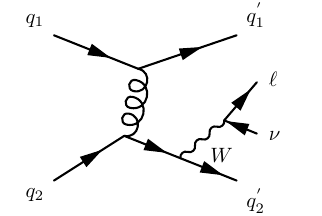}
\caption{Subset of the Feynman diagrams for the \qcd \Wjj production at Born level
constituting the dominant background when extracting the \ewk \Wjj signal from the data.}
\label{fig:qcd_diag}
\end{figure}

The \ewk \Wjj processes exhibit a specific experimental signature including
two jets with large invariant mass, oriented at small angles with respect to each proton
beam and thus well separated in rapidity, referred to as `tag'
jets.
The final state is characterised by suppressed hadronic activity within
the rapidity gap between the two tag jets, due to the absence of colour
exchange between the interacting partons.
The main background for this process comes from mixed strong-and-electroweak \Wjj
production at $\mathcal{O}(\alpha_{\text{EW}}^{2}, \alphas^{2})$, involving at least two \qcd
vertices at tree level as illustrated in Figure~\ref{fig:qcd_diag}, and is referred to
as \qcd \Wjj production.
The other backgrounds considered originate from \qcd and \ewk $Z$ boson production,
top-quark pair and single-top production, and $WW$ and $WZ$ production, as well as
from misidentified leptons.
Only events where the $W$ boson decays into an electron and a neutrino, or
a muon and a neutrino, are treated as signal and these are combined into a single sample.
As a consequence, \qcd and \ewk \Wjj production with a $W$ boson decaying
into a $\tau$-lepton and a neutrino are considered as background. This includes
events with a $\tau$-lepton decaying into an electron or a muon, and a neutrino.

Measurements of \ewk \Wjj and \Zjj processes have been made by the
ATLAS and CMS experiments based at the LHC. Using the Run~1 dataset,
ATLAS measured fiducial
cross-sections for \Zjj production at $\sqrt{s}=8\,\TeV$~\cite{STDM-2013-02},
as well as for \Wjj production at $\sqrt{s}=7$ and $8\,\TeV$ along with differential
cross-sections in a variety of phase spaces~\cite{STDM-2014-11}.
CMS has measured fiducial \Wjj cross-sections at
$\sqrt{s}=8\,\TeV$~\cite{CMS-SMP-13-012}, and fiducial \Zjj and \Wjj cross-sections at
$\sqrt{s}=13\,\TeV$~\cite{CMS-SMP-16-018,CMS-SMP-17-011},
using the LHC Run~1 dataset and a partial Run~2 dataset, respectively.
ATLAS measured differential \Zjj cross-sections in four kinematic
observables~\cite{STDM-2017-27} at $\sqrt{s}=13\,\TeV$ using the complete Run~2 dataset.
Both experiments have used these processes and an effective field theory approach
to extract limits on anomalous triple-gauge-boson couplings.

This paper reports a measurement of the fiducial cross-section for \ewk \Wjj
production in an enhanced VBF phase space, and differential
measurements of six observables characteristic of the signal process that are
unfolded to particle level.
These observables are the transverse momentum ($\pt$) of the
leading-\pt jet ($\pt^{j1}$), the \pt of the charged lepton ($\pt^{\ell}$)
originating from the $W$ boson decay, and four variables computed from the
tag jets defined as the two leading-\pt jets, namely their rapidity-ordered azimuthal angle difference ($\Delta\phi_{jj}$),\footnote{$\Delta \phi_{jj}=\phi_{j_{1}}-\phi_{j_{2}}$ with $\phi_{j_{1}}$ and $\phi_{j_{2}}$ the azimuthal angle of the tag jet with the largest and second largest rapidity, respectively.} their invariant mass ($m_{jj}$), their transverse momentum ($\pt^{jj}$), and their absolute rapidity difference ($|\Delta y_{jj}|$).
The best discrimination between \qcd and \ewk \Wjj production is provided by
$m_{jj}$, while $|\Delta y_{jj}|$ is particularly sensitive to \ewk--\qcd
interference. The $\Delta \phi_{jj}$ variable %
is especially sensitive to signals beyond the Standard Model (SM),
and $\pt^{jj}$, $\pt^{j1}$ and $\pt^{\ell}$ complete the
description of the $W$ and jet system.

The results are interpreted in the context of an
effective field theory (EFT) in order to study anomalous dimension-six
triple-gauge-boson couplings~\cite{Degrande_2013}.
EFT operators that are even in charge (C) times parity (P), corresponding to the
Wilson coupling coefficients $c_{W}$ and $c_{HWB}$, are investigated along with
their CP-odd counterparts $\tilde{c}_{W}$ and $\tilde{c}_{HWB}$.


%
\section{ATLAS detector}
%
%

%
\newcommand{\AtlasCoordFootnote}{%
ATLAS uses a right-handed coordinate system with its origin at the nominal interaction point (IP)
in the centre of the detector and the \(z\)-axis along the beam pipe.
The \(x\)-axis points from the IP to the centre of the LHC ring,
and the \(y\)-axis points upwards.
Polar coordinates \((r,\phi)\) are used in the transverse plane,
\(\phi\) being the azimuthal angle around the \(z\)-axis.
The pseudorapidity is defined in terms of the polar angle \(\theta\) as \(\eta = -\ln \tan(\theta/2)\) and is equal to the rapidity
$ y = (1/2) \ln[(E + p_z)/(E - p_z)] $ in the relativistic limit.
Angular distance is measured in units of \(\Delta R \equiv \sqrt{(\Delta y)^{2} + (\Delta\phi)^{2}}\).}

The ATLAS experiment~\cite{PERF-2007-01} at the LHC is a multipurpose particle detector
with a forward--backward symmetric cylindrical geometry and nearly \(4\pi\) coverage in
solid angle.\footnote{\AtlasCoordFootnote}
It consists of an inner tracking detector surrounded by a thin superconducting solenoid
providing a \qty{2}{\tesla} axial magnetic field, electromagnetic and hadronic calorimeters, and a muon spectrometer.
The inner tracking detector covers the pseudorapidity range \(|\eta| < 2.5\) and consists of silicon pixel, silicon microstrip, and transition radiation tracking detectors.
Lead/liquid-argon (LAr) sampling calorimeters provide electromagnetic (EM) energy measurements
with high granularity within the region \(|\eta|< 3.2\).
A steel/scintillator-tile hadronic calorimeter covers the central pseudorapidity range (\(|\eta| < 1.7\)).
The endcap and forward regions are instrumented with LAr calorimeters
for EM and hadronic energy measurements up to \(|\eta| = 4.9\).
The muon spectrometer surrounds the calorimeters and is based on
three large superconducting air-core toroidal magnets with eight coils each.
The field integral of the toroids ranges between \num{2.0} and \qty{6.0}{\tesla\metre}
across most of the detector.
The muon spectrometer includes a system of precision tracking chambers up to \(|\eta| = 2.7\) and fast detectors for triggering up to \(|\eta| = 2.4\).

The luminosity is measured mainly by the LUCID-2~\cite{LUCID2} detector, which is located close to the beampipe.
A two-level trigger system was used to select events~\cite{TRIG-2016-01}.
The first-level trigger is implemented in hardware and used a subset of the detector information
to accept events at a rate close to \qty{100}{\kHz}.
This is followed by a software-based trigger that
reduced the accepted rate of complete events to \qty{1.25}{\kHz} on average
depending on the data-taking conditions.
A software suite~\cite{SOFT-2022-02} is used in data simulation, in the reconstruction
and analysis of real and simulated data, in detector operations, and in the trigger and data acquisition
systems of the experiment.


%
\section{Data and simulated events}
\label{sec:samples}
%

%
This analysis is performed using data from $pp$ collisions at a centre-of-mass energy of $\sqrt{s}=\SI{13}{TeV}$, collected with the ATLAS detector between 2015--2018 during Run~2 of the LHC. The total integrated luminosity of this dataset, after offline data-quality checks, is $\runtwolumi$~\cite{DAPR-2021-01, DAPR-2018-01}.
The corresponding events were recorded by requiring a logical OR
of single-electron and single-muon triggers~\cite{TRIG-2018-05,TRIG-2018-01}.
The triggered electrons and muons considered were required to satisfy a
$\pt$ threshold ranging from $24\,\GeV$ to $26\,\GeV$ and from $20\,\GeV$ to
$26\,\GeV$, respectively.
Furthermore, these leptons must satisfy specific
lepton-quality requirements, depending on the data-taking period.
Additional triggers with looser lepton-quality criteria, and with
$\pt$ thresholds of $60\,\GeV$ to $300\,\GeV$
for electrons and $50\,\GeV$ for muons, are used to increase the efficiency of the event selection.

Monte Carlo (MC) event generators are used to model the signal and all background processes apart from the background originating from misidentified or non-prompt leptons, which is estimated using a data-driven approach. In addition, the event yields of the dominant \qcd \Wjj background are constrained by data in dedicated control regions, to better model the distribution shapes of the observables of interest, which are not necessarily predicted well at high dijet mass.

MC predictions of SM processes were obtained using the ATLAS simulation infrastructure~\cite{SOFT-2010-01}, which incorporates a complete \GEANT~\cite{Agostinelli:2002hh} simulation of the ATLAS detector. The effect of multiple interactions in the same or nearby bunch crossings was simulated by generating a set of inelastic proton--proton collision events using \PYTHIA[8]~\cite{Sjostrand:2007gs} with the A3 set of tuned parameters~\cite{ATL-PHYS-PUB-2016-017} and the \NNPDF[2.3lo] set~\cite{Ball:2012cx} of parton distribution functions (PDFs). These events were overlaid on the original hard-scatter signal and background events, and weighted so that the distribution of the average number of proton--proton interactions per bunch crossing in simulation matches that observed in the data. Differences between data and simulation in lepton trigger, reconstruction and isolation efficiencies are corrected on an event-by-event basis using lepton kinematic-dependent scale factors for each selected lepton~\cite{EGAM-2018-01,MUON-2018-03}.

Modelling of \ewk \Wjj production was provided by the \POWHEGBOX simulation framework~\cite{Nason:2004rx,Frixione:2007vw,Alioli:2010xd,Schissler:2013nga} for the electron and muon decays of the $W$ boson,
with the matrix element (ME) calculated at next-to-leading order (NLO) in perturbative \qcd for two partons.
The \NNPDF[3.0nlo] PDF set~\cite{Ball:2014uwa} was used as the baseline PDF description~\cite{Butterworth:2015oua}. The event generation was performed in the `VBF approximation'~\cite{Jager:2012xk}, which excludes any overlap with diboson topologies.
To simulate the parton showering, \POWHEG was interfaced with \PYTHIA[8]~\cite{Sjostrand:2014zea}, where the A14 tune~\cite{ATL-PHYS-PUB-2014-021} and \NNPDF[2.3lo] PDF set were used and the dipole recoil was activated.
Particle-level scale factors computed as a function of $m_{jj}$ were
provided by theorists to reweight this sample to NLO \ewk
accuracy~\cite{Lindert_2023}. These corrections were only used for comparison of
the measurement with the particle-level prediction in $m_{jj}$, and for the
interpretation of the results in Section~\ref{sec:eft}.

An alternative \ewk \Wjj sample is included in the analysis to assess the sensitivity of the measurement to the showering model. In this sample, the same ME events originally generated with \POWHEG for the nominal \ewk \Wjj estimate were showered using \Herwig[7.2.1]~\cite{Bahr:2008pv,Bellm:2015jjp} with the angle-ordered model and the default hadronisation and underlying-event tune.
In addition, an alternative prediction from \Sherpa[2.2.11]~\cite{Gleisberg:2008ta,Bothmann:2019yzt} is used to assess systematic uncertainties in the unfolding method and for comparison with
the unfolded measurement at particle level.
The \Sherpa[2.2.11] \ewk \Wjj simulation used MEs calculated in the `VBF approximation' at leading order (LO) in perturbative \qcd, including the first jet emission in addition to the two tag jets. The \AtlasMC{Comix}~\cite{Gleisberg:2008fv} and \OPENLOOPS~\cite{Denner:2016kdg,Cascioli:2011va} libraries provided the MEs used to produce the parton predictions, while the CKKW matching procedure~\cite{Hoeche:2009rj} was used to merge final-state topologies.
The \NNPDF[3.0nnlo] PDF set~\cite{Ball:2014uwa} was used for the calculation, and the default parameters of \Sherpa were used for the parton showering, hadronisation, and underlying-event activity.

The nominal \qcd \Wjj prediction was modelled with \POWHEGBOX for the electron and muon decays of the $W$ boson~\cite{Nason:2004rx,Frixione:2007vw,Alioli:2010xd,Campbell:2013vha}.
The prediction was computed at NLO in perturbative \qcd for two partons.
The baseline PDF description was provided by the \NNPDF[3.0nlo] PDF set. The generator was interfaced with \PYTHIA[8] for parton showering, which used the CTEQ6L1 PDF set~\cite{Lai:2010vv} along with the AZNLO tune~\cite{STDM-2012-23}.
Additionally, the \POWHEG \qcd \Wjj simulation includes a Born (parton-level) suppression factor which enhances the event statistics in the EW signal phase space by reducing the number of events where the two leading-\pt jets have a dijet mass below $\SI{200}{GeV}$ or either jet has \pt below
$\SI{30}{GeV}$~\cite{Nason:2013uba,Re_2012}.

The sensitivity of the \qcd \Wjj prediction to the showering model is
assessed by comparing the nominal simulation with one where events were generated with \POWHEG
but showered using \Herwig[7.2.1] with the angle-ordered model and the default
hadronisation and underlying-event tune.

The production of \Wjj events was modelled via distinct simulations of its
\ewk and \qcd components. The effect of interference between the \ewk and
\qcd diagrams is suppressed
by the selection of candidate events, and therefore treated as an uncertainty
in the \ewk signal measurement. Events arising from the interference term
were generated by \MADGRAPH{}~\cite{Alwall:2014hca} at LO in
perturbative \qcd, while \PYTHIA[8]
was used with the \NNPDF[2.3lo] PDF set and A14
tune to simulate the parton showering and subsequent hadronisation.

The other background processes considered include \qcd and \ewk \zjets production (both simulated using \Sherpa[2.2.11]), top-quark pair and single-top
production %
modelled using \POWHEGBOX[v2]~\cite{Frixione:2007nw}, and strong diboson production with fully leptonic and semileptonic final states simulated with  \Sherpa[2.2.12] and \Sherpa[2.2.11], respectively. The \ewk and \qcd $W(\rightarrow\tau\nu)\!+\!\mathrm{jets}$ events were simulated via \Sherpa[2.2.11], with the \ewk contribution predicted to make up less than 1\% of the event yield in the analysed phase space.
The simulation of these backgrounds was performed at NLO accuracy in perturbative \qcd, except for the \ewk \Zjj events, which were simulated at LO.
In addition, the \qcd \zjets and $W(\rightarrow\tau\nu)\!+\!\mathrm{jets}$ samples allow up to three extra partons at LO.

All samples are normalized to the integrated luminosity of the data using the cross-sections computed by the generators considered, except for top-quark production. The production of top-quark pairs is normalised to the cross-section computed at NNLO accuracy in perturbative \qcd, including the resummation of next-to-next-to-leading logarithmic (NNLL) soft gluon terms calculated with \TOPpp[2.0]~\cite{Czakon:2011xx}.
The associated production of top quarks with $W$ bosons is corrected to the cross section at NLO+NNLL~\cite{Kidonakis:2021vob}, while the $s$- and $t$-channel production is corrected to NNLO~\cite{Liu:2018gxa,Campbell:2020fhf}.

A summary of the simulated processes and their generator settings is provided in Table~\ref{tab:samples_and_generators}.

\begin{table}
\begin{center}
\caption{Summary of generators used for MC simulation.
For each sample, the parton showering algorithm and the tune used
are specified.
The generated jet multiplicity at the highest accuracy in
perturbative \qcd for the ME computation is stated, along
with the PDF used.
}
\label{tab:samples_and_generators}
\scalebox{0.85}{
\begin{tabular}{lllll}
\hline\hline
Process & Simulation & ME accuracy & PDF & Tune \\ \hline\hline
\ewk \Wjj
& \makecell[l]{ Nominal: \POWPY[8] \\ Alt.: \POWHER[7.2.1] \\ Alt.: \Sherpa[2.2.11] }
& \makecell[l]{ 2j@NLO \\ 2j@NLO \\ 2,3j@LO }
& \makecell[l]{ \NNPDF[3.0nlo] \\ \NNPDF[3.0nlo] \\ \NNPDF[3.0nnlo] }
& \makecell[l]{ A14 \\ Default \\ Default }
\\ \hline

\qcd \Wjj
& \makecell[l]{ Nominal: \POWPY[8] \\ Alt.: \POWHER[7.2.1] }
& \makecell[l]{ 2j@NLO \\ 2j@NLO }
& \makecell[l]{ \NNPDF[3.0nlo] \\ \NNPDF[3.0nlo] }
& \makecell[l]{ AZNLO \\ Default }
\\ \hline

\qcd $W(\rightarrow\tau\nu)\!+\!\mathrm{jets}$ & \Sherpa[2.2.11] & 0,1,2j@NLO or 3,4,5j@LO & \NNPDF[3.0nnlo] & Default \\ \hline
\qcd $\zjets$ & \Sherpa[2.2.11] & 0,1,2j@NLO or 3,4,5j@LO  & \NNPDF[3.0nnlo]  & Default    \\ \hline
Top & \POWPY[8] & NLO  & \NNPDF[3.0nlo]  & A14  \\ \hline
Diboson  &
\Sherpa[2.2.12] \& 2.2.11
& 0,1j@NLO or 2,3j@LO
& \NNPDF[3.0nnlo]
& Default  \\ \hline

Other \ewk
& \Sherpa[2.2.11]
& 2,3j@LO
& \NNPDF[3.0nnlo]
& Default
\\ \hline\hline
\end{tabular}
}
\end{center}
\end{table}


%
\section{Event selection}
\label{sec:selection}
%

A set of event selection criteria define the fiducial region
for the differential cross-section measurements.
The candidate events are composed of $W$ bosons decaying into an electron or muon, plus a neutrino, produced
in association with at least two jets.

\subsection{Event reconstruction}

Only events containing a reconstructed primary vertex from the hard-scattering
interaction with at least two associated tracks are kept~\cite{ATL-PHYS-PUB-2015-026}.
The primary vertex is selected from the reconstructed vertices as the one with the highest
$\sum\pt^2$ of the associated tracks found in the inner detector with
$\pt>500\,\MeV$.

Reconstructed electrons, muons and jets are assigned to either a baseline-
or signal-object category. The first constitutes the lowest quality
considered in the analysis, while the second includes additional tighter
requirements to refine the definition of candidate events, control
regions, and observables of interest.

Electrons are reconstructed from energy deposits in the EM calorimeter that are
matched to inner-detector tracks~\cite{EGAM-2018-01}. Their energy is calibrated to account for detector response affecting the energy scale and resolution, and they are required to
have absolute pseudorapidity $|\eta| < 2.47$, excluding the transition
region $1.37<|\eta|<1.52$ between the barrel and endcaps of the EM calorimeter.
Muons are reconstructed by combining the tracking information from the inner
detector and the muon spectrometer~\cite{MUON-2018-03}. Their momentum is calibrated to account for detector response affecting the momentum scale and resolution,
and they are required to have $|\eta|<2.5$.
Leptons of each flavour are required to originate from the primary vertex, which
requires the longitudinal impact parameter $z_0$ of the tracks associated
with the leptons to satisfy $|z_0\sin\theta|<0.5\,\mathrm{mm}$.
Baseline electrons and muons must in addition have a $\pt$ that exceeds
$10\,\GeV$, and fulfil the \texttt{Loose} identification
criteria~\cite{PERF-2017-01,MUON-2018-03}.

Signal electrons and muons are required to be isolated from other physics
objects, using information from the
tracks in the inner detector and the energy clusters in the calorimeters
in a cone of lepton-\pt-dependent angular size $\Delta R$ around the lepton.
The \texttt{Tight\_VarRad}~\cite{EGAM-2021-01} criteria are used for electrons,
while muons are isolated according to the \texttt{PflowTight\_VarRad}
criteria~\cite{PERF-2015-10}.
They are also required to have their transverse impact
parameter significance
fulfil $|d_0/\sigma_{d_0}|<5$ for electrons, and $<3$ for muons.
Furthermore, signal electrons and muons must have a $\pt$ that exceeds
$27\,\GeV$, and fulfil their respective \texttt{Tight} identification
criteria~\cite{PERF-2017-01,MUON-2018-03}.

Jets are reconstructed using the anti-$k_{t}$ clustering
algorithm~\cite{Cacciari:2008gp} with a radius parameter of $R=0.4$,
as implemented in the \textsc{FastJet} software package~\cite{Fastjet}.
The algorithm is run on particle-flow objects, combining calorimeter
energy deposits and tracks to determine jet properties~\cite{PERF-2015-09}.
They are calibrated to account for the detector response, including
corrections for the energy scale and energy resolution~\cite{JETM-2018-05}.
Baseline jets are calibrated jets satisfying $\pt>20\,\GeV$ and $|\eta|<4.5$.
Jets with $\pt>20\,\GeV$ and $|\eta|<2.5$ originating from the decay
of a $b$-hadron are identified ($b$-tagged) using the \texttt{DL1r} $b$-tagging
algorithm~\cite{ATL-PHYS-PUB-2017-013,FTAG-2019-07} at the 85\%-efficiency
working point.

To suppress jets not originating from the hard scatter that produces the
$W$ boson,
jets with $\pt<60\,\GeV$ and $|\eta|<2.4$ are required to
satisfy the \texttt{Tight} jet-vertex tagger (\texttt{JVT}) working
point~\cite{PERF-2014-03}, while jets with $20<\pt<120\,\GeV$ and
$2.5<|\eta|<4.5$ are required to pass the \texttt{Tight} forward jet-vertex
tagger (\texttt{fJVT}) working point~\cite{ATL-PHYS-PUB-2019-026}.
Signal jets are baseline jets that fulfil the \texttt{JVT} and \texttt{fJVT}
requirements, and
have a $\pt$ that exceeds $30\,\GeV$, in order to further suppress jets from
additional proton--proton interactions (pile-up) and to reduce the uncertainties due to these
pile-up jets.

To ensure that charged leptons are not also reconstructed as jets, and
vice versa, overlaps between baseline leptons and jets are removed by retaining
leptons within a distance $\Delta R<0.2$ from a jet, while discarding the overlapping
jet, after which any jets within a distance $0.2<\Delta R<0.4$
from a lepton are retained and the lepton is discarded.

The \pt of the final-state neutrinos can be inferred from the missing
transverse momentum with magnitude
\ensuremath{p_{\mathrm{T}}^{\mathrm{miss}}\xspace},
which is reconstructed from the baseline leptons and baseline jets
that fulfil the JVT and fJVT requirements.
Tracks recorded using the inner detector that are not associated with a
physics object are also considered in the determination of
\ensuremath{p_{\mathrm{T}}^{\mathrm{miss}}\xspace}~\cite{JETM-2020-03}.

\subsection{Event selection and categorisation}

Events including exactly one signal lepton and no other baseline leptons are selected.
This requirement of exactly one lepton helps to suppress the
background due to the production of a $Z$ boson followed by its decay into a pair of leptons.
Since the events of interest are selected by single-lepton
triggers, the signal lepton is matched to the lepton selected by the trigger.
In addition, any event containing a $b$-tagged jet is vetoed to help suppress the
contribution from background processes involving top quarks.

Events containing a leptonically decaying $W$ boson are further targeted by
requiring $\ensuremath{p_{\mathrm{T}}^{\mathrm{miss}}\xspace}>30\,\GeV$ and the transverse mass of the $\ell$--\ensuremath{p_{\mathrm{T}}^{\mathrm{miss}}\xspace} system ($\mtw$)
to be greater than $50\,\GeV$.
\mtw is calculated as

\begin{equation}
\mtw = \sqrt{2\pt^{\ell}\ensuremath{p_{\mathrm{T}}^{\mathrm{miss}}\xspace}[1-\cos(\Delta\phi_{\ell\ensuremath{p_{\mathrm{T}}^{\mathrm{miss}}\xspace}})]},
\label{eq:selection:mtw}
\end{equation}

where $\Delta\phi_{\ell\ensuremath{p_{\mathrm{T}}^{\mathrm{miss}}\xspace}}$ is the azimuthal angle difference between the selected lepton and the \ensuremath{p_{\mathrm{T}}^{\mathrm{miss}}\xspace} two-vector.
While these selections affect the signal acceptance, they significantly reduce the
background arising from misidentified or non-prompt leptons, which can only
be estimated with substantial systematic uncertainties.

Only events with at least two signal
jets  are selected. The leading and subleading signal jets, defined to be the tag jets, must have
transverse momenta $\pt^{j_1}$ and $\pt^{j_2}$ larger than $80\,\mathrm{GeV}$
and $60\,\mathrm{GeV}$, respectively. This suppresses the
contribution of pile-up events, and increases the relative contribution of
\ewk \Wjj signal events. The latter is also enhanced by imposing a
$1\,\mathrm{TeV}$ threshold on the invariant mass of the dijet system,
$m_{jj}$. Furthermore, a large separation between the two tag jets is
required, and events where their absolute rapidity difference satisfies
$|\Delta y_{jj}|<2$ are rejected. This enables selection
of the \ewk \Wjj event topology and reduces the impact of
interference between the \ewk and \qcd \Wjj processes.

The requirements above define the
inclusive phase-space region, which contains contributions from both \ewk and \qcd \Wjj production. In VBF-$W$ production, the two forward tag jets are expected to recoil against the emitted $W$ boson. As a result, the charged lepton originating from the $W$ boson candidate is expected to be seen in the central part of the detector, while minimal hadronic activity is expected within the rapidity interval spanned by the two tag jets. These features are exploited to subdivide the inclusive region into four mutually exclusive regions and to define an \ewk \Wjj-enhanced signal region.

\begin{figure}[ht!]
\centering
\includegraphics[width=\textwidth]{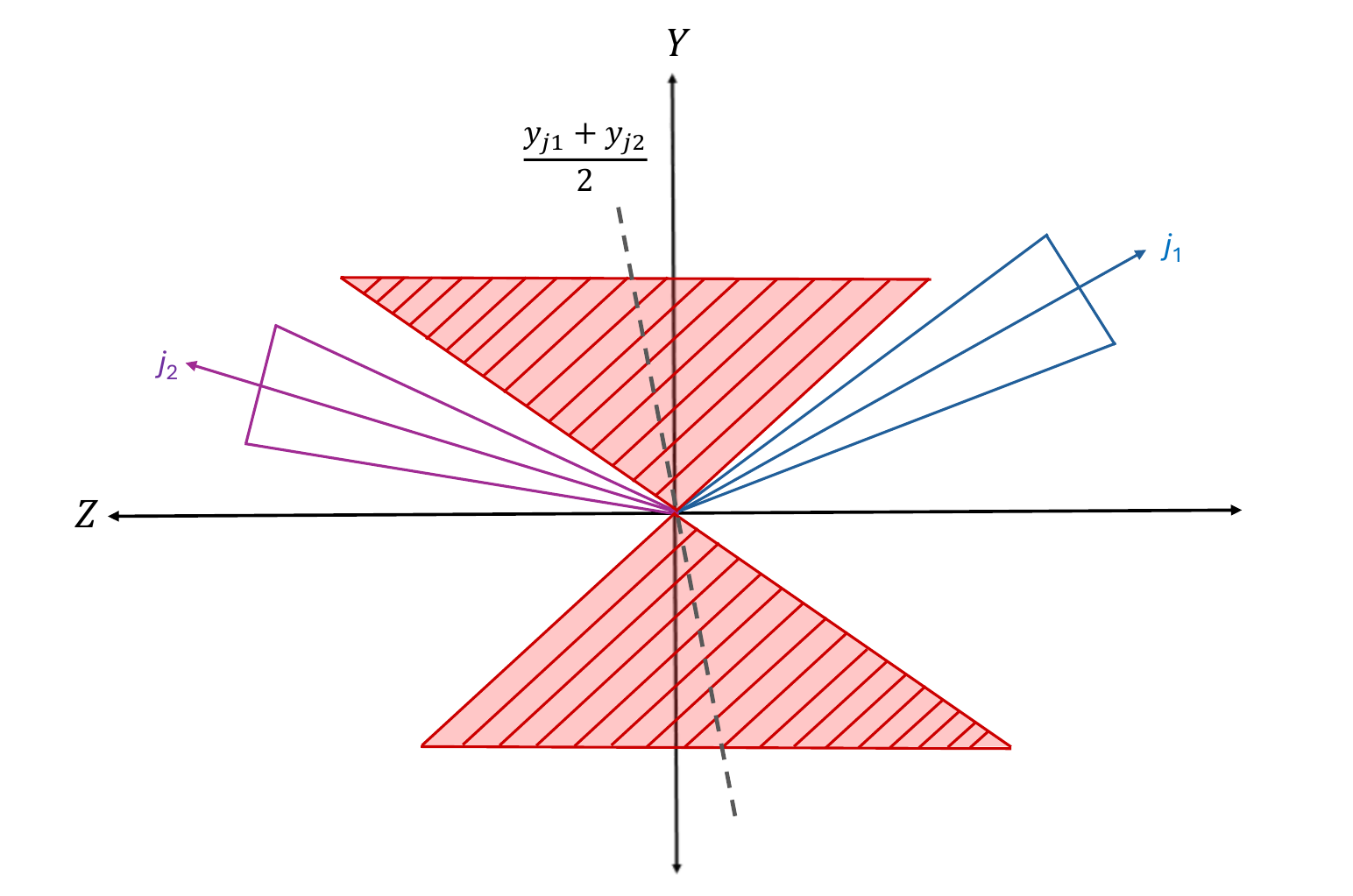}
\caption{General illustration of the central region, highlighted in red and defined with respect to the two tagged jets.
The dashed line represents the middle of the central region defined by $(y_{j_1} + y_{j_2})/2$.}
\label{fig:selection:centrality}
\end{figure}

The signal region (SR) is designed to be enriched in \ewk \Wjj events, while three
orthogonal control regions (CR0, CR1, CR2)
are constructed to constrain the dominant \qcd \Wjj background.
These regions are derived from the inclusive region by imposing further requirements on the centrality of the selected lepton and of any additional signal jets with respect to the two tag jets. The centrality of a lepton ($\ell$) or jet ($j$), denoted by $C_{\ell(j)}$, is defined as

\begin{equation}
C_{\ell(j)} = \left | \frac{y_{\ell(j)} - \frac{y_{j_1} + y_{j_2}}{2}}{y_{j_1} - y_{j_2}} \right |, \\
\label{eq:selection:centrality}
\end{equation}

where $y_{\ell(j)}$, $y_{j_1}$, and $y_{j_2}$ are the rapidities of the lepton (or jet) and the two tag jets, respectively. This variable quantifies the position of an object relative to the tag jets. Figure~\ref{fig:selection:centrality} highlights the central regions for both the lepton and any additional jets. Objects with $C_{\ell(j)}<0.4$ are classified as central with respect to the tag jets, while those with $C_{\ell(j)}\geq 0.4$ are forward. The number of central jets, $N_{\text{jets}}^{\text{cent}}$, is defined as the number of additional jets with $C_{j}<0.4$. The value of 0.4 is chosen in order to optimise the statistical significance of the \ewk \Wjj contribution in the signal region.

\begin{figure}[ht!]
\centering
\includegraphics[trim={5.1cm 0.75cm 6cm 1.5cm},clip,width=0.75\textwidth]{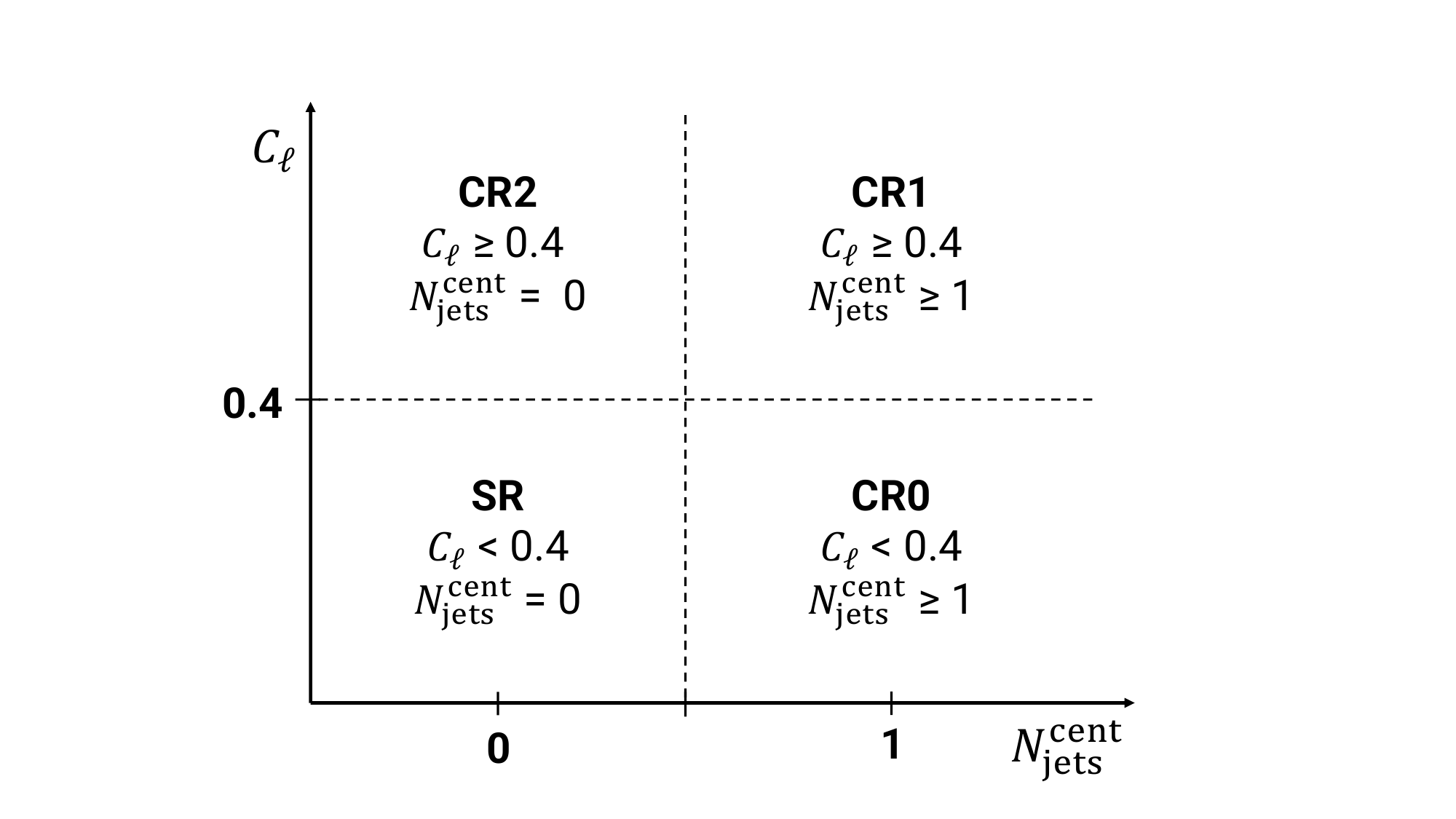}
\caption{Definition of the four orthogonal regions which
constitute the inclusive region. The different regions are distinguished
by specific requirements on the lepton centrality and number of central jets.}
\label{fig:selection:regions}
\end{figure}

As illustrated in Figure~\ref{fig:selection:regions}, the SR requires the selected lepton to be central ($C_{\ell} < 0.4$) and any additional jets to be forward ($N_{\text{jets}}^{\text{cent}} = 0$), while control region CR2 differs from the SR by requiring the lepton to be forward ($C_{\ell} \geq 0.4$). Control regions CR1 and CR0 are populated by events with at least one central jet and the selected lepton being forward or central, respectively.

%
\section{Background determination and signal extraction}
\label{sec:bkg-determination}
%

%
All background processes are simulated using MC samples.
However, data-driven estimates of backgrounds arising from misidentified or non-prompt leptons
and the dominant \qcd-induced \Wjj production process are needed to reduce systematic uncertainties.
The signal is then extracted by performing an extended binned maximum-likelihood fit.
Uncertainties, described in Section~\ref{sec:uncertainties}, are propagated to the final result by repeating the fit for each individual source of uncertainty. The effects of deviations from the nominal prediction are mitigated by the varied simulation being constrained to data via the likelihood fit.

\subsection{Multijet background estimate}

Events where the selected lepton originates from a jet or photon misidentified as a lepton, or from a hadron decay are collectively referred to as multijet background.
Such processes are difficult to model using simulation because of limited sample size and dependence on detector effects. As a result, data events from multijet-enriched kinematic regions are used to estimate the contribution from this category.

The multijet background contribution in each signal and control region is estimated by using two complementary data-driven methods:\ the template-fit method, which provides the nominal estimate and associated uncertainties, and the matrix method. The matrix method is used to validate the nominal multijet estimate as well as to refine the estimation of the systematic uncertainty of the multijet contribution. The description of the uncertainty treatment follows in Section~\ref{sec:uncertainties}.

The template-fit method, used in the previous ATLAS \ewk \Wjj measurement~\cite{STDM-2014-11}, relies on a set of data events selected using the same triggers as those described in Section~\ref{sec:samples}. However, the leading lepton is allowed to satisfy the looser, \texttt{Medium}, identification criteria and is explicitly required to fail the track or calorimeter isolation requirements imposed on signal-category leptons.
The resulting data sample is enriched in misidentified leptons from multijet events.
In addition, the $\ensuremath{p_{\mathrm{T}}^{\mathrm{miss}}\xspace}>\SI{30}{GeV}$ cut is removed to enhance the number of multijet events.
To form the multijet template, the contamination from prompt leptons is subtracted from binned data using the MC simulation.
Next, this multijet template, along with MC events containing signal-category leptons, is fitted to the data \ensuremath{p_{\mathrm{T}}^{\mathrm{miss}}\xspace} distribution in each signal and control region with the $\ensuremath{p_{\mathrm{T}}^{\mathrm{miss}}\xspace}>\SI{30}{GeV}$ requirement removed, while also allowing the \qcd \Wjj normalization to float. A binned profile likelihood fit~\cite{TRExFitter} is used to determine a normalisation factor which scales the multijet-enriched lepton template to its expected contribution in the sample with a signal category lepton.
For the electron channel, these normalisation factors are derived by performing the fitting procedure, binned in each kinematic observable under study within each signal and control region. This provides improved multijet modelling as a function of each observable. For the muon channel, where the multijet contribution is smaller, one normalisation factor is derived for each of the signal and control regions and applied uniformly to all bins of the multijet template for each observable.

The matrix method uses the efficiencies with which prompt and misidentified leptons, selected with the baseline lepton criteria, satisfy the signal lepton criteria.
High-purity samples of prompt lepton pairs from processes involving $Z$ boson decay are used to determine the prompt-lepton efficiencies.
The estimation of misidentified-lepton efficiency uses data and MC events in control regions enriched in misidentified leptons, defined by inverting the \ensuremath{p_{\mathrm{T}}^{\mathrm{miss}}\xspace} and \mtw cuts, and using a set of triggers with looser lepton-quality requirements.
In the electron channel, the selections are $\ensuremath{p_{\mathrm{T}}^{\mathrm{miss}}\xspace}<\SI{50}{GeV}$ and $\mtw < \SI{50}{GeV}$, and for the muon channel, where misidentified muons arise mainly from heavy-flavour decays which involve neutrinos, they become $\ensuremath{p_{\mathrm{T}}^{\mathrm{miss}}\xspace}<\SI{100}{GeV}$ and $\mtw<\SI{50}{GeV}$.
For each lepton flavour, the efficiencies of prompt and misidentified leptons are binned as functions of lepton $\pt$ and $|\eta|$.
Finally, the number of events with misidentified leptons falling into the signal category, constituting the multijet background, can then be estimated by solving a system of linear equations depending on both the number of events containing each category of lepton and the prompt and misidentified lepton efficiencies~\cite{EGAM-2019-01}.

Multijet background events represent about 6\% of the total background in the inclusive region, before constraining the \qcd \Wjj background and extracting the \ewk \Wjj signal. Nearly 85\% of the multijet background events contribute to the electron channel, with the others contributing to the  muon channel.

\subsection{\ewk and \qcd \Wjj constraints}
\label{sec:bkg-determination:lh}

\begin{figure}[h!]
\centering
\subfloat[]{
\includegraphics[width=0.50\textwidth]{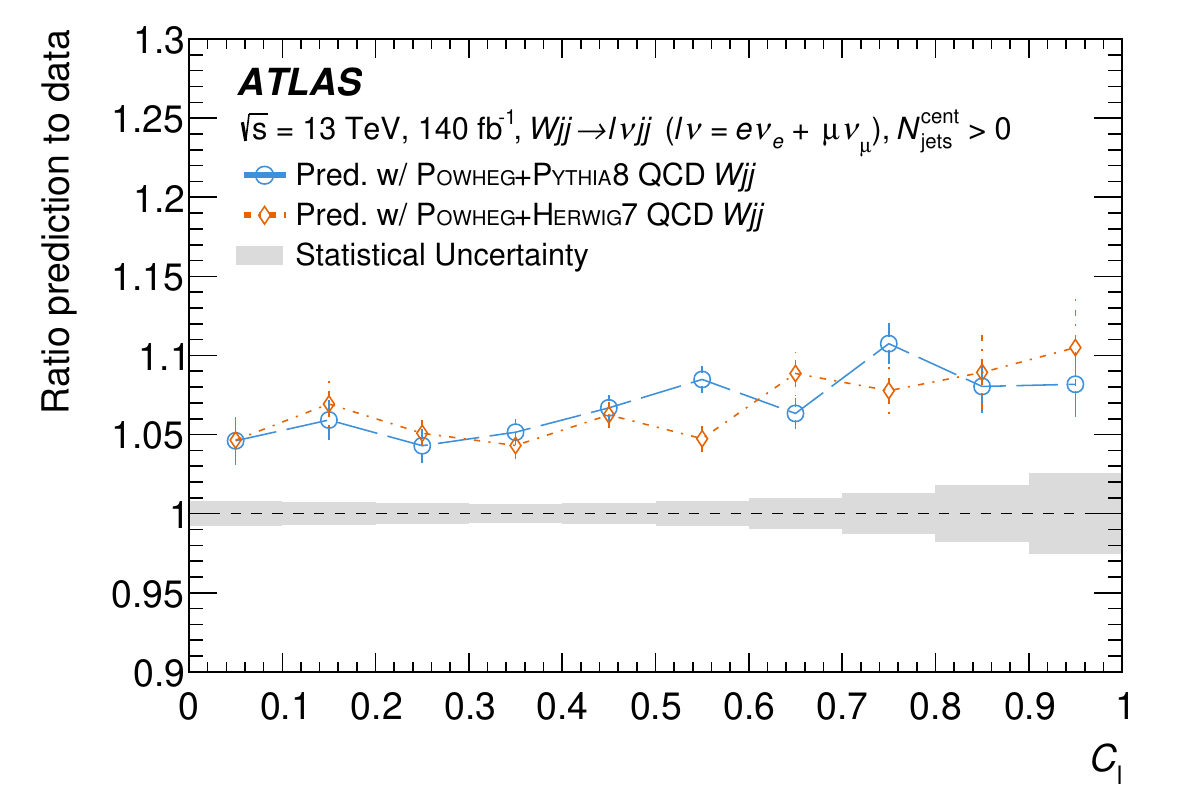}
}
\subfloat[]{
\includegraphics[width=0.50\textwidth]{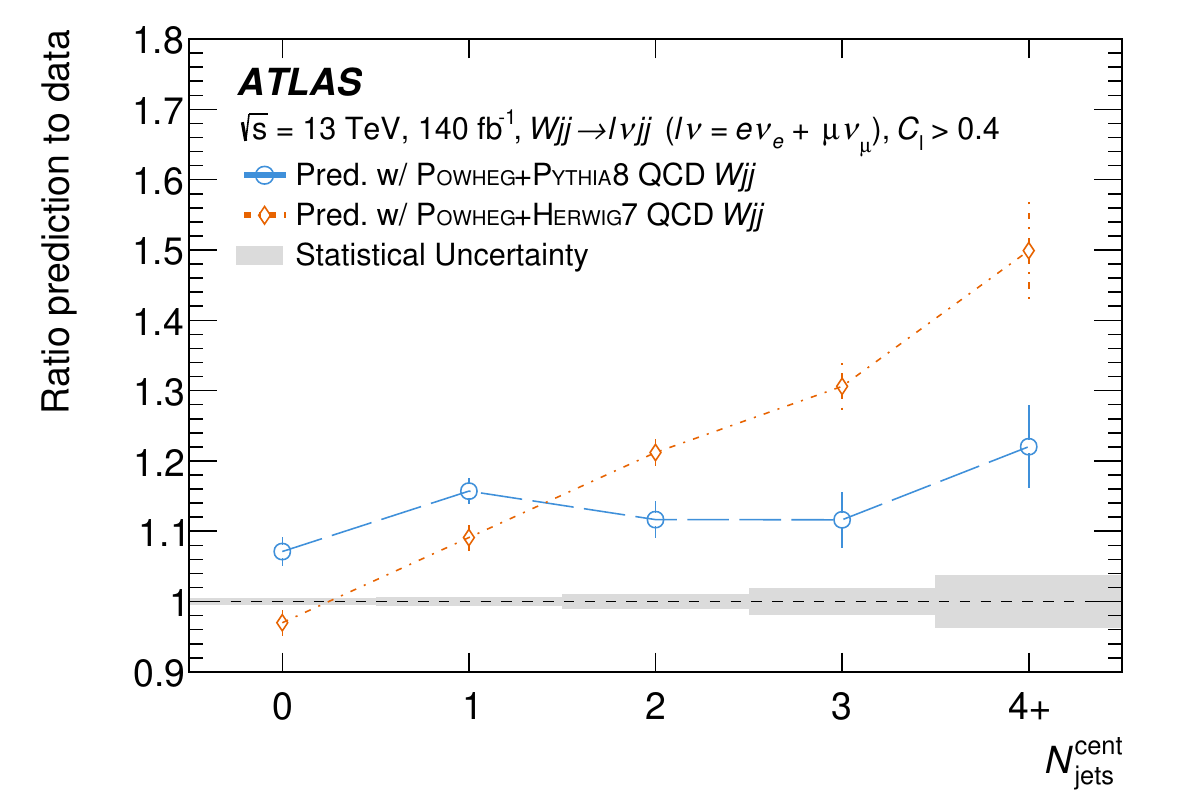}
}
\caption{Ratio of the total predictions to the data in the inclusive region using different generators for \qcd \Wjj events, as a function of (a) $C_{\ell}$ and (b) $N^{\text{cent}}_{\text{jets}}$. The \qcd \Wjj expectations from \POWPY[8] (circle) and \POWHER[7.2.1] (diamond) are considered. Error bars on the predictions represent the statistical uncertainty of the MC events, while the grey band represents the statistical uncertainty of the data. Comparisons are made prior to the signal extraction fit.}
\label{fig:DataPredRatiosVariousQCDWjj}
\end{figure}

After the multijet background is determined, the modelling of the signal and background contributions provides a consistent description of the data in both the electron and muon channels. As a result, the two channels are analysed simultaneously by adding their events  together.
Apart from the dominant \qcd \Wjj background, the other sources of background are the leptonic decay modes in \ewk and \qcd \zjets production, the $\tau$-lepton decay mode in \ewk and \qcd \wjets production, and top-quark and diboson production. They are estimated using MC simulation based on their theoretically predicted contributions and are collectively referred to as `Other~MC' in the following.
In contrast, the large expected background from \qcd \Wjj production is estimated using constraints from data in CR0, CR1, and CR2 to reduce systematic uncertainties.
These regions differ from the SR in the selection applied to $C_{\ell}$ and $N^{\text{cent}}_{\text{jets}}$, so the estimation of \qcd \Wjj production is impacted by the modelling of these observables in MC simulation. Figure~\ref{fig:DataPredRatiosVariousQCDWjj} considers two \qcd \Wjj models and shows the ratio of the total prediction (including the multijet estimate) to the data as a function of both $C_{\ell}$ and $N^{\text{cent}}_{\text{jets}}$. The comparisons only consider the ranges $N^{\text{cent}}_{\text{jets}} > 0$ and $C_{\ell} > 0.4$, respectively, omitting the \ewk \Wjj signal region.
The nominal \POWPY[8] prediction provides better overall agreement with data across $C_{\ell}$ and $N^{\text{cent}}_{\text{jets}}$,
whereas \POWHER[7.2.1], used as an alternative generator, gives a good description of the data in $C_{\ell}$, but overestimates it by up to 50\% at high $N^{\text{cent}}_{\text{jets}}$.

The analysis strategy is designed to simultaneously measure the \ewk \Wjj signal and determine the \qcd \Wjj background. This is achieved by using simulated-event templates, together with the data-driven estimate of the background from misidentified leptons.
The measurement is performed simultaneously for the six observables of interest and builds upon the techniques used in the analogous ATLAS \ewk \Zjj measurement~\cite{STDM-2017-27}.
The procedure utilises the three control regions to derive correction factors from the data to further constrain the shape and normalisation of the \qcd \Wjj background yield in the \ewk \Wjj signal region. It is performed as an extended binned maximum-likelihood fit, minimising the following negative log-likelihood function for each observable:
\begin{align}
- \ln L &= \sum_{r,i}\nu_{r,i} - \sum_{r,i}N^{\text{Data}}_{r,i}\ln \nu_{r,i}. \nonumber
\end{align}

The index $r$ spans the regions $r \in \left\{\text{SR},\text{CR0},\text{CR1},\text{CR2}\right\}$, and $i$ indexes the bins of each observable. Here, $N^{\text{Data}}_{r,i}$ represents the number of events in data, while $\nu_{r,i}$ is the event count predicted by the model in the $i$-th bin of region $r$. The fitted estimator $\nu_{r,i}$ is expanded as follows:
\begin{align}
\nu_{r,i} &= \mu^{\text{EW}}_{i}\nu^{\text{EW,MC}}_{r,i} + \nu^{\text{QCD}}_{r,i} + \nu^{\text{Other,MC}}_{r,i} + \nu^{\text{Multijet}}_{r,i}, \nonumber
\end{align}

where $\mu^{\text{EW}}_{i}$ denotes the \ewk \Wjj signal strength in bin $i$, and $\nu^{\text{EW,MC}}_{r,i}$ and $\nu^{\text{Other,MC}}_{r,i}$ are the predictions for the \ewk \Wjj signal and `other' background processes, respectively.
The term $\nu^{\text{Multijet}}_{r,i}$ represents the contribution from multijet background, as described above.
The \qcd \Wjj prediction, $\nu^{\text{QCD}}_{r,i}$, is decomposed into its region-dependent contributions according to:
\begin{align}
\label{eq:fit_qcd_est}
\begin{aligned}
\nu^{\text{QCD}}_{\text{CR0},i} &= \rho^{\text{central}}_{i}\nu^{\text{QCD,MC}}_{\text{CR0},i}, \\
\nu^{\text{QCD}}_{\text{CR1},i} &= \rho^{\text{forward}}_{i}\nu^{\text{QCD,MC}}_{\text{CR1},i},\\
\nu^{\text{QCD}}_{\text{CR2},i} &= \rho^{\text{forward}}_{i}f\left(x\right)\nu^{\text{QCD,MC}}_{\text{CR2},i},\\
\nu^{\text{QCD}}_{\text{SR},i} &= \rho^{\text{central}}_{i}f\left(x\right)\nu^{\text{QCD,MC}}_{\text{SR},i}.
\end{aligned}
\end{align}

The binned parameters $\rho^{\text{central}}_{i}$ and $\rho^{\text{forward}}_{i}$ are used to
determine the normalisation in each bin
of the \qcd \Wjj background in the central ($C_{\ell} < 0.4$) and forward ($C_{\ell} \geq 0.4$) lepton centrality regions, respectively.
These parameters allow better agreement between data and prediction in CR0 and CR1.
An additional correction, $f(x)$, is introduced to reduce the dependence on the MC model by allowing the normalisations in CR2 and SR to differ from those in CR0 and CR1. Here, $x$ denotes the observable being measured.
This function, mainly constrained in CR2, serves as a residual correction that ensures a proper match of the \qcd background to the data.

The function $f(x)$ is evaluated at the bin centres of the observable under study and is defined by two parameters. For $m_{jj}$, $|\Delta y_{jj}|$, and $\Delta\phi_{jj}$, $f(x)$ takes the form of a first-order polynomial, while for $p_{\text{T}}^{jj}$, $p_{\text{T}}^{j1}$, and $p_{\text{T}}^{\ell}$, it is defined as a logarithmic function. The latter choice is motivated by its provision of a better description of the steeply falling distributions that are characteristic of these momentum-related observables.

\begin{figure}[h!]
\centering
\includegraphics[width=0.49\textwidth]{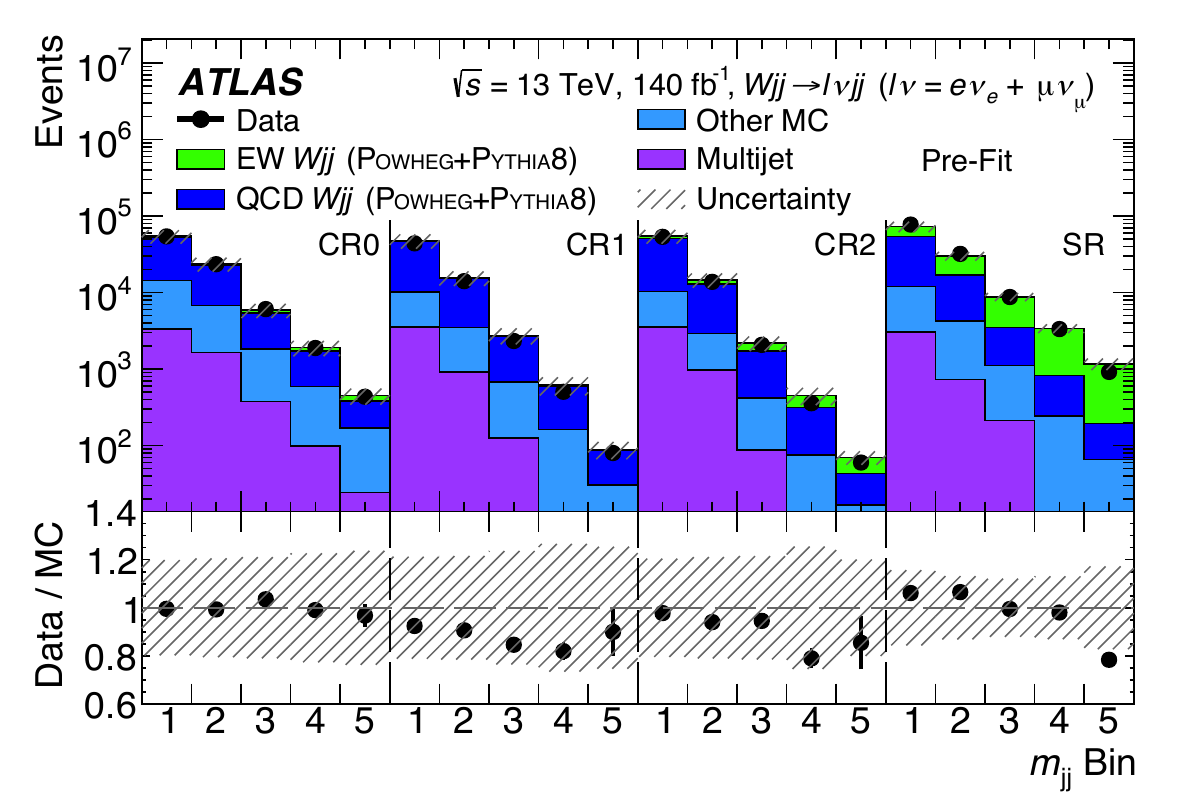}
\includegraphics[width=0.49\textwidth]{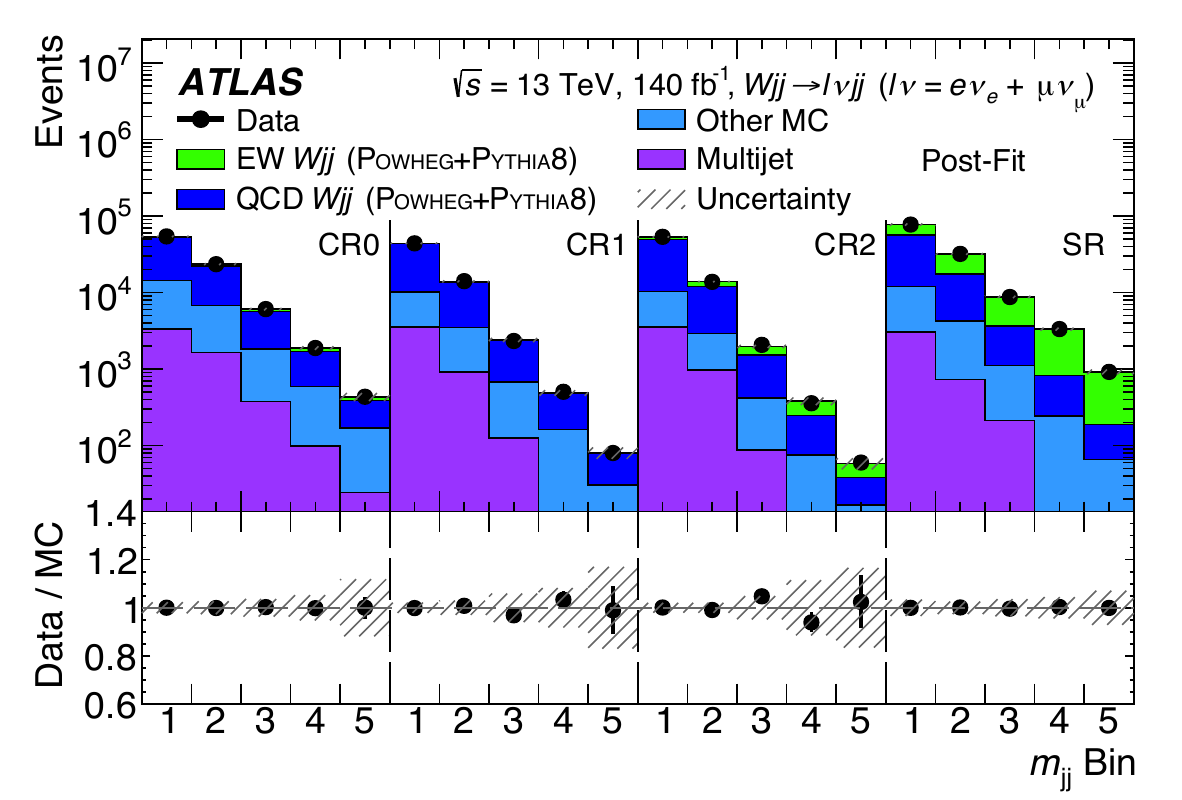} \\
\includegraphics[width=0.49\textwidth]{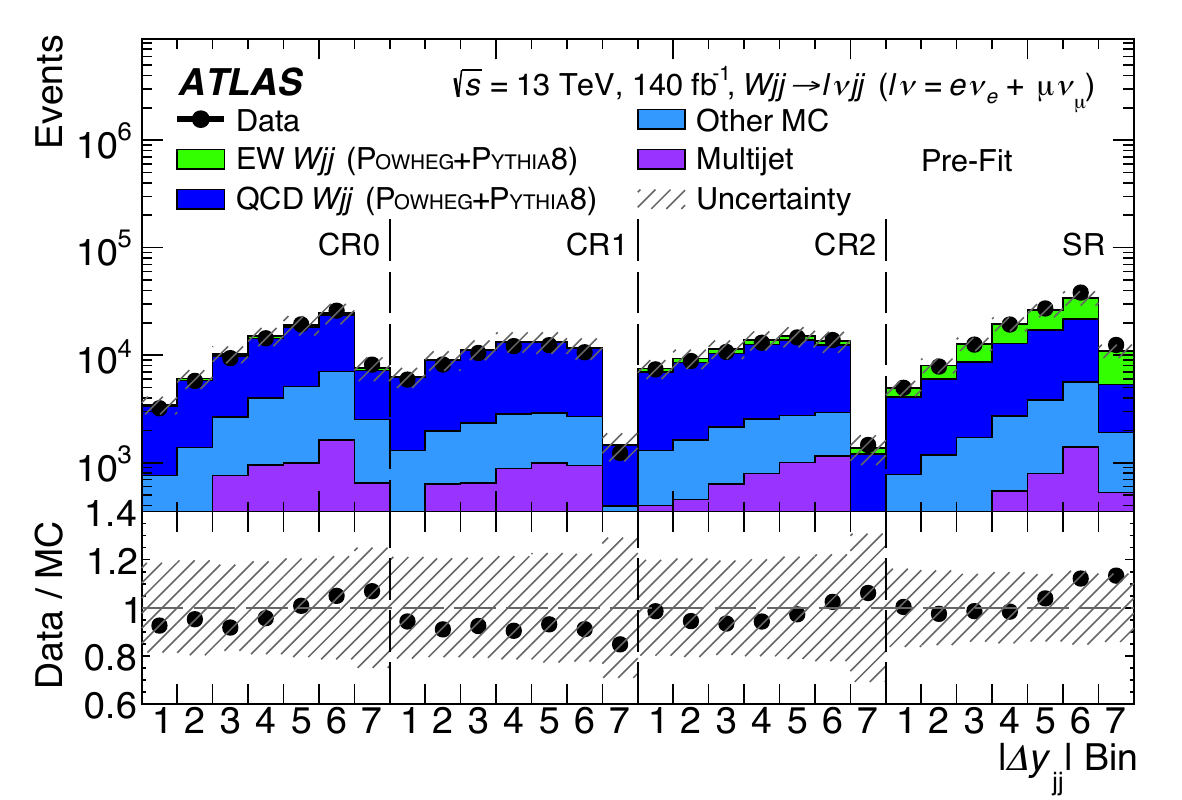}
\includegraphics[width=0.49\textwidth]{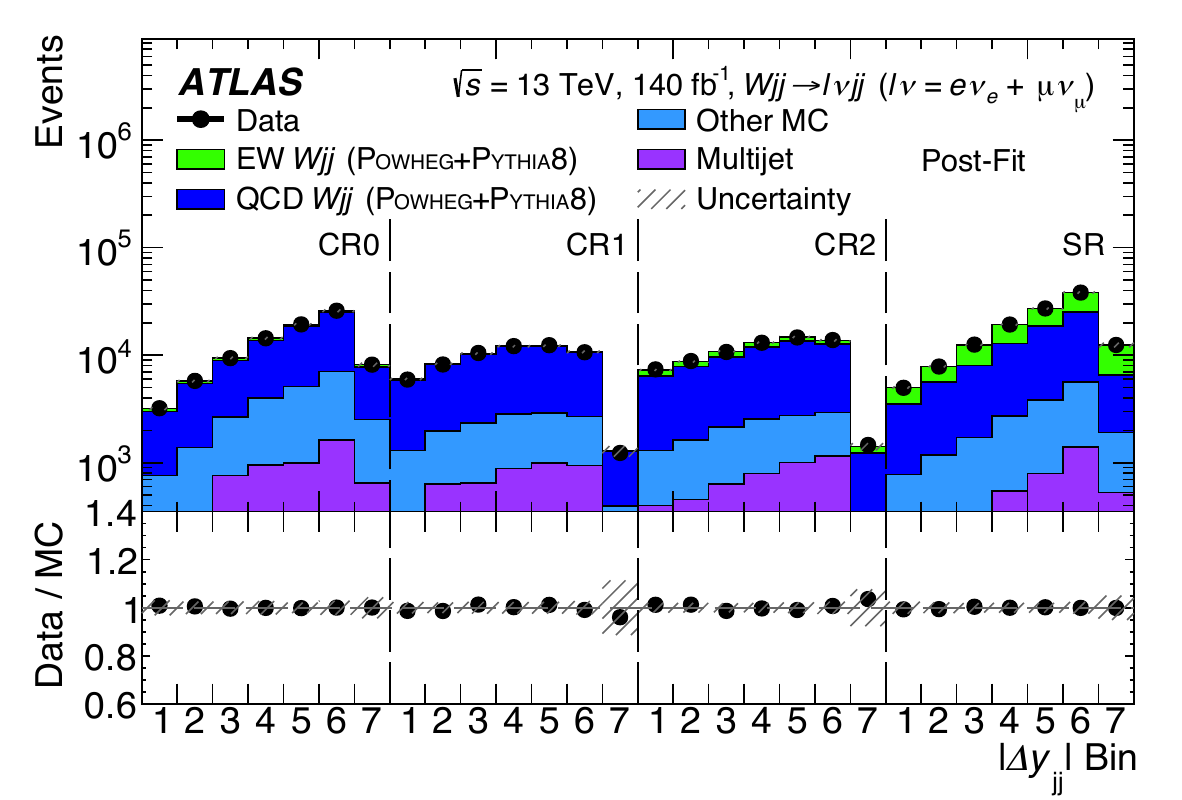} \\
\includegraphics[width=0.49\textwidth]{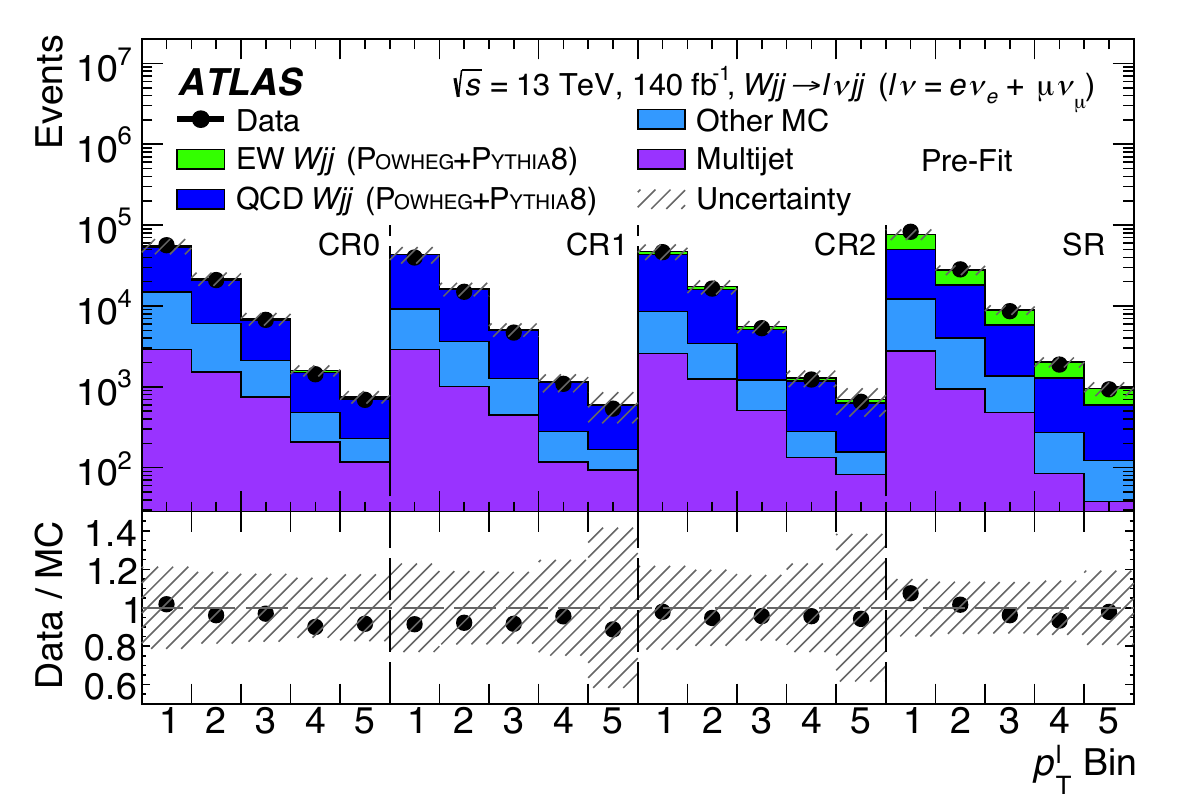}
\includegraphics[width=0.49\textwidth]{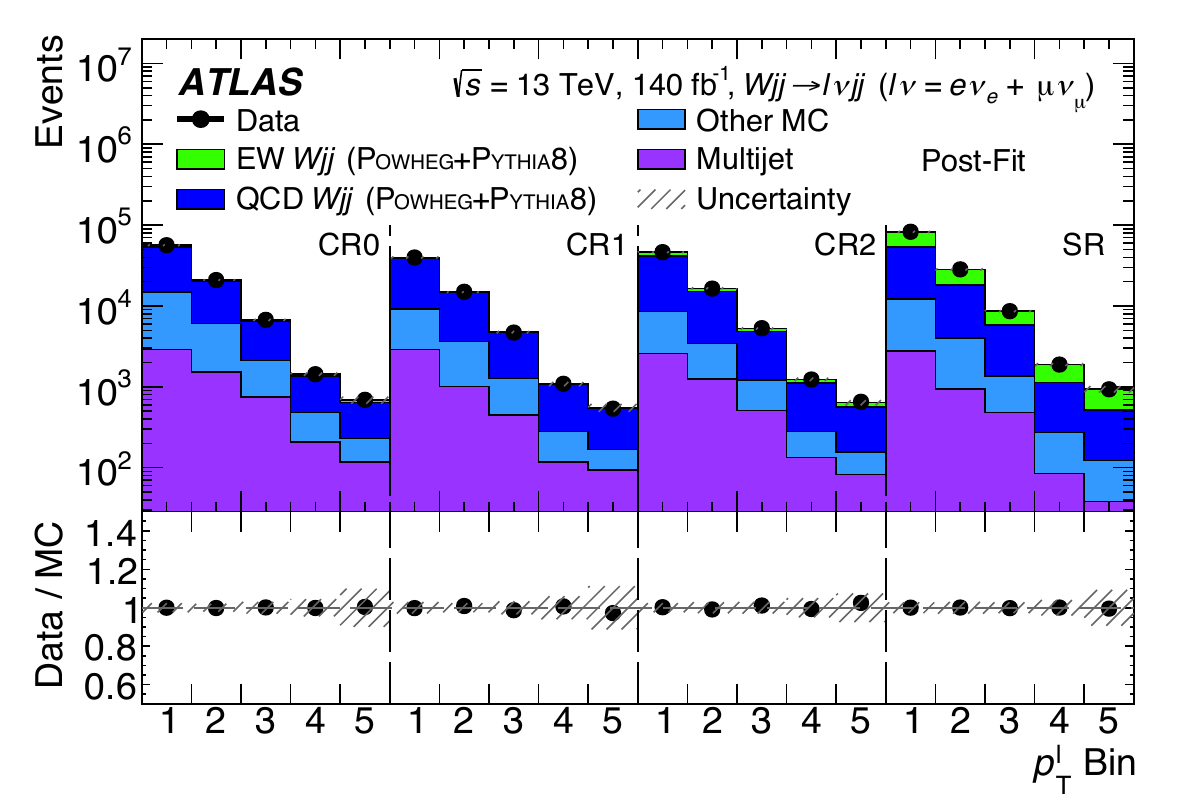}
\caption{Comparison of data and the prediction prior to (left) and following (right) the likelihood fit for $m_{jj}$ (top), $|\Delta y_{jj}|$ (middle), and $\pt^{\ell}$ (bottom). `Other MC' refers to \ewk{}+\qcd \zjets, \ewk{}+\qcd $W(\rightarrow\tau\nu)\!+\!\mathrm{jets}$, top, and diboson processes. The bin edges for $m_{jj}$, $|\Delta y_{jj}|$, and $\pt^{\ell}$ are $\left(1000, 1500, 2250, 3000, 4000, 8000\,\GeV\right)$, $\left(2.0, 2.6, 3.2, 3.8, 4.4, 5.0, 6.0, 8.3\right)$, and $\left(27, 80, 150, 250, 350, 1200\,\GeV\right)$, respectively.}
\label{fig:PrePostFitComparison}
\end{figure}

In the likelihood fit, the free parameters comprise $\mu^{\text{EW}}_{i}$, $\rho^{\text{central}}_{i}$, $\rho^{\text{forward}}_{i}$, and the two parameters defining $f(x)$, resulting in a total of $3N_{\text{bins}} + 2$ parameters fitted to $4N_{\text{bins}}$ data bins. Here, $N_{\text{bins}}$ denotes the number of bins in the observable under study in the four regions of interest (SR, CR0, CR1, and CR2).

Figure~\ref{fig:PrePostFitComparison} shows the pre-fit predictions and their comparison with data for $m_{jj}$, $|\Delta y_{jj}|$, and $\pt^{\ell}$, along with the corresponding post-fit results.
The fit significantly improves the level of agreement between the data and the predicted signal and backgrounds, with a significant decrease in the associated uncertainties. The uncertainties affecting the measurement are propagated to the final result by repeating the fit for each individual uncertainty source, described in Section~\ref{sec:uncertainties}.
Table~\ref{tab:PostFitEvntYields} summarises the post-fit yields after integrating the $m_{jj}$ distribution, including the \ewk and \qcd \Wjj contributions obtained from the signal extraction, the multijet estimate, and all other background processes.
In the SR, the contribution of `other' backgrounds to the total event yield is 11\%, composed of 4\% \qcd \zjets events, 3\% top-quark events, 2\% \wjets events with $\tau$-lepton decay, 1\% non-signal \ewk events, and 1\% diboson events. Multijet events constitute 3\% of the SR, while the \ewk and \qcd \Wjj contributions represent 36\% and 50\% of the SR, respectively.

\begingroup
\renewcommand{\arraystretch}{1.25}
\begin{table}[h!]
\centering
\caption{Post-fit event yields integrated from the $m_{jj}$ distribution for individual processes, their total, and the yield measured in data, in each region. `Other MC' refers to the \ewk{}+\qcd \zjets, \ewk{}+\qcd $W(\rightarrow\tau\nu)\!+\!\mathrm{jets}$, top, and diboson processes. The uncertainties provided represent the total uncertainty of the yields following the signal extraction.}
\label{tab:PostFitEvntYields}
\begin{tabular}{lllll}
\hline\hline
\multicolumn{1}{c}{\multirow{2}{*}{Process}}      &              \multicolumn{4}{c}{Event yields}      \\
&     \multicolumn{1}{c}{SR}    &    \multicolumn{1}{c}{CR0}    &    \multicolumn{1}{c}{CR1}    &    \multicolumn{1}{c}{CR2}     \\ \hline\hline

EW \Wjj                 &  ~~$ 44140 \pm 4190 $ & ~~$  4030 \pm  780 $ & ~~~~$   690 \pm  230 $ &  ~~$  6750 \pm  740 $  \\ \hline
QCD \Wjj                &  ~~$ 60750 \pm 6450 $ & $ 58500 \pm 5420 $ & $ 45780 \pm 4130 $ &  $ 49410 \pm 3490 $  \\ \hline
Other MC background               &  ~~$ 13610 \pm 1700 $ & $ 18150 \pm 2940 $ & ~~$  9940 \pm  2472 $ &  ~~$  9090 \pm  1760 $  \\ \hline
Multijet                &  ~~~~$  4020 \pm 1730 $ & ~~$  5480 \pm 2130 $ & ~~$  4630 \pm 1740 $ &  ~~$  4640 \pm 2070 $  \\ \hline
\hline
Total SM expectation         & $122520 \pm 3380 $ & $ 86160 \pm 2070 $ & $ 61040 \pm 1350 $ &  $ 69890 \pm 1290 $  \\ \hline
Data                    &  122507            &   86163            &   61038            &    69877             \\ \hline\hline

\end{tabular}
\end{table}
\endgroup


%
\section{Unfolding to particle level}
\label{sec:unfolding}
%

The \ewk \Wjj differential cross-sections at particle level are determined in
the signal region by performing an iterative Bayesian unfolding (IBU)
procedure~\cite{DAgostini:1994fjx, Adye:2011gm} on the measured \ewk \Wjj yields, removing effects from detector response. Each observable distribution of interest
is unfolded separately and requires the construction of a response matrix which maps
the yields per bin at particle level to those at the detector level.
The response matrix is constructed by using the \ewk \Wjj prediction from \POWPY[8].

The definition of the particle-level fiducial phase space closely follows
the event selection outlined in Section~\ref{sec:selection}, with the
differences being mostly in the criteria applied to analysis objects.
Details of the motivation for the particle-level object definitions can be found in Ref.~\cite{ATL-PHYS-PUB-2015-013}.
Such particles are required to be stable, with $c\tau > \SI{10}{mm}$.
The selected particle-level leptons are `dressed' by adding the four-momenta of photons ($\gamma$) not originating from hadron decays and satisfying
$\Delta R(\ell, \gamma)<0.1$. The resulting dressed leptons are required to have $|\eta|<2.5$, with no particle-level isolation requirement applied.
Particle-level jets are constructed using the anti-$k_t$
algorithm and considering all interacting final-state particles other than
dressed leptons not originating from hadron decays. Selected jets must fulfil $\pt>30\,\GeV$ and $|\eta|<4.5$.
A particle-level overlap removal procedure is applied to leptons and jets in order
to mimic the detector-level reconstruction. For $\Delta R(\ell, \mathrm{jet})<0.2$,
the lepton is retained, while the jet is retained for $0.2<\Delta R(\ell, \mathrm{jet})<0.4$.
Events are required to include one lepton with $\pt>27\,\GeV$, and no additional one with $\pt>10\,\GeV$.
Each event must contain at least two jets, with the leading and subleading ones having \pt exceeding
$80\,\gev$ and $60\,\gev$, respectively.
The \ensuremath{p_{\mathrm{T}}^{\mathrm{miss}}\xspace} at particle level is computed as the magnitude of the sum of the transverse momenta of
all stable, non-interacting particles, and must be greater than $30\,\GeV$.
The transverse mass of the lepton and \ensuremath{p_{\mathrm{T}}^{\mathrm{miss}}\xspace}, calculated as in Eq.~(\ref{eq:selection:mtw}), has to exceed
$50\,\GeV$, while events in which the two leading jets have an absolute rapidity difference below $2$ or an invariant mass below $1\,\TeV$ are
discarded.
Events with jets which contain `ghost-associated'~\cite{Cacciari_2008} $b$-hadrons are also vetoed.
The SR is constructed in the same way at particle level and reconstruction level, requiring exactly one central lepton and no central jets.
Table~\ref{tab:unfolding:table_fidSel} summarises the definition of the particle-level fiducial SR.

\begin{table}
\centering
\caption{Summary of selections defining the fiducial signal region at
particle level.}
\label{tab:unfolding:table_fidSel}
\begin{tabular}{ l l }
\hline\hline& \\[-2.ex]
Region Name & Requirements \\[0.5ex]
\hline\hline& \\[-2.ex]
Preselection & $1$ electron or muon with:
$\pt > 27\,\GeV$, $|\eta| < 2.5$ \\[0.5ex]
& No additional electron or muon with:
$\pt > 10\,\GeV$, $|\eta| < 2.5$ \\[0.5ex]
& At least 2 jets: $\pt > 30\,\GeV$, $|\eta| < 4.5$ \\[0.5ex]
\hline& \\[-2.ex]
Fiducial SR   & $\ensuremath{p_{\mathrm{T}}^{\mathrm{miss}}\xspace}>30\,\gev$ (stable, non-interacting particles) \\[0.5ex]
& $\mtw>50\,\gev$ \\[0.5ex]
& $\pt^{j_1}>80\,\gev$ \\[0.5ex]
& $\pt^{j_2}>60\,\gev$ \\[0.5ex]
& $|\Delta y(j_1,j_2)| > 2.0$ \\[0.5ex]
& $m_{jj}>1\,\tev$ \\      [0.5ex]
& No event with a `ghost-associated' $b$-tagged jet \\[0.5ex]
& $N_{\ell}^{\text{cent}} = 1, N_{\text{jets}}^{\text{cent}} = 0$
(see Eq.~(\ref{eq:selection:centrality})) \\[0.5ex]
\hline\hline& \\[-2.ex]
\end{tabular}
\end{table}

The binning for each observable and the number of unfolding iterations were optimised in the IBU method.
The former was chosen in order to mitigate migrations of events between bins at
particle level and at reconstruction level, and the latter was tuned to minimise the total uncertainty of the measurement and be close enough to the procedure's point of convergence.
The bin boundaries for each observable's distribution are chosen to
maintain a purity\footnote{
The purity is defined as
$P_i=N^{\text{reco$_i$, particle$_i$}}/N^{\text{reco$_i$}}$,
where $N^{\text{reco$_i$, particle$_i$}}$ represents the number of events
reconstructed at detector level in bin $i$, where the corresponding
particle-level event is also in bin $i$, and $N^{\text{reco$_i$}}$ represents
the total number of reconstructed detector-level events in bin $i$.
}
higher than 60\% in most of the bins.
The number of iterations is chosen to be two for each
observable, as this was found to be a suitable compromise that reduces the
dependence on MC simulation without significantly increasing the total uncertainty in the
unfolded distributions.


%
\section{Uncertainties}
\label{sec:uncertainties}
%

Uncertainties in the measurement of the \ewk \Wjj cross-section originate
from finite event yields in the data and simulation used, and from experimental and
theoretical sources of systematic uncertainty. All considered uncertainties are
propagated through the entire analysis, including the parameters in
Eq.~(\ref{eq:fit_qcd_est}) and the signal prediction in the
response matrix.
The complete set of uncertainties used in determining the unfolded
differential cross-sections are detailed below, together with
their characteristic values.

The uncertainty due to finite event yields in data and in simulation is
computed via pseudo-experiments~\cite{ATL-PHYS-PUB-2021-011}.
For each dataset used, a set of 1000 pseudo-datasets are
constructed, where each event is weighted according to
a Poisson distribution with a mean of unity. For a given distribution, the statistical uncertainty is taken as the RMS of the results obtained from these pseudo-datasets. The statistical uncertainty due to the data is typically below 12\% across almost all observables. However, in regions with limited event yields, such as at central $\Delta\phi_{jj}$values and high values of $\pt^{\ell}$, the uncertainty increases to 24\% and 18\%, respectively. Similarly, the statistical uncertainty due to the finite number of MC events is generally below 10\% across the observables, except at low values of $|\Delta y_{jj}|$ and again in the central $\Delta\phi_{jj}$ region, where it increases to 12\% and 16\%, respectively.

The integrated luminosity for the data collected in 2015, 2016, 2017 and 2018 is computed using
the LUCID-2 detector for the primary luminosity measurements, complemented by
input from the inner detector and calorimeters. It is estimated with
an uncertainty of \runtwolumiuncertainty~\cite{DAPR-2021-01}.

The dominant experimental uncertainties are related to
the calibration of jets, namely the jet energy scale and resolution~\cite{JETM-2018-05}, the
jet-vertex taggers~\cite{PERF-2014-03,PERF-2016-06}, and the $b$-tagging efficiency~\cite{FTAG-2019-07}. Additionally, an
uncertainty in \ensuremath{p_{\mathrm{T}}^{\mathrm{miss}}\xspace} takes into account the uncertainties in the aforementioned
objects' momenta as well as scale and resolution uncertainties in the track-based
soft term~\cite{JETM-2020-03}. The combined jet and \ensuremath{p_{\mathrm{T}}^{\mathrm{miss}}\xspace} uncertainties have an effect of up to 23\% at central values of $\Delta\phi_{jj}$,
while the remaining observables exhibit effects of up to 13\% in the tails of their spectra.
Other sources of experimental uncertainty are the modelling of
pile-up, which is varied to account for the uncertainty in the ratio of predicted
to measured inelastic proton--proton cross-sections~\cite{STDM-2015-05}, and the calibration of the electron or muon, which affects their
identification and isolation efficiencies~\cite{EGAM-2021-01,MUON-2018-03},
their energy resolution and momentum scales~\cite{EGAM-2021-02,MUON-2022-01},
and their trigger efficiencies~\cite{TRIG-2018-01,TRIG-2018-05}. These uncertainties have an effect that is largest for $\pt^{\ell}$,
reaching 7\% at high values, whereas for jet-related observables their effect is less than 4\%.

Theoretical uncertainties in the total MC prediction arise mainly from the modelling of \qcd \Wjj production.
The parton shower uncertainty is dominant, and is estimated
independently for the two main processes, namely \ewk and \qcd \Wjj
production simulated by \POWHEG.
It is evaluated as the effect of applying the parton showering algorithm in \Herwig[7.2.1] as an alternative to the one in \PYTHIA[8],
using samples described in Section~\ref{sec:samples}.
For \ewk \Wjj production, the effect is
less than 5\% across the six observables.
In contrast, the largest effects of the parton shower variation for \qcd \Wjj production are observed at low $|\Delta y_{jj}|$ and high $\pt^{\ell}$, reaching 34\% and 30\%, respectively, while the other observables exhibit relatively smaller effects of up to 14\%.

To estimate the uncertainty due to missing higher-order \qcd corrections, the
renormalisation and factorisation scales are raised and lowered independently
by factors of two, avoiding opposite variations.
The impact on \ewk and \qcd \Wjj production, as well as the other MC
backgrounds when not negligible, is treated as uncorrelated across the
simulated samples but correlated between the regions analysed.
The envelope of the effects of the different scale variations is then used as an uncertainty.
These uncertainties are found to be less than 2\% and 4\% for the \ewk \Wjj
and other MC process variations, respectively, with the high lepton-\pT region being affected the most.
The effects of PDF uncertainties are estimated by using a reweighting procedure at the
matrix-element level for 100 variations of the PDF set used for each
simulated event sample, and by comparing the use of two alternative PDF sets,
namely CT14~\cite{Dulat:2015mca} and MMHT2014~\cite{Harland-Lang:2014zoa}.
The uncertainties in the strong coupling constant \alphas cover the
experimental errors in its determination and the
truncation at a fixed order in perturbation theory. Their effects are computed by
using the same set of PDFs evaluated with two different values of \alphas,
and are added in quadrature to the effects of PDF uncertainties. This combined `PDF\,+\,\alphas'
uncertainty is less than 6\% for most of the observables, with larger effects
observed at high lepton $\pt$ and in the central $\Delta\phi_{jj}$ region,
reaching 8\% and 12\%, respectively.
Interference between \ewk and \qcd \Wjj production, which is treated as
an uncertainty in the \ewk \Wjj signal, is assessed by comparing the
results considering signal with or without the interference term. This
uncertainty has an effect of up to 10\% at
low values of $|\Delta y_{jj}|$, and of up to 6\% at high values of jet $\pt$.
For the other observables considered, this uncertainty is found to be
less than 2\%.

\begin{figure}
\centering
\subfloat[]{
\includegraphics[width=0.48\textwidth]{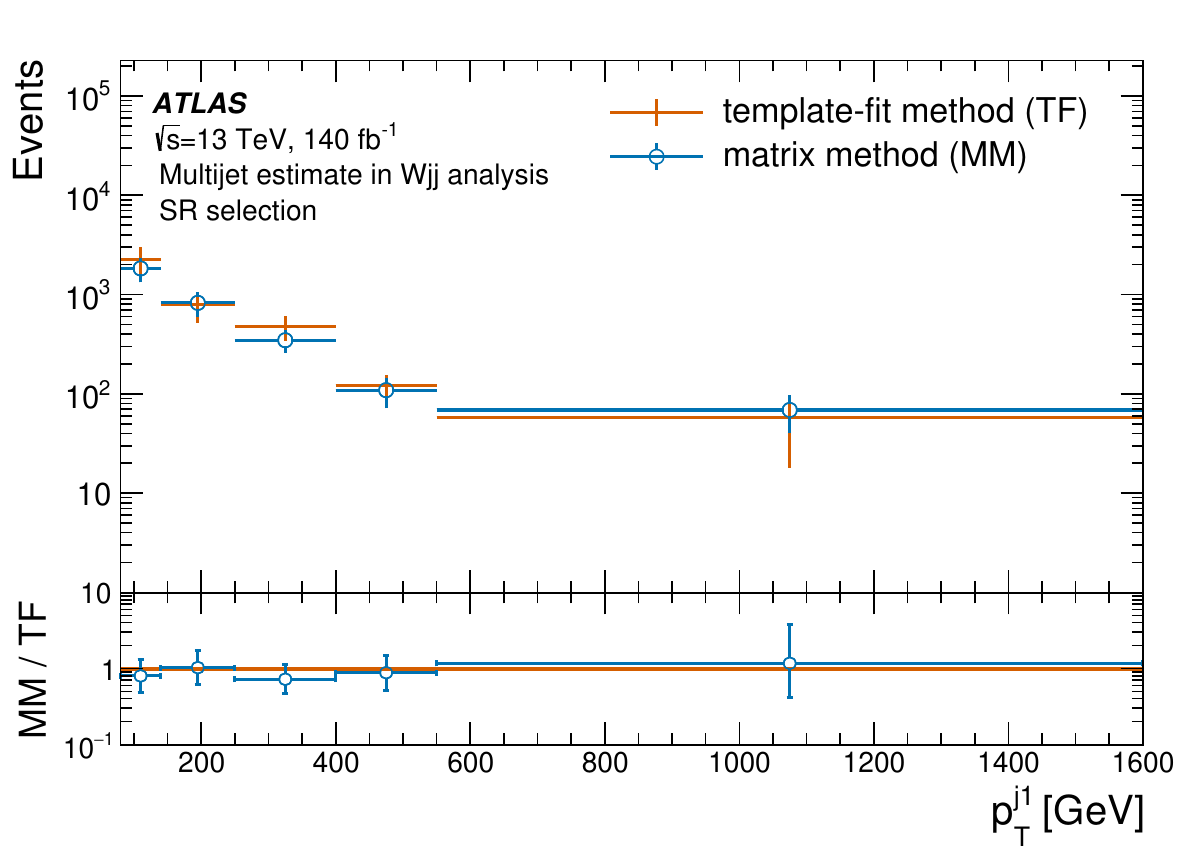}
}
\subfloat[]{
\includegraphics[width=0.48\textwidth]{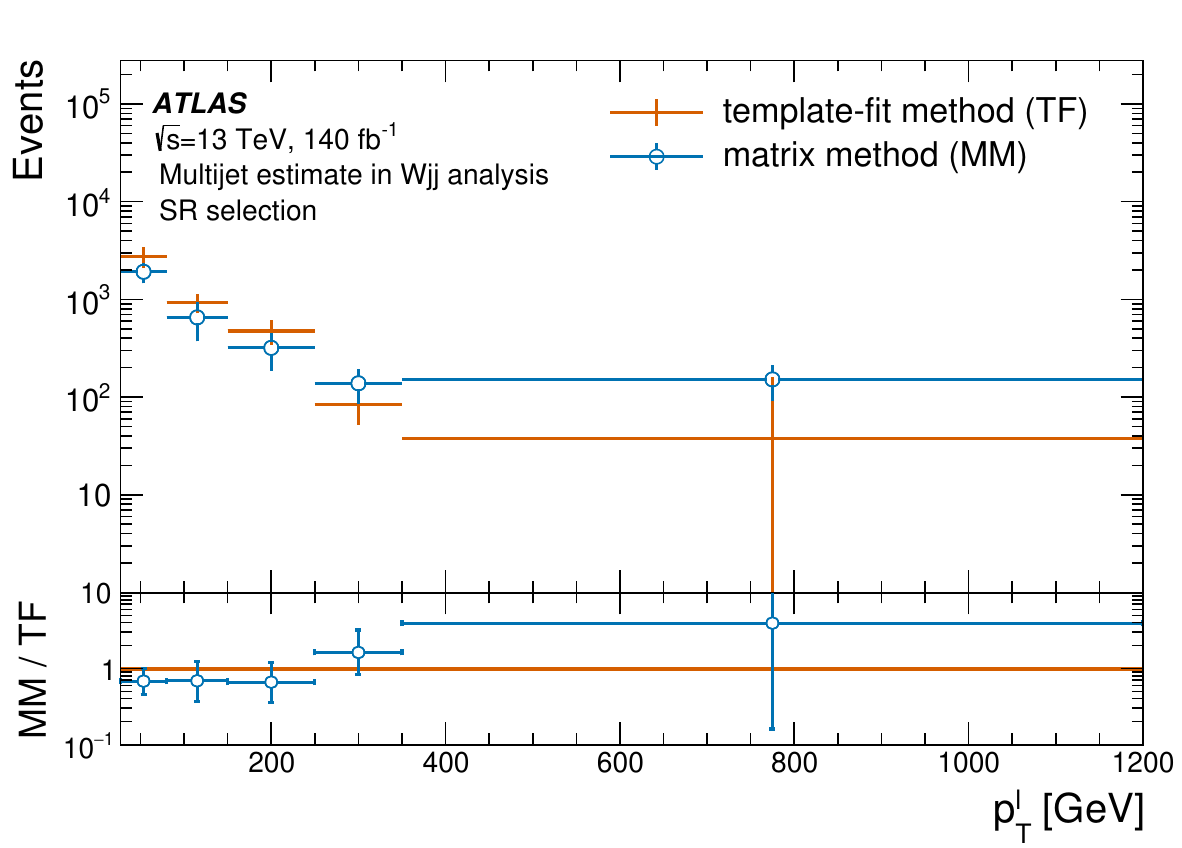}
}
\caption{Comparison of multijet estimates determined by the template-fit (TF) and the matrix method (MM), for (a) the leading-jet \pt and (b) lepton $\pt$ in the SR. Good agreement is generally observed within statistical and systematic uncertainties. The uncertainties displayed in this figure correspond to the sum in quadrature of the statistical uncertainty and method-based systematic uncertainties, for each method respectively.}
\label{fig:syst:cmp_multijet}
\end{figure}

The determination of the multijet background is dominated by systematic
uncertainties originating from the estimation method. For the template-fit
method, these are assessed by varying its defining parameters,
namely the choice of observable used for fitting, the fit range, and the bin
boundaries, as well as the granularity of the multijet normalisation factors
as a function of each unfolded observable. In addition, all the reconstruction
and theoretical uncertainties affecting the simulated processes with a
prompt lepton are propagated to the multijet estimate.
All uncertainties are evaluated separately for the electron and muons channels,
with the relative total uncertainty amounting to 30\%--50\% across the signal
and control regions, dominated by the electron channel yield. Analogous uncertainties are
independently derived for the alternative matrix method. The template-fit-method and
matrix-method multijet estimates are found to agree within their respective
uncertainties for nearly all of the measured phase space, as depicted
in Figure~\ref{fig:syst:cmp_multijet} for the leading-jet and lepton $\pt$ in
the SR.

The compatibility of the two multijet background estimates is further tested by comparing their impact on post-fit \ewk \Wjj distributions. In this comparison, the matrix-method estimate is used in place of the template-fit-method estimate and the resulting post-fit \ewk \Wjj predictions are checked for compatibility, given their uncertainties.
In the bins of observables where the difference is not covered by the total post-fit uncertainty, the difference between the central values from the two methods is taken as the total uncertainty in the nominal multijet estimate.
This increases the relative uncertainty of the multijet
background in some sparsely populated parts of the signal region to
as much as 12\% at high $m_{jj}$ values.

Deriving the \qcd \Wjj estimate directly from the data
suggests the control regions chosen to receive the $\rho_{i}$ and $f(x)$ corrections could be switched without changing the results.
To verify this, the likelihood fit was reparameterised such that CR0 supplied the residual correction $f(x)$, while the $\rho_{i}$ were derived from CR1 and CR2.
The resulting \ewk \Wjj measurements were found to be consistent with those from the nominal configuration, given the statistical uncertainty. In addition, alternative functional forms for $f(x)$ were tested, including a second-order polynomial and an exponential function. These variations also yielded results consistent with those obtained using the nominal parameterisation, given the statistical uncertainty. Therefore, no additional systematic uncertainty is assigned to the signal extraction method.

The unfolding procedure propagates statistical, reconstruction, and theoretical uncertainties from the input post-fit \ewk yields. The response matrix is also updated to include the corresponding uncertainty variations. However, it includes additional sources of uncertainty, due to the method itself.

The sensitivity of the unfolding to distribution shape differences between the data and the \ewk \Wjj prediction from \POWPY[8] is assessed by comparing the MC generator-level and unfolded detector-level simulations. Before this comparison, both predictions are reweighted using weights computed such that the detector-level yields match the data.
In addition, a potential bias arising from missing dependence on variables other than the one being unfolded is similarly evaluated using weights derived from  distribution shape differences between the particle-level \ewk \Wjj predictions from \POWPY[8] and \Sherpa[2.2.11].
The uncertainties from these two sources are attributed to the unfolding method and referred to as unfolding uncertainties.
They are usually small, being less than 7\% across the majority of measured observables, although sizeable effects of up to 11\% and 13\% are observed at higher values of $|\Delta y_{jj}|$ and $m_{jj}$, respectively.
These larger effects are due to out-of-fiducial migrations.


%
\section{Measured cross-sections}
\label{sec:results}
%

\begin{figure}[h]
\centering
\subfloat[]{
\includegraphics[width=0.5\textwidth]{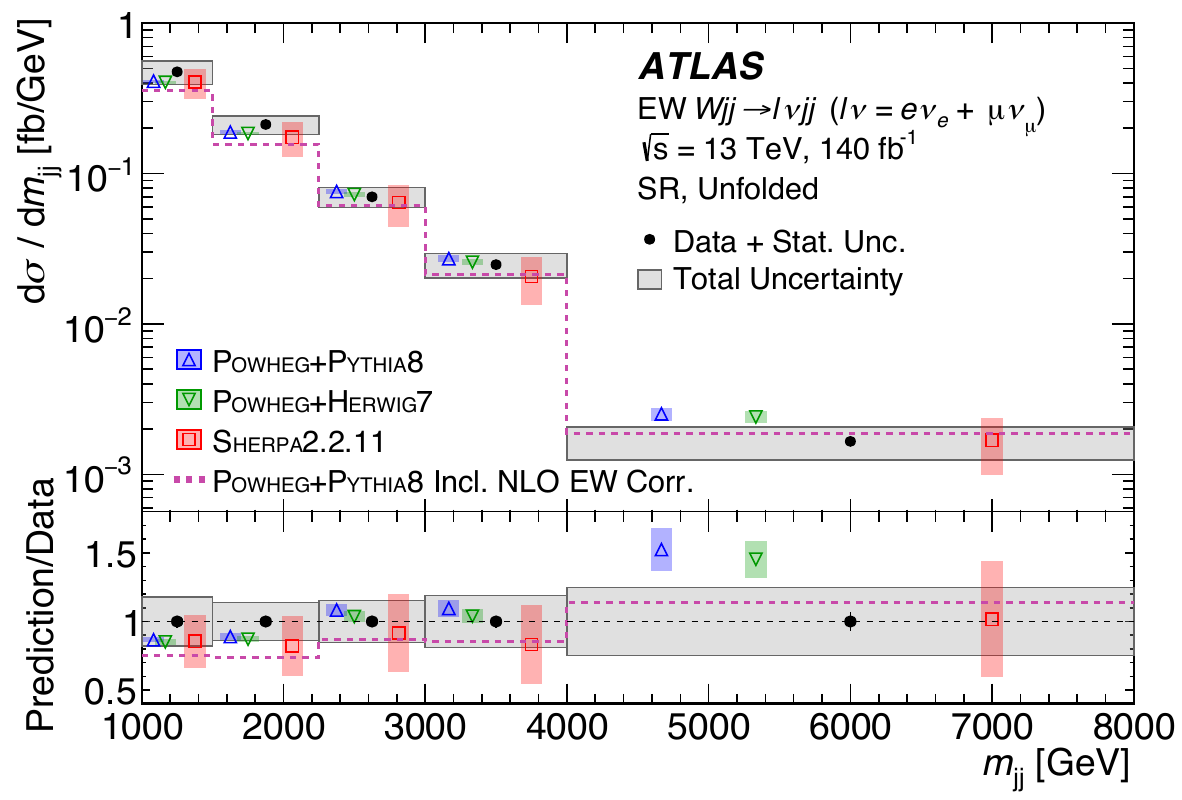}
}
\subfloat[]{
\includegraphics[width=0.5\textwidth]{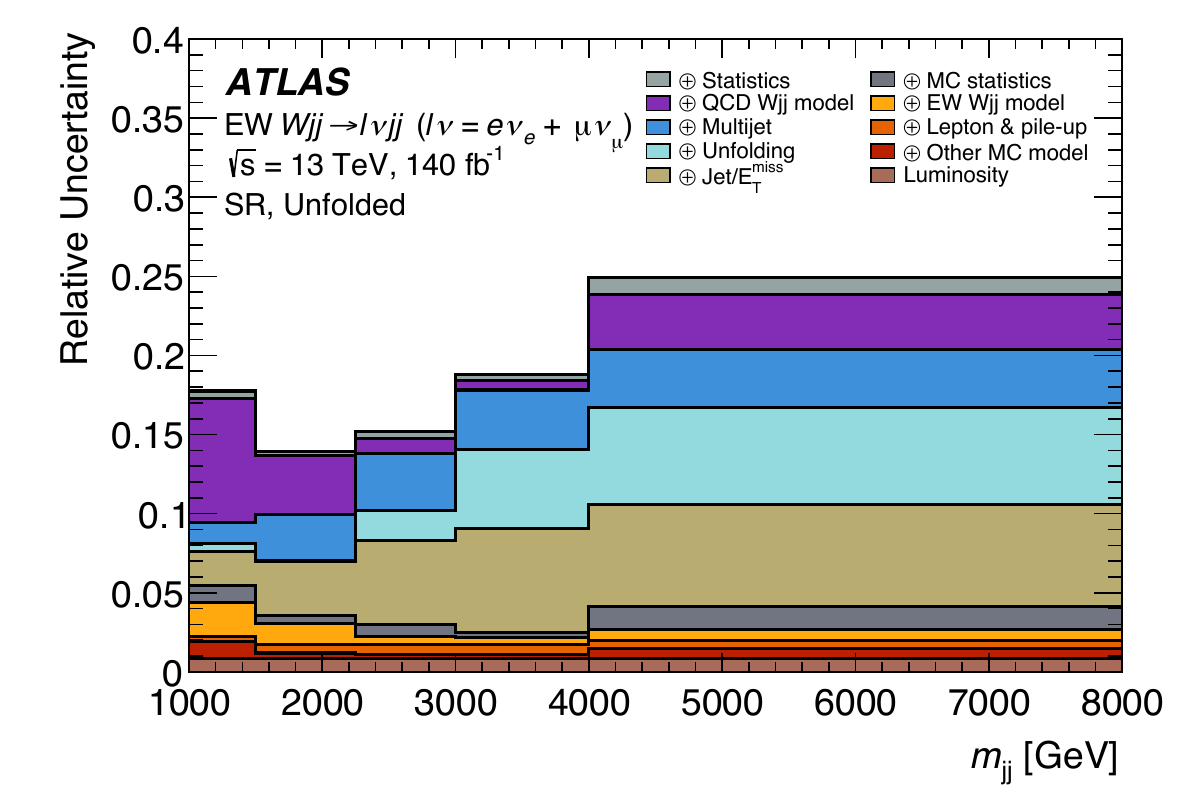}
}
\caption{(a) Unfolded differential cross-section for \ewk \Wjj production and (b) associated statistical, experimental, and theoretical uncertainties
as a function of $m_{jj}$.
Measurements are compared with predictions at LO and NLO accuracy
in perturbative \qcd from \Sherpa[2.2.11] and \POWHEG{}, respectively. The
latter is showered with either \PYTHIA[8] or \Herwig[7.2.1].
Theoretical uncertainties in the SM expectations are included. Uncertainties on the measurement are grouped into categories and summed in quadrature ($\oplus$) to yield the total uncertainty.
}
\label{fig:results:unf_mjj}
\end{figure}

\begin{figure}[h]
\centering
\subfloat[]{
\includegraphics[width=0.5\textwidth]{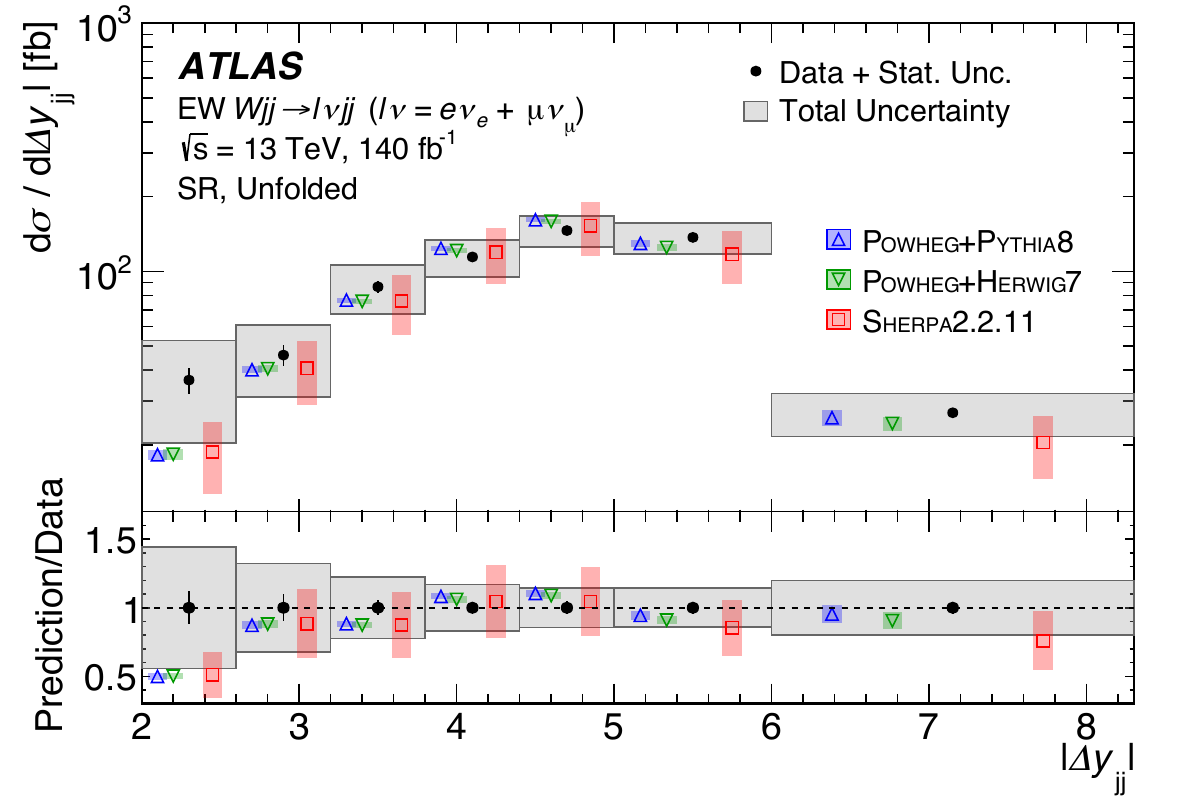}
}
\subfloat[]{
\includegraphics[width=0.5\textwidth]{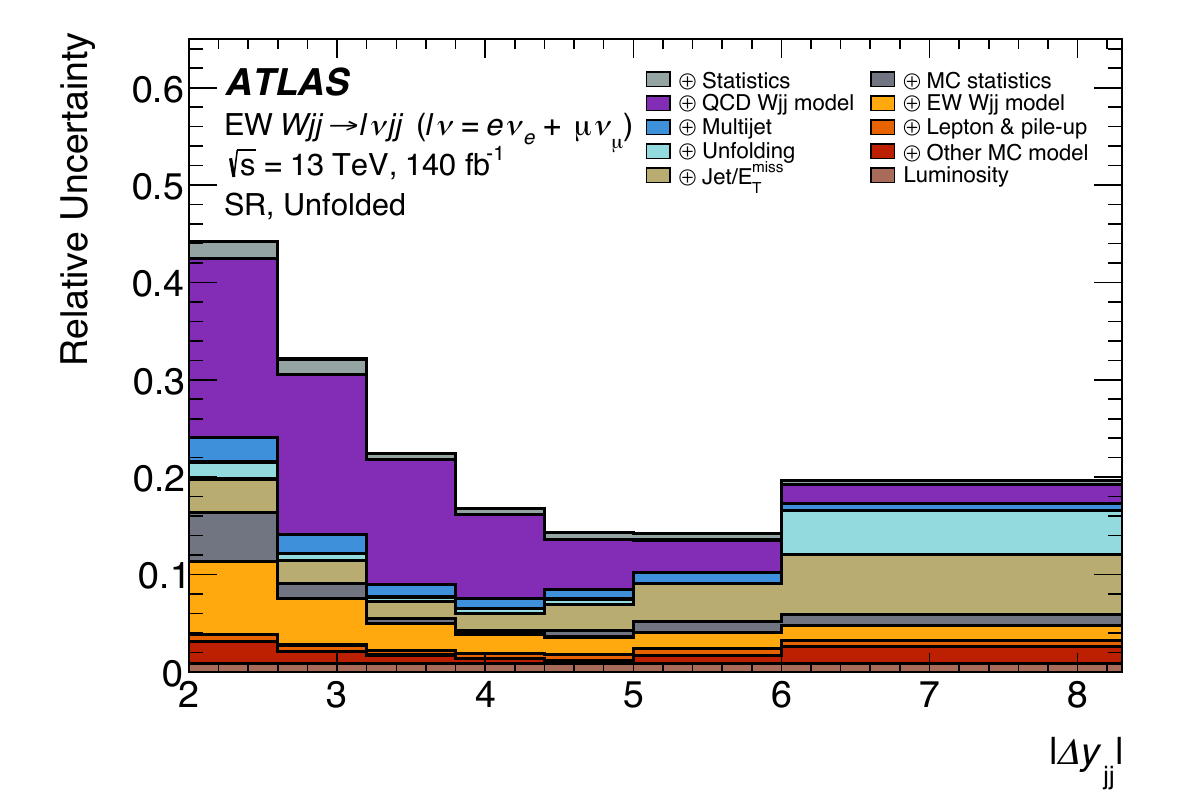}
}
\caption{(a) Unfolded differential cross-section for \ewk \Wjj production and (b) associated statistical, experimental, and theoretical uncertainties
as a function of $|\Delta y_{jj}|$.
Measurements are compared with predictions at LO and NLO accuracy
in perturbative \qcd from \Sherpa[2.2.11] and \POWHEG{}, respectively. The
latter is showered with either \PYTHIA[8] or \Herwig[7.2.1].
Theoretical uncertainties in the SM expectations are included. Uncertainties on the measurement are grouped into categories and summed in quadrature ($\oplus$) to yield the total uncertainty.
}
\label{fig:results:unf_dyjj}
\end{figure}

\begin{figure}[h]
\centering
\subfloat[]{
\includegraphics[width=0.5\textwidth]{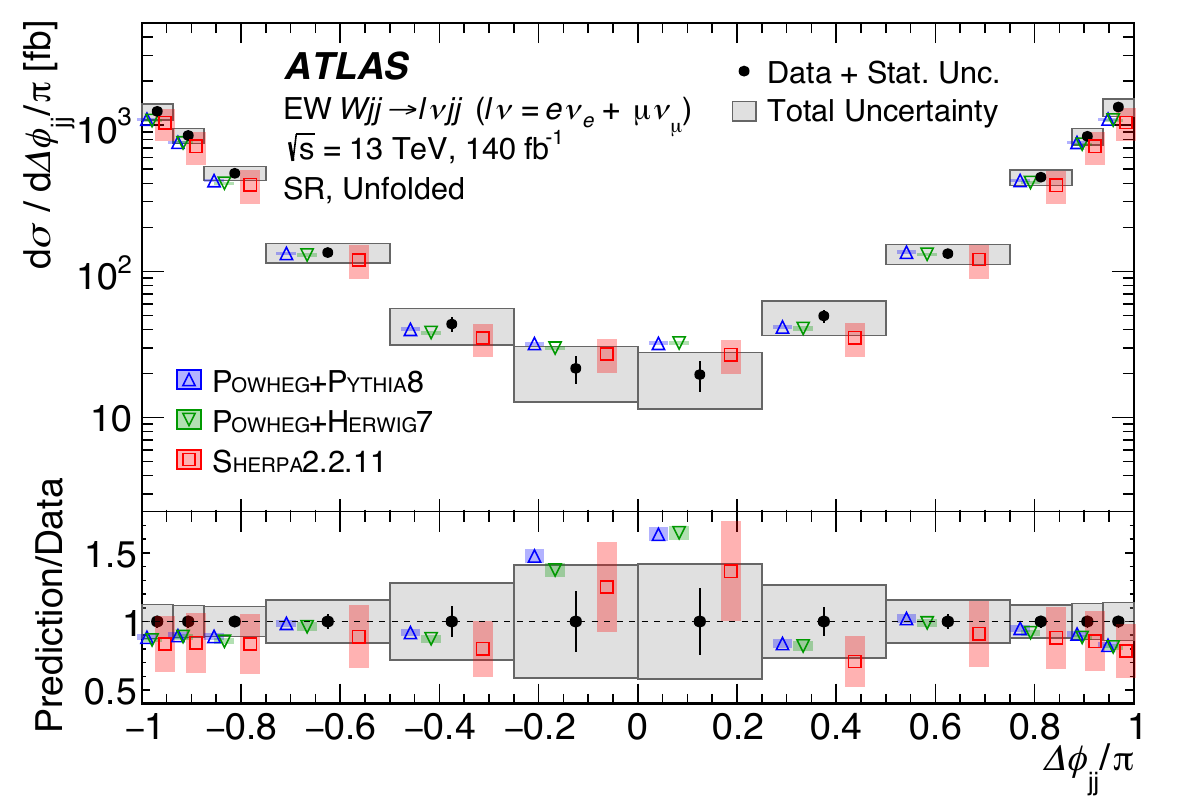}
}
\subfloat[]{
\includegraphics[width=0.5\textwidth]{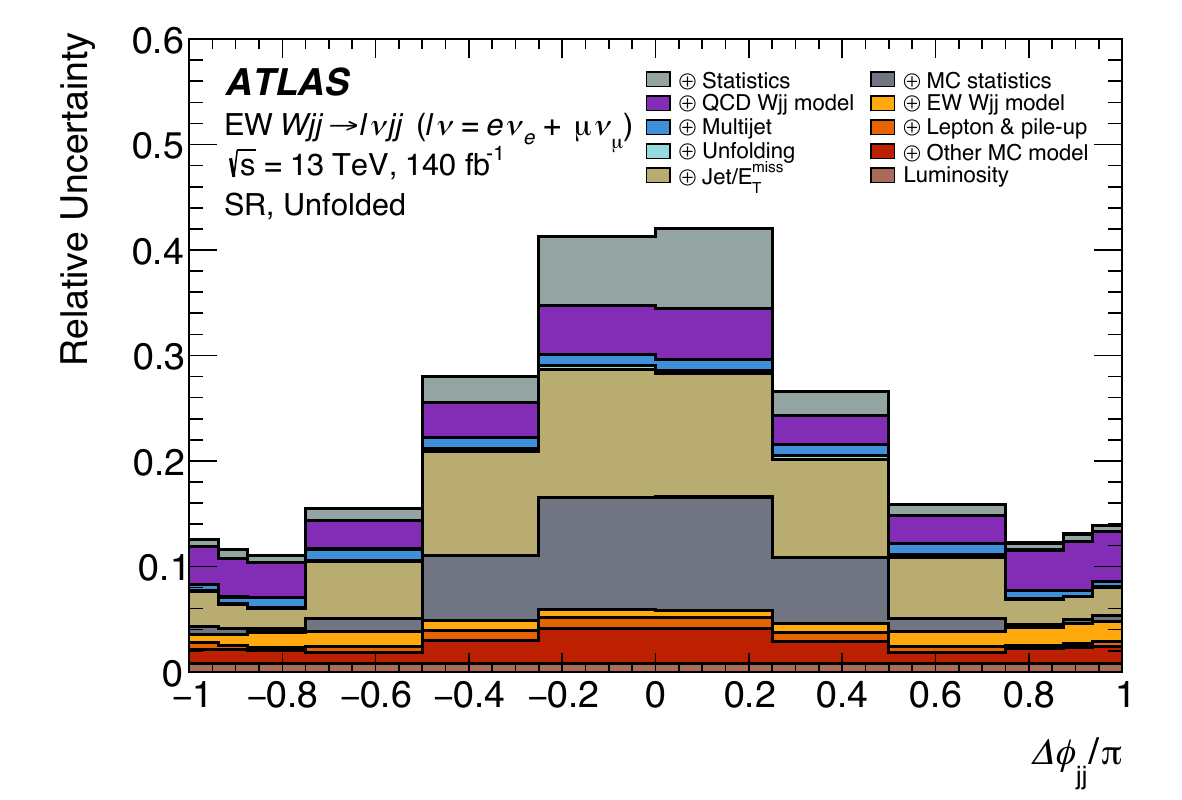}
}
\caption{(a) Unfolded differential cross-section for \ewk \Wjj production and (b) associated statistical, experimental, and theoretical uncertainties
as a function of $\Delta\phi_{jj}$.
Measurements are compared with predictions at LO and NLO accuracy
in perturbative \qcd from \Sherpa[2.2.11] and \POWHEG{}, respectively. The
latter is showered with either \PYTHIA[8] or \Herwig[7.2.1].
Theoretical uncertainties in the SM expectations are included. Uncertainties on the measurement are grouped into categories and summed in quadrature ($\oplus$) to yield the total uncertainty.
}
\label{fig:results:unf_dphijj}
\end{figure}

The \ewk \Wjj production cross-sections are unfolded to particle level
as functions of $m_{jj}$, $\pt^{jj}$, $\pt^{\ell}$, $\pt^{j1}$, $|\Delta y_{jj}|$,
and $\Delta\phi_{jj}$.
The results are presented in Figures~\ref{fig:results:unf_mjj}--\ref{fig:results:unf_dphijj},
which also include the associated uncertainties summed in quadrature and grouped according to their commonalities or relative impact.
The measurements are compared with particle-level predictions from different
generators, together with their associated theoretical uncertainties, which
consider variations of the PDF, \alphas, and the \qcd scales.
In the case of $m_{jj}$, the
measurement is also compared with the nominal SM \ewk \Wjj prediction from
\POWPY[8] that is reweighted to NLO \ewk accuracy with
$m_{jj}$-dependent $K$-factors made available in Ref.~\cite{Lindert_2023}.
These corrections decrease the predicted \ewk \Wjj production cross-section by up to
16\%.

For all observables, the data are seen to be well described by
the \ewk \Wjj predictions from \POWPY[8] and
\POWHER[7.2.1], which are both accurate to NLO in perturbative \qcd.
Regions in the extreme tails of the distributions
are, however, more subject to statistical fluctuations,
as can be seen in low $|\Delta y_{jj}|$, central $\Delta\phi_{jj}$, and high
energy regions.
The prediction from  \Sherpa[2.2.11], accurate to LO in perturbative \qcd, also
agrees with data, although within relatively large intrinsic errors, dominated by
theoretical uncertainties in the cross-section.
As expected, the $m_{jj}$ and $|\Delta y_{jj}|$ distributions for \ewk \Wjj events in
Figures~\ref{fig:results:unf_mjj} and~\ref{fig:results:unf_dyjj} are harder than those for \qcd \Wjj events,
which are included in Figures~\ref{fig:results:unf_incl_mjj} and~\ref{fig:results:unf_incl_dyjj} in the appendix.

The dominant uncertainty originates from the theoretical \qcd
\Wjj modelling, while the effect of the \ewk \Wjj modelling uncertainty is smaller.
The next largest uncertainties are due to the jet calibration and resolution, and
in some regions of the phase space the assumptions of the unfolding method.

The cross-section for \ewk \Wjj production measured in the particle-level fiducial volume by integrating the $m_{jj}$ distribution is found to be:
\begin{equation*}
\sigma_{EW} = 480 \pm 12 \, (\text{stat.}) \pm 39 \, (\text{exp.}) \pm 53 \, (\text{theory}) \; \text{fb},
\end{equation*}
and agrees with the values measured for all observables within the uncertainty.
The statistical and systematic bin-to-bin correlations resulting from the \ewk \Wjj measurement are
accounted for in this computation. The measured cross-section is found to agree well with predictions of
$431 \pm 7 \; \text{fb}$, $420 \pm 6 \; \text{fb}$, and $402 \pm 60 \; \text{fb}$
provided by \POWPY[8], \POWHER[7.2.1], and \Sherpa[2.2.11], respectively. The uncertainties in the predictions cover the impact of
varying the PDF, the \qcd factorisation and renormalisation scales, and the strong coupling constant, \alphas. The latter are especially large for the \ewk \Wjj prediction
from \Sherpa[2.2.11] as it is LO in perturbative \qcd.

\begin{figure}[h]
\centering
\subfloat[]{
\includegraphics[width=0.5\textwidth]{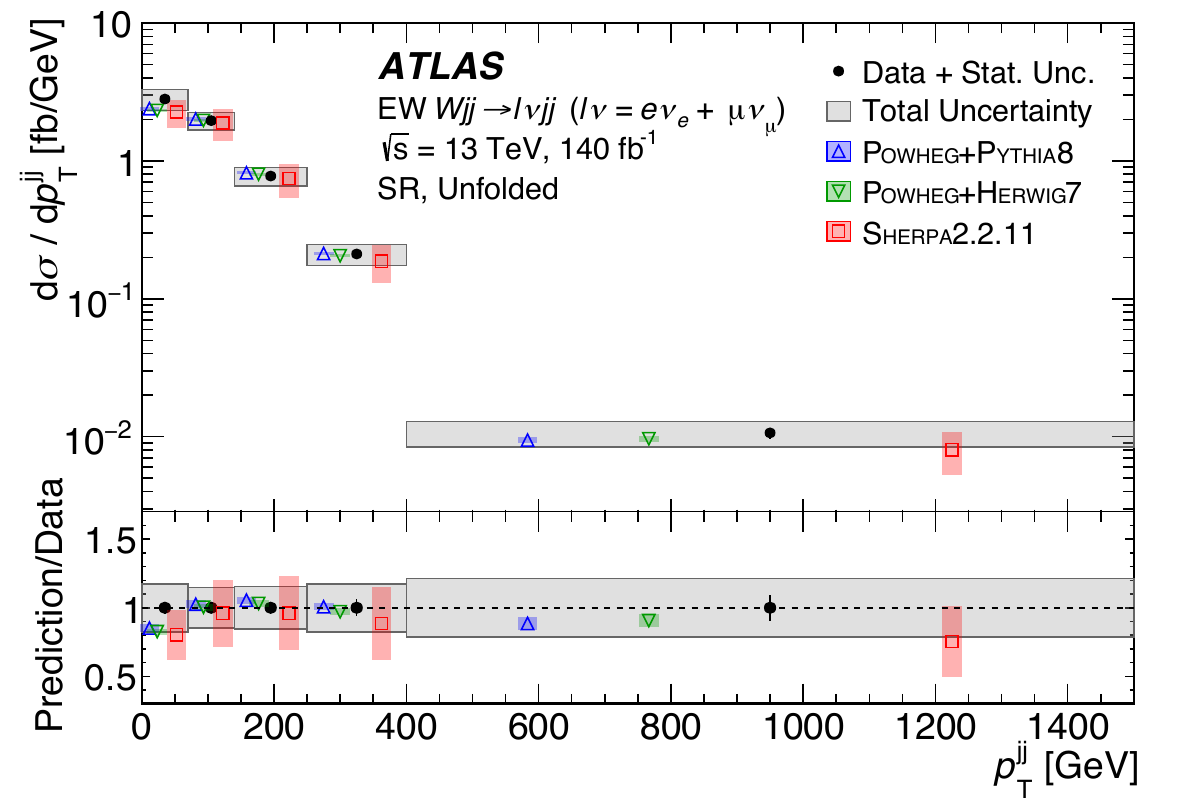}
}
\subfloat[]{
\includegraphics[width=0.5\textwidth]{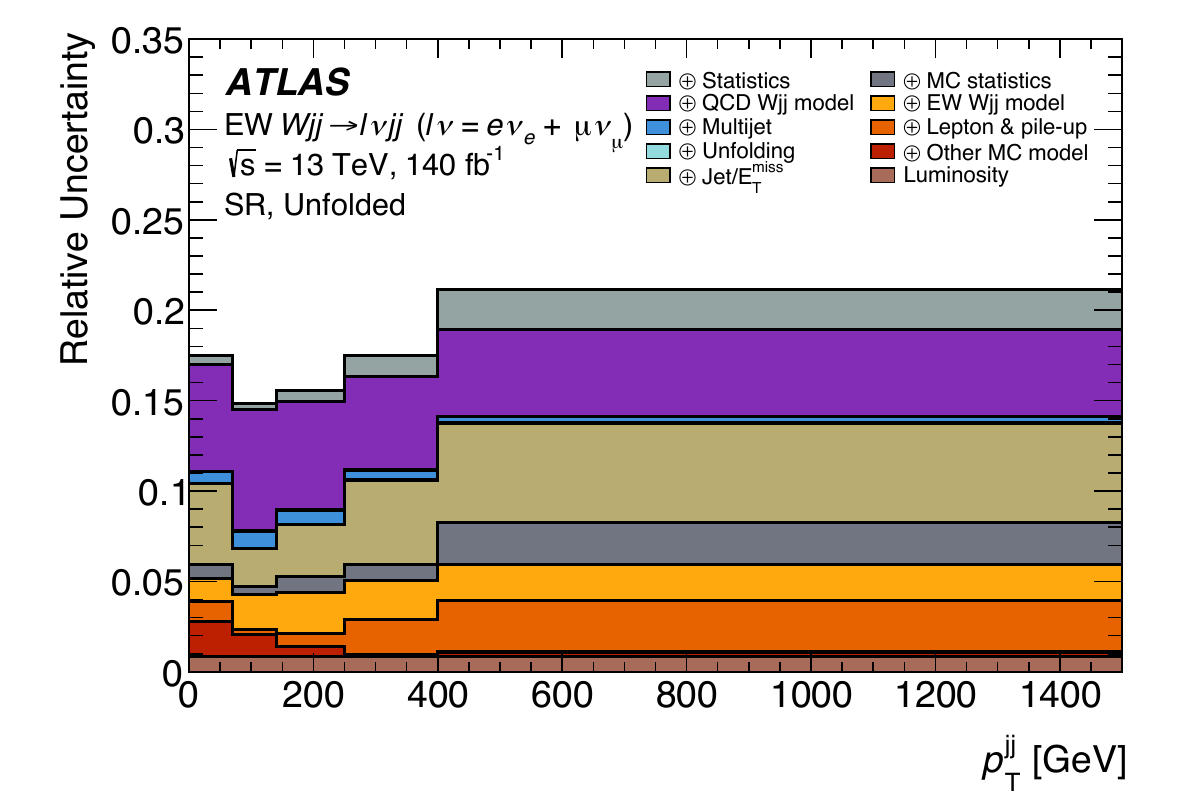}
}
\caption{(a) Unfolded differential cross-section for \ewk \Wjj production and (b) associated statistical, experimental, and theoretical uncertainties
as a function of $\pt^{jj}$.
Measurements are compared with predictions at LO and NLO accuracy
in perturbative \qcd from \Sherpa[2.2.11] and \POWHEG{}, respectively. The
latter is showered with either \PYTHIA[8] or \Herwig[7.2.1].
Theoretical uncertainties in the SM expectations are included. Uncertainties on the measurement are grouped into categories and summed in quadrature ($\oplus$) to yield the total uncertainty.
}
\end{figure}

\begin{figure}[h]
\centering
\subfloat[]{
\includegraphics[width=0.5\textwidth]{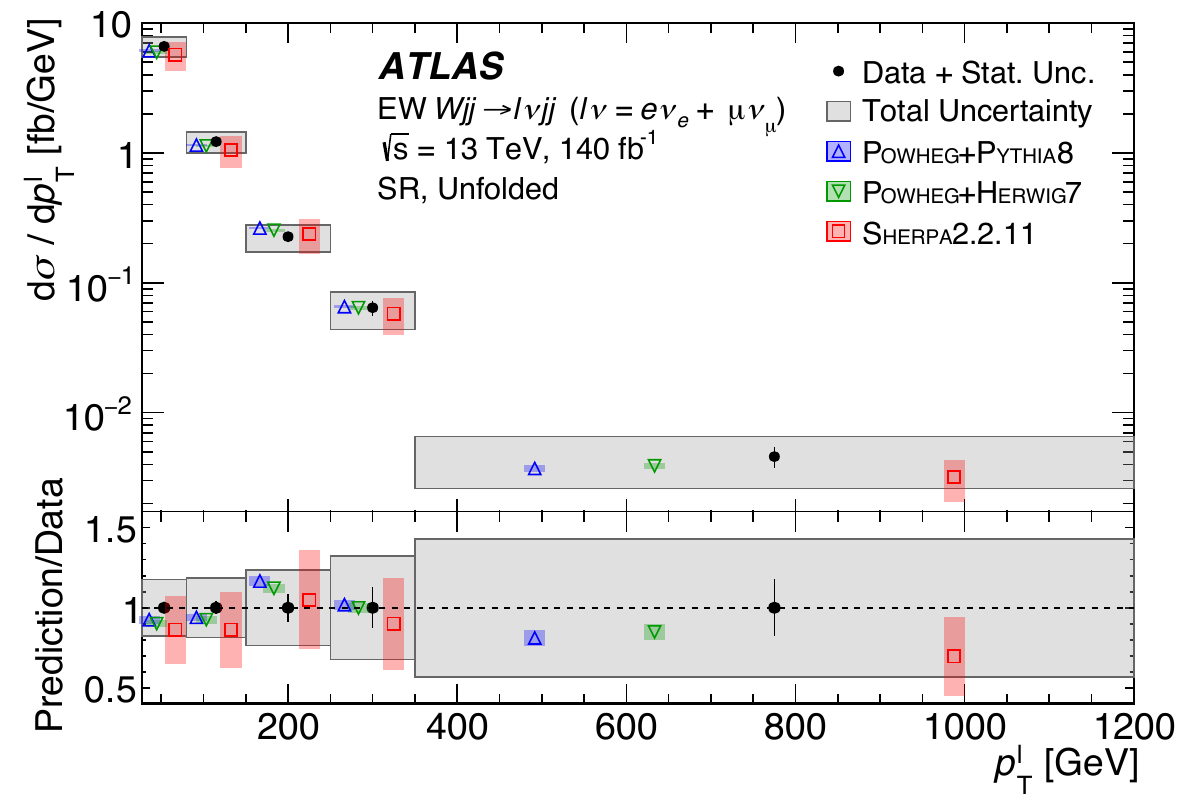}
}
\subfloat[]{
\includegraphics[width=0.5\textwidth]{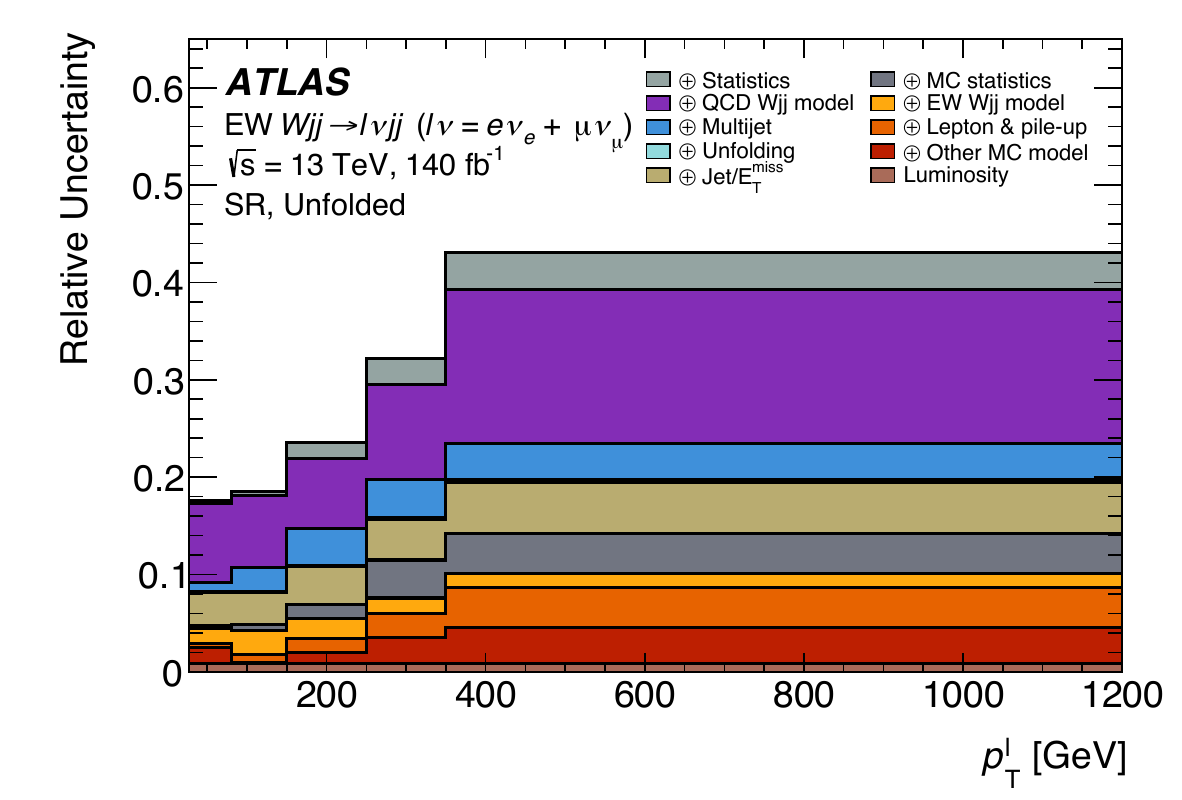}
}
\caption{(a) Unfolded differential cross-section for \ewk \Wjj production and (b) associated statistical, experimental, and theoretical uncertainties
as a function of $\pt^{\ell}$.
Measurements are compared with predictions at LO and NLO accuracy
in perturbative \qcd from \Sherpa[2.2.11] and \POWHEG{}, respectively. The
latter is showered with either \PYTHIA[8] or \Herwig[7.2.1].
Theoretical uncertainties in the SM expectations are included. Uncertainties on the measurement are grouped into categories and summed in quadrature ($\oplus$) to yield the total uncertainty.
}
\end{figure}

\begin{figure}[h]
\centering
\subfloat[]{
\includegraphics[width=0.5\textwidth]{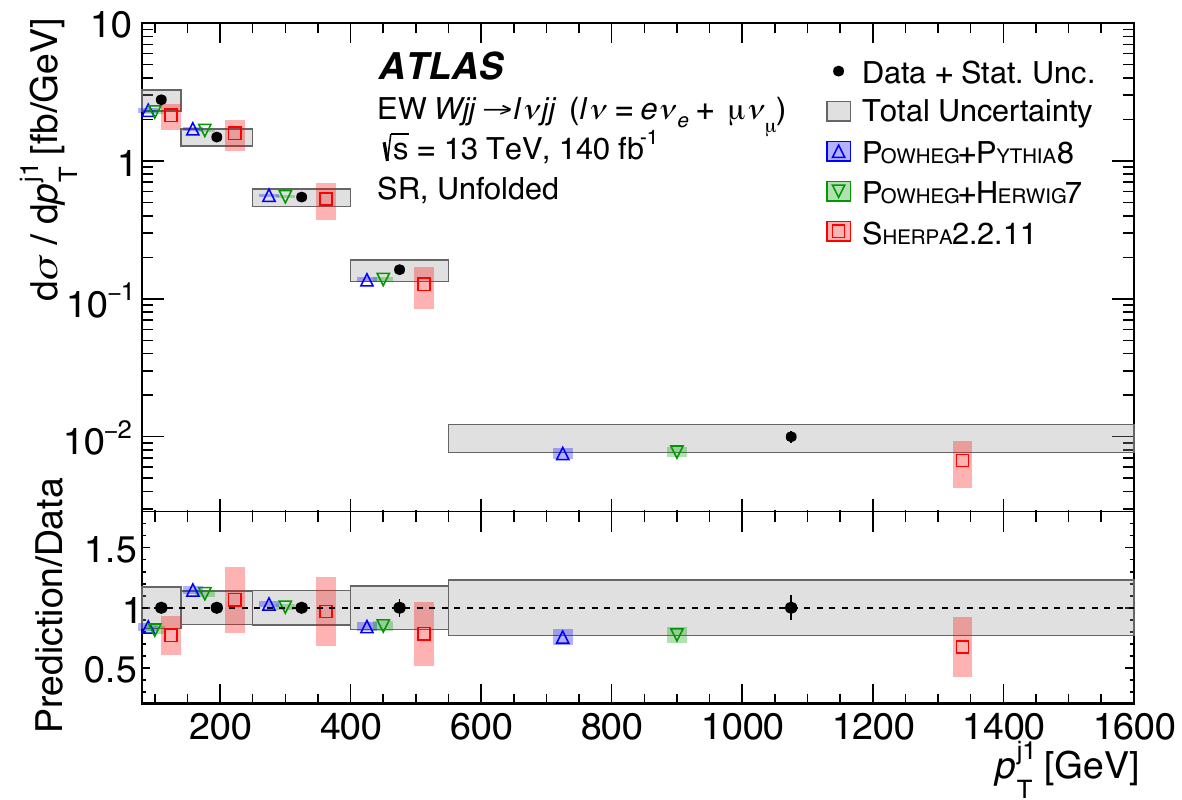}
}
\subfloat[]{
\includegraphics[width=0.5\textwidth]{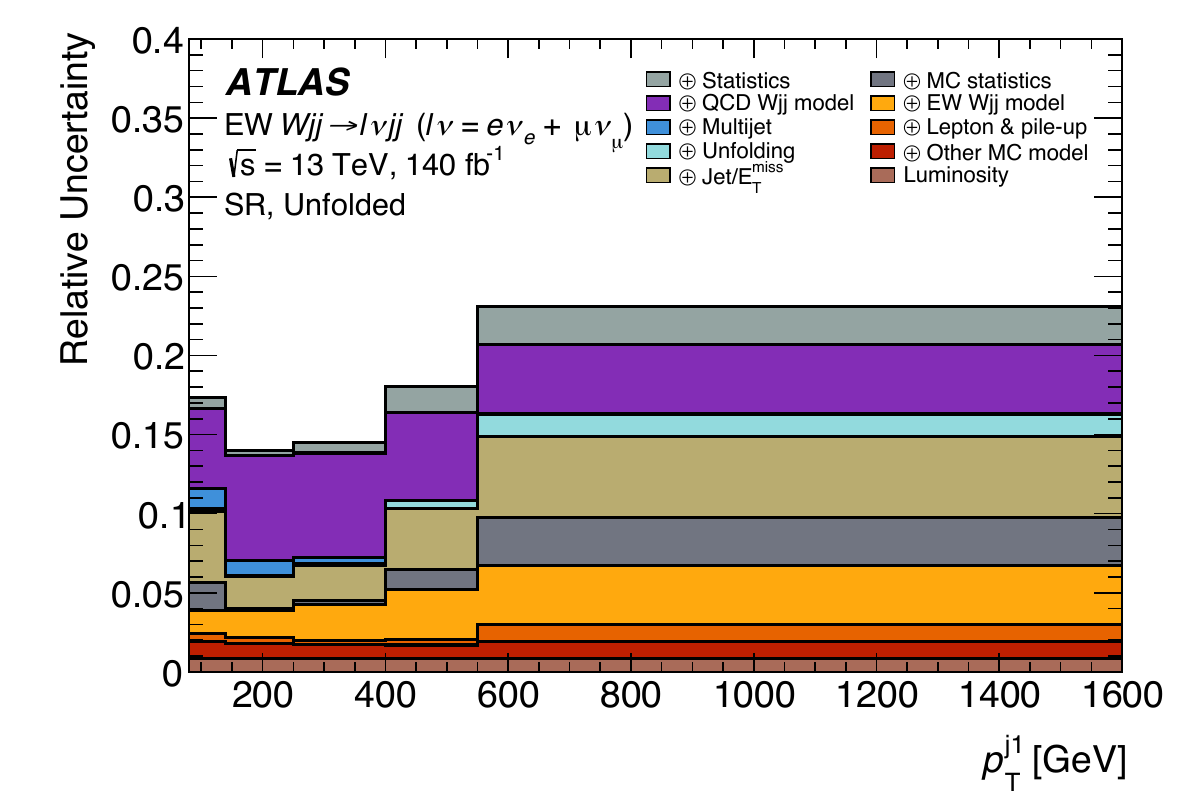}
}
\caption{(a) Unfolded differential cross-section for \ewk \Wjj production and (b) associated statistical, experimental, and theoretical uncertainties
as a function of $\pt^{j1}$.
Measurements are compared with predictions at LO and NLO accuracy
in perturbative \qcd from \Sherpa[2.2.11] and \POWHEG{}, respectively. The
latter is showered with either \PYTHIA[8] or \Herwig[7.2.1].
Theoretical uncertainties in the SM expectations are included. Uncertainties on the measurement are grouped into categories and summed in quadrature ($\oplus$) to yield the total uncertainty.
}
\end{figure}

%
\section{Constraints on anomalous weak-boson self-interactions}
\label{sec:eft}
%

The unfolded differential cross-section for \ewk \Wjj production is interpreted in the
context of an EFT using the SMEFT model~\cite{Brivio_2021, Brivio_2017}.
The robustness of the
likelihood fit described in Section~\ref{sec:bkg-determination} and the
response matrix computed in Section~\ref{sec:unfolding}
was successfully tested in the presence of physics beyond the SM by
replacing data with pseudo-data.
The likelihood fit was performed on pseudo-data constructed from the simulated
SM background together with the \ewk \Wjj and EFT predictions, yielding
post-fit signal yields compatible with the corresponding \ewk \Wjj plus EFT
expectation. Similarly, unfolding the simulated detector-level \ewk \Wjj plus EFT
differential cross-section produced results consistent with the \ewk \Wjj plus EFT expectation
at particle level.

In the SMEFT model considered, the Warsaw basis~\cite{Grzadkowski_2010} is
used to extend the SM Lagrangian with additional dimension-six operators
describing the effective low-energy behaviour of an unknown theory that is UV-complete
above a scale $\Lambda=1\,\mathrm{TeV}$.
The resulting Lagrangian, $\mathcal{L}_\text{SMEFT}$, is given by
Eq.~(\ref{eq:eft:lagrangian}),
where $\mathcal{L}_\text{SM}$ is the Lagrangian from the SM, $\mathcal{O}_{i}$ are
the additional dimension-six operators, and $c_{i}$ their associated Wilson
coefficients:
\begin{equation}
\mathcal{L}_\text{SMEFT}=\mathcal{L}_\text{SM}+\sum_{i}\frac{c_{i}}{\Lambda^{2}}\mathcal{O}_{i}^{(6)}.
\label{eq:eft:lagrangian}
\end{equation}
The $U(3)^{5}$ flavour-symmetric model
is assumed, along with massless fermions and the $\hat{m}_{W}$ input parameter
scheme~\cite{Brivio_2017}.

Both the CP-conserving and CP-violating dimension-six EFT operators are included
in the model, limiting the interpretation to those to which the \ewk \Wjj
measurement is sensitive.
Consequently, the EFT operators that are tested are the CP-even
$\mathcal{O}_{W}$ and $\mathcal{O}_{HWB}$, and their CP-odd counterparts
$\tilde{\mathcal{O}}_{W}$ and $\tilde{\mathcal{O}}_{HWB}$, which are sensitive
to triple-gauge-boson couplings.
The corresponding Wilson coupling coefficients are denoted by their UFO
(Universal FeynRules Output) labels~\cite{Degrande_2012} $c_{W}$,
$c_{HWB}$, $\tilde{c}_{W}$, and $\tilde{c}_{HWB}$.

Only one EFT coupling coefficient at a time is assumed to be non-zero, with
the others set to 0. As a consequence, the cross-section for \ewk signal
production, assuming the extended SM Lagrangian, can be expressed as in
Eq.~(\ref{eq:eft:xsec}), with $\sigma_{\text{SM}}$ the predicted SM cross-section for the
\ewk process, and $\sigma_{\text{lin}}^{i}$ and $\sigma_{\text{quad}}^{i}$
the cross-sections corresponding to the SM--EFT linear interference and purely quadratic terms
of the EFT operator $i$:

\begin{equation}
\sigma_{\text{SM+EFT}} = \sigma_{\text{SM}} + \sigma_{\text{lin}}^{i}\frac{c_{i}}{(\Lambda/\TeV)^{2}}  +  \sigma_{\text{quad}}^{i}\frac{c_{i}^{2}}{(\Lambda/\TeV)^{4}}.
\label{eq:eft:xsec}
\end{equation}

EFT event samples including the operators
tested were generated at particle level via \MADGRAPH and showered by
\PYTHIA[8] using the \NNPDF[2.3lo] PDF set and A14 tune.
The ratio $c_{i}/\Lambda^{2}$ was set to $1\,\TeV^{-2}$, and the dipole recoil
was activated in \PYTHIA for the showering.
Enough events were generated to ensure that their statistical uncertainty does not affect the results.

\MADGRAPH{} generated events for the EFT samples at LO in perturbative \qcd. These events
are reweighted to NLO in perturbative \qcd using the bin-by-bin ratio of the SM \ewk \Wjj
expectation from \POWPY[8] to the SM prediction from \MADGRAPH{}, and this
increases the normalisation by 9\% overall.
To mitigate the effect of missing NLO \ewk corrections with respect to NLO in perturbative
\qcd, additional corrections from Ref.~\cite{Lindert_2023} are applied
bin-by-bin to the nominal SM \ewk \Wjj and reweighted EFT simulations.
Since these corrections were calculated in a phase space slightly
different from that used in this measurement, and were made available and applied
only as function of $m_{jj}$, an uncertainty equal to the size of those
corrections is included when propagating them to another
observable.

\begin{figure}
\captionsetup[subfigure]{labelformat=empty}
\centering
\subfloat[]{
\includegraphics[width=0.75\textwidth]{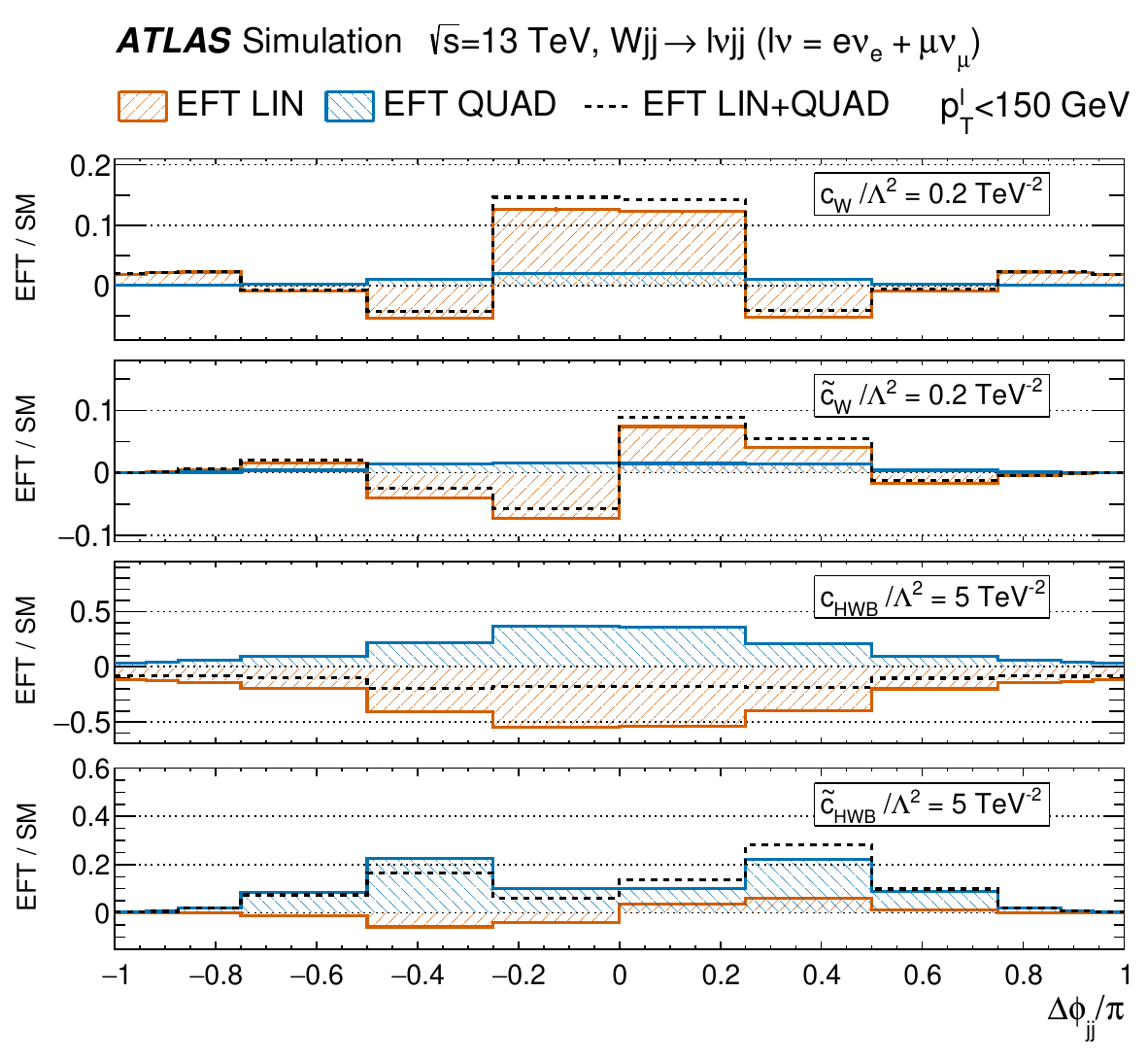}
}
\caption{EFT signal sensitivity relative to SM \ewk \Wjj predictions
as a function of $\Delta\phi_{jj}$, for the low $\pt^{\ell}$ region.
Linear SM--EFT terms and purely quadratic EFT terms are shown separately
and summed for
CP-even and CP-odd EFT predictions assuming
$c_{W}/\Lambda^{2}=0.2\,\TeV^{-2}$ or $c_{HWB}/\Lambda^{2}=5\,\TeV^{-2}$,
and $\tilde{c}_{W}/\Lambda^{2}=0.2\,\TeV^{-2}$ or $\tilde{c}_{HWB}/\Lambda^{2}=5\,\TeV^{-2}$, respectively.}
\label{fig:eft:sensitivity_lowpt}
\end{figure}

\begin{figure}
\captionsetup[subfigure]{labelformat=empty}
\centering
\subfloat[]{
\includegraphics[width=0.75\textwidth]{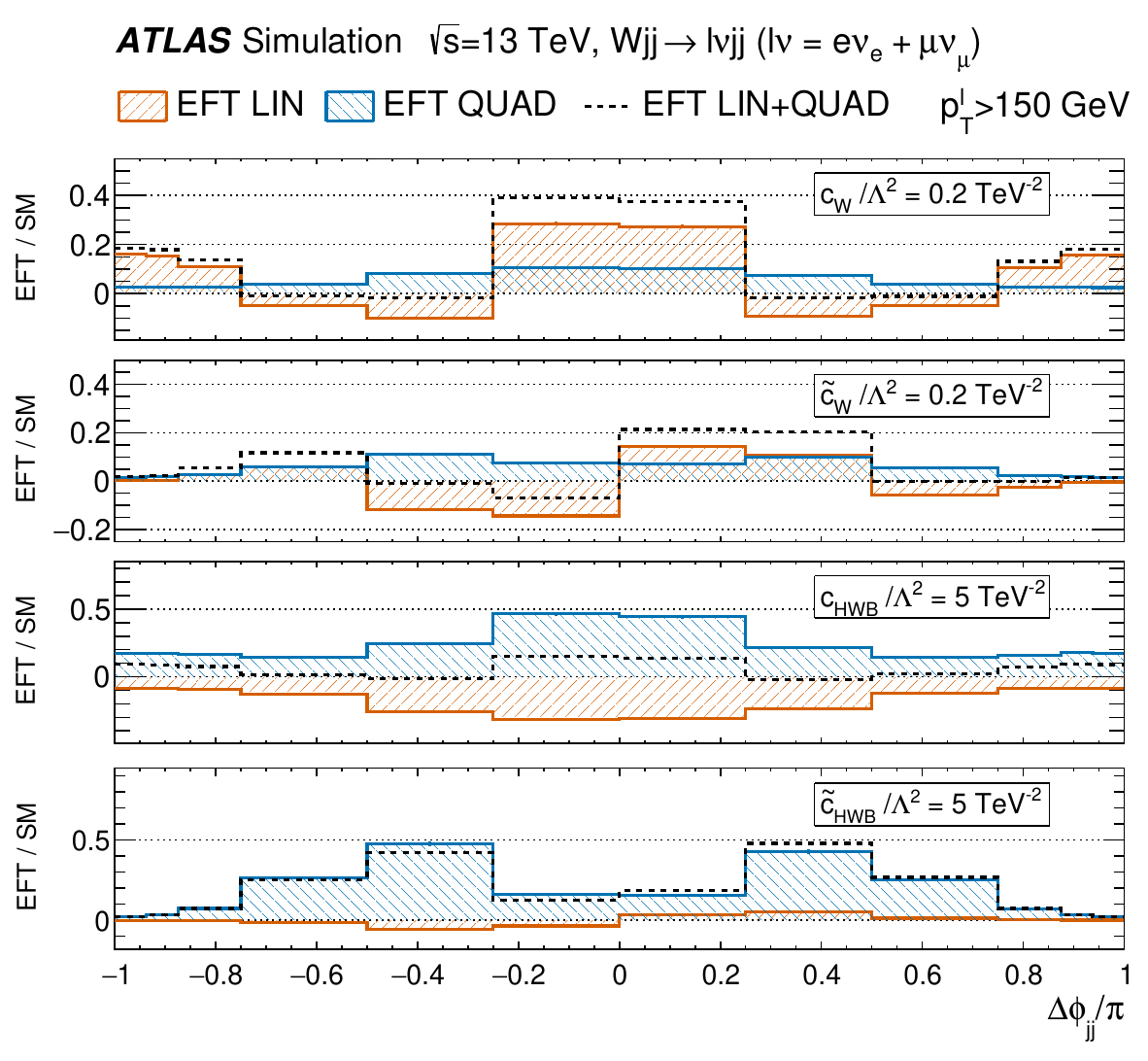}
}
\caption{EFT signal sensitivity relative to SM \ewk \Wjj predictions
as a function of $\Delta\phi_{jj}$, for the high $\pt^{\ell}$ region.
Linear SM--EFT terms and purely quadratic EFT terms are shown separately
and summed for
CP-even and CP-odd EFT predictions assuming
$c_{W}/\Lambda^{2}=0.2\,\TeV^{-2}$ or $c_{HWB}/\Lambda^{2}=5\,\TeV^{-2}$,
and $\tilde{c}_{W}/\Lambda^{2}=0.2\,\TeV^{-2}$ or $\tilde{c}_{HWB}/\Lambda^{2}=5\,\TeV^{-2}$, respectively.}
\label{fig:eft:sensitivity_highpt}
\end{figure}

The 95\% confidence interval limits on the coefficients of the
EFT operators are computed via a fit of the SM \ewk plus EFT
prediction to the unfolded SM \ewk measurement. The fit relies on a
frequentist statistical approach based on a binned profiled likelihood-ratio
test statistic.
The statistical model is given by a multivariate Gaussian distribution for the measurement
uncertainties, which are treated as correlated between bins. The most sensitive
observable for performing the fit was found to be $\Delta\phi_{jj}$ for
`linear-only' limits on both the CP-even and CP-odd coefficients. However, in order
to further increase the sensitivity of the interpretation, information about
the energy of the process is added by splitting $\Delta\phi_{jj}$ into two bins of
lepton $\pt$ separated by a cut at $150\,\GeV$. These two bins are then
combined in the EFT fit.
The choice of $\pt^{\ell}$ is motivated by a neural-network-based study
comparing the importance of different observables, while the
cut at $150\,\GeV$ ensures that sufficient statistics are available for
a robust background determination.
Distributions of $\Delta\phi_{jj}$ depicting the ratios of an EFT signal to the
SM \ewk prediction are shown in Figures~\ref{fig:eft:sensitivity_lowpt}
and~\ref{fig:eft:sensitivity_highpt}, for the cases $\pt^{\ell}<150\,\GeV$ and
$\pt^{\ell}>150\,\GeV$, respectively.
These figures illustrate the impact of an EFT signal versus
$\Delta\phi_{jj}$ when considering its linear and quadratic components separately or
summed, and using values of the Wilson coefficients that are close to the expected
limits in the linear-plus-quadratic case, namely
$c_{W}/\Lambda^{2}=0.2\,\TeV^{-2}$,
$\tilde{c}_{W}/\Lambda^{2}=0.2\,\TeV^{-2}$,
$c_{HWB}/\Lambda^{2}=5\,\TeV^{-2}$, and
$\tilde{c}_{HWB}/\Lambda^{2}=5\,\TeV^{-2}$.
The figures also illustrate that in most cases the high $\pt^{\ell}$ region
is significantly more impacted by the EFT signal.

\begin{figure}
\centering
\subfloat[]{
\includegraphics[width=0.5\textwidth]{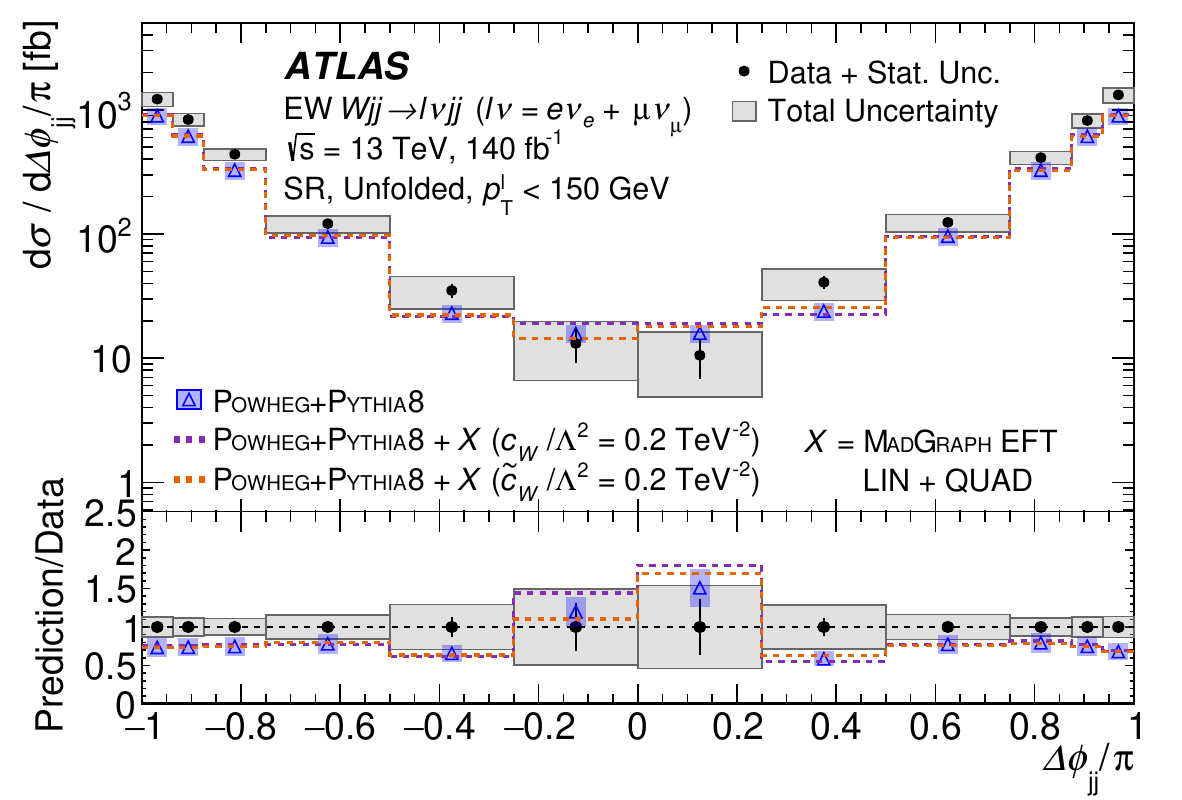}
}
\subfloat[]{
\includegraphics[width=0.5\textwidth]{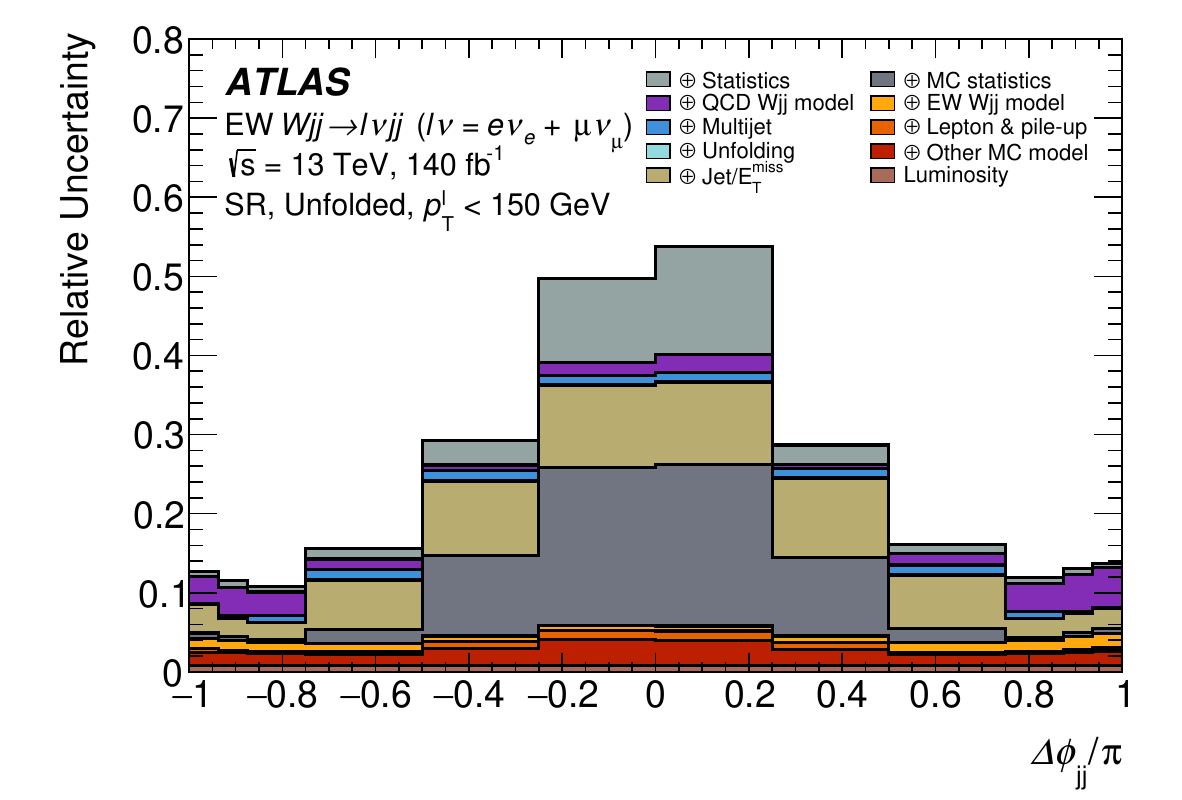}
}
\caption{Unfolded differential cross-section for \ewk \Wjj production
as a function of $\Delta\phi_{jj}$ for $\pt^{\ell}<150\,\GeV$ (a).
Measurements are compared with the SM prediction from
\POWPY[8] and its theoretical uncertainties.
Nominal EFT predictions from \MADGRAPH{} added to the SM expectation
from \POWPY[8] are also shown, assuming either
$c_{W}/\Lambda^{2}=0.2\,\TeV^{-2}$ (violet) or
$\tilde{c}_{W}/\Lambda^{2}=0.2\,\TeV^{-2}$ (orange), with both the linear
(LIN) and quadratic (QUAD) terms.
The \POWPY[8] and \MADGRAPH{} simulations, which achieve
NLO (LO) and LO (LO) accuracy in perturbative \qcd (\ewk) theory,
are reweighted to
NLO (NLO) precision, respectively. Components of the statistical,
experimental, and theoretical uncertainties in the measured
unfolded distribution are shown (b).
}
\label{fig:results:unf_dphijj_lowPt}
\end{figure}

\begin{figure}
\centering
\subfloat[]{
\includegraphics[width=0.5\textwidth]{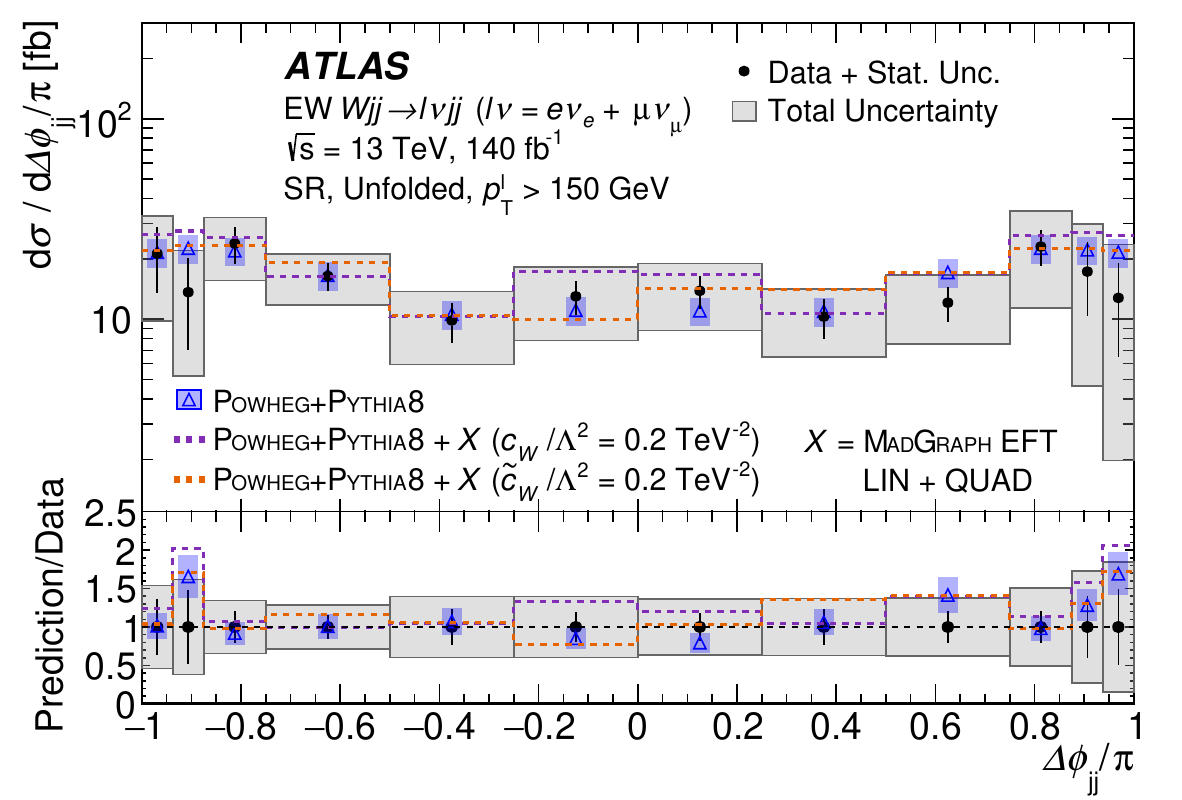}
}
\subfloat[]{
\includegraphics[width=0.5\textwidth]{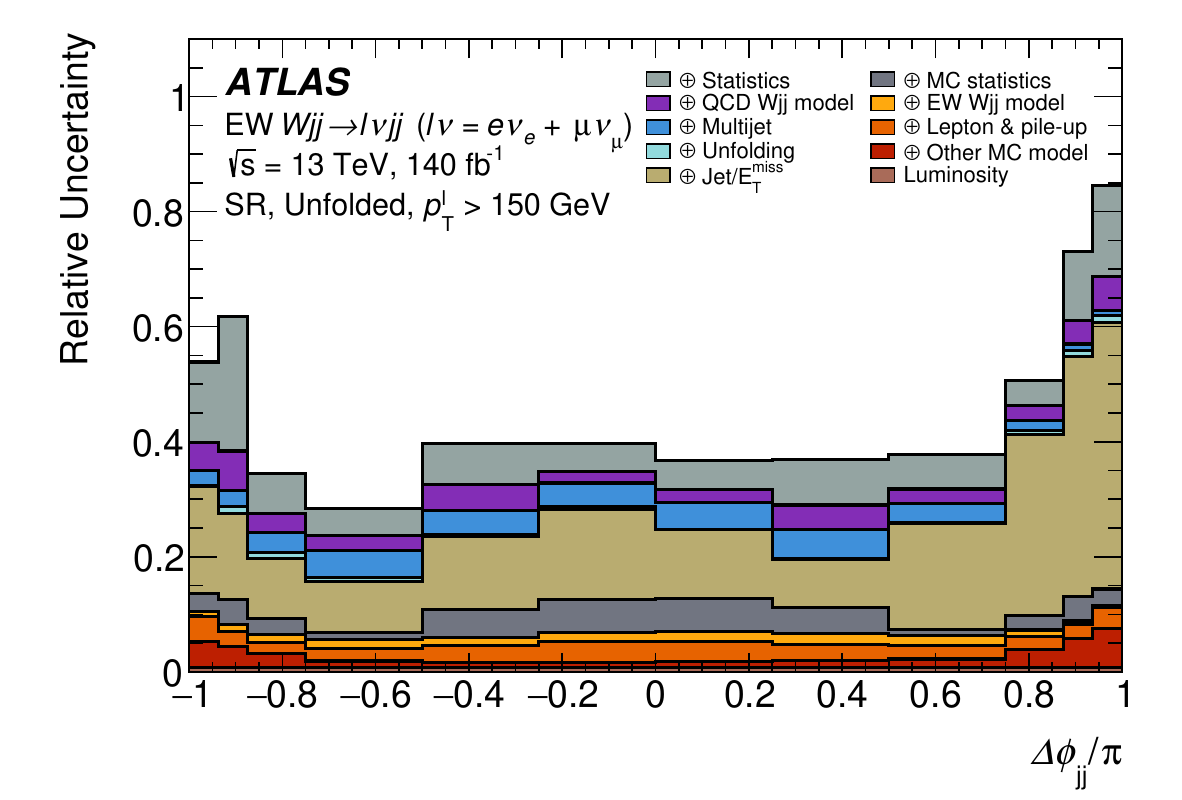}
}
\caption{Unfolded differential cross-section for \ewk \Wjj production
as a function of $\Delta\phi_{jj}$ for $\pt^{\ell}>150\,\GeV$ (a).
Measurements are compared with the SM prediction from
\POWPY[8] and its theoretical uncertainties.
Nominal EFT predictions from \MADGRAPH{} added to the SM expectation
from \POWPY[8] are also shown, assuming either
$c_{W}/\Lambda^{2}=0.2\,\TeV^{-2}$ (violet) or
$\tilde{c}_{W}/\Lambda^{2}=0.2\,\TeV^{-2}$ (orange), with both the linear
(LIN) and quadratic (QUAD) terms.
The \POWPY[8] and \MADGRAPH{} simulations, which achieve
NLO (LO) and LO (LO) accuracy in perturbative \qcd (\ewk) theory, are
reweighted to NLO (NLO) precision, respectively.
Components of the statistical, experimental, and theoretical
uncertainties on the measured unfolded distribution are shown (b).
}
\label{fig:results:unf_dphijj_highPt}
\end{figure}

Measurements of $\Delta\phi_{jj}$ in the low and high $\pt^{\ell}$ regions
are shown together with categorised uncertainties in
Figures~\ref{fig:results:unf_dphijj_lowPt}
and~\ref{fig:results:unf_dphijj_highPt}, respectively.
These constitute the data to which the SM \ewk plus EFT expectation is fitted
to derive EFT limits. The prediction tested is also illustrated
in the case of CP-even and CP-odd operators,
assuming $c_{W}/\Lambda^{2}=0.2\,\TeV^{-2}$ and
$\tilde{c}_{W}/\Lambda^{2}=0.2\,\TeV^{-2}$.

All systematic uncertainties from the unfolded $\Delta\phi_{jj}$
measurement are considered in the EFT fit, along with the theoretical
uncertainties in the SM and EFT predictions, which are summarised in
Table~\ref{tab:eft:inputSigUncert}. The statistical bin-to-bin correlations
are taken into account, while each source of systematic
uncertainty, from both the measurement and the predictions,
is treated as fully correlated among the bins of
$\Delta\phi_{jj}$. Systematic uncertainties are also treated as
fully correlated between the two $\pt^{\ell}$ regions.

\begin{table}
\centering
\caption{Ranges of theoretical uncertainties in the \ewk \Wjj SM and EFT
prediction across the low and high $\pt^{\ell}$ regions.
}
\label{tab:eft:inputSigUncert}
\begin{tabular}{lc}
\hline\hline
Source & Relative uncertainty [\%] \\
\hline\hline
\qcd scale         & 1.4--3.5   \\
PDF               & 1.7--4.0   \\
\alphas           & $<1\%$ \\
Parton shower     & 4.9--8.1        \\
NLO \ewk correction & 16--17      \\
\hline\hline
\end{tabular}
\end{table}

The expected and observed 95\% confidence interval limits when
combining the $\Delta\phi_{jj}$ distributions in the low and high
$\pt^{\ell}$ regions are shown in Table~\ref{tab:eft:dphijj_limits_2d}.
Expected limits are computed by fitting the SM \ewk plus EFT
prediction to Asimov data, which correspond to the SM \ewk prediction
and its associated measurement uncertainties and statistical correlations.
While the addition of quadratic terms significantly improves the limits
on $c_{HWB}$ and $\tilde{c}_{HWB}$, those on $c_{W}$ and $\tilde{c}_{W}$
remain much the same, highlighting their lower sensitivity to missing higher-order
operators in the EFT expansion.

The extracted limits on $c_{HWB}$ and $\tilde{c}_{HWB}$ are weaker than
the strongest constraints ATLAS has obtained from its
$H\gamma\gamma$~\cite{HIGG-2020-16} and
$W\gamma$~\cite{STDM-2022-15} analyses.
When considering only the linear EFT term, the limits placed on
$c_{W}$ are slightly tighter than the strongest ATLAS bounds from the
$H\gamma\gamma$, $W\gamma$, and \Zjj measurements, and are complementary
to those from the CMS $W\gamma$ analysis~\cite{CMS-SMP-20-005}.
When adding the quadratic EFT term, the computed limits on $c_{W}$ are weaker
than the tightest bounds ATLAS has derived from its $W\gamma$ measurement.
The limits placed on $\tilde{c}_{W}$ are complementary
to those from the ATLAS $WZ$~\cite{STDM-2019-26}, $W\gamma$ and \Zjj
analyses, which have the tightest of the already existing bounds.

\begin{table}
\centering
\caption{Expected and observed 95\% confidence intervals
$[\TeV^{-2}]$
for EFT Wilson coefficients in the linear-only and linear-plus-quadratic
cases, when combining the $\Delta\phi_{jj}$ distributions in the low and high
$\pt^{\ell}$ regions.}
\label{tab:eft:dphijj_limits_2d}
\begin{tabular}{lcccc}
\hline\hline

\multicolumn{1}{c}{} & \multicolumn{2}{c}{Expected interval $[\TeV^{-2}]$ }
& \multicolumn{2}{c}{Observed interval $[\TeV^{-2}]$ }
\\

Wilson coefficient & Linear-only & Linear+quadratic
& Linear-only & Linear+quadratic \\ \hline\hline

$c_{W}/\Lambda^{2}$
&  $[-0.17, 0.17]$  &  $[-0.17, 0.16]$
&  $[-0.17, 0.12]$  &  $[-0.15, 0.10]$ \\
$\tilde{c}_{W}/\Lambda^{2}$
&  $[-0.19, 0.19]$  &  $[-0.18, 0.19]$
&  $[-0.083, 0.25]$  &  $[-0.079, 0.18]$ \\
$c_{HWB}/\Lambda^{2}$
& $[-5.6, 4.3]$  &  $[-3.4, 5.4]$
& $[-4.1, 4.4]$  &  $[-2.2, 2.9]$ \\
$\tilde{c}_{HWB}/\Lambda^{2}$
& $[-10, 10]$  &  $[-4.1, 4.1]$
& \;$[-11, 7.3]$  & $[-2.8, 2.5]$ \\

\hline\hline
\end{tabular}
\end{table}


%

%
%
\FloatBarrier

\section{Conclusion}
%
%

Differential cross-section measurements of the \ewk production of a $W$
boson in association with two jets are reported for several observables, using
$\runtwolumi$ of data from proton--proton collisions at
$\sqrt{s}=13\,\TeV$ recorded by the ATLAS detector at the LHC.
Both the electron and muon channels are considered as signal and their contributions are summed.
These are the first  \ewk \Wjj differential cross-sections to
be measured in $13\,\TeV$ proton--proton collisions.

The analysis estimates the dominant background,
from the \qcd \Wjj process,
by using a prediction at NLO accuracy in perturbative QCD, which is
further constrained by data in dedicated control regions to reduce
systematic uncertainties.
The background due to events with misidentified
leptons is estimated using two independent methods, relying on data
in phase-space regions enriched in such events.

The fiducial cross-section for \ewk \Wjj production is found to be
$480 \pm 12 \, (\text{stat.}) \pm 39 \, (\text{exp.}) \pm 53 \, (\text{theory}) \; \text{fb}$ at particle level, and is in agreement with theoretical predictions.
Six observables are considered for the differential measurements, namely $m_{jj}$,
$\pt^{jj}$, $|\Delta y_{jj}|$, $\pt^{j1}$, and $\pt^{\ell}$, as well as
$\Delta\phi_{jj}$ in the $\pt^{\ell}<150\,\GeV$ and
$\pt^{\ell}>150\,\GeV$ regions and inclusively.
These observables characterise angular properties of the two tag jets, the
kinematics of the $W$ boson, and the $\pt$ of the
highest-energy jet and the charged lepton.
The data are corrected for detector inefficiency and resolution effects using an iterative Bayesian unfolding method,
and are compared with  \ewk \Wjj predictions from \POWHEG, which is accurate
to NLO in perturbative \qcd.
Two parton showering scenarios, using either \PYTHIA[8] or \Herwig[7.2.1],
are explored for the \POWHEG-generated events.
Both provide a good description of the data, given the uncertainties,
except in the high-energy tails of the distributions.
A simulation from \Sherpa[2.2.11], which is only accurate to LO in perturbative \qcd,
also shows good agreement with the data, but with large theoretical uncertainties.

The sensitivity of the measured differential cross-sections to the $WWZ$ and
$WW\gamma$ vertices enables tests of the SM, and in particular of
triple-gauge-boson coupling operators in the context of a
dimension-six EFT. Both CP-even and CP-odd operators were tested,
allowing 95\% confidence interval limits to be placed on the Wilson
coefficients
$c_{W}/\Lambda^{2}$ and $c_{HWB}/\Lambda^{2}$, and their CP-odd counterparts $\tilde{c}_{W}/\Lambda^{2}$ and
$\tilde{c}_{HWB}/\Lambda^{2}$. Limits on the EFT coefficients are obtained for
the linear EFT term alone and for the sum of the linear and purely quadratic EFT
terms. By combining results from the sensitive angular observable
$\Delta\phi_{jj}$ in low and high lepton-\pt regions, the `linear-only'
limits placed on $c_{W}/\Lambda^{2}$, corresponding to expected and observed
95\% confidence intervals of $[-0.17, 0.17]$ and $[-0.17, 0.12]\,\TeV^{-2}$,
are among the most stringent to date.
In addition, the computed `linear-only' and `linear-plus-quadratic' limits placed on
$\tilde{c}_{W}/\Lambda^{2}$ are complementary to the strongest already existing
limits, with expected 95\% confidence intervals of $[-0.19, 0.19]$ and $[-0.18, 0.19]\,\TeV^{-2}$, and
observed intervals of $[-0.083, 0.25]$ and $[-0.079, 0.18]\,\TeV^{-2}$, respectively.


%
%
\appendix
\part*{Appendix}
\addcontentsline{toc}{part}{Appendix}
\section*{Combined \ewk{}+\qcd \Wjj production}
%

In addition to the \ewk \Wjj results, the combined \ewk{}+\qcd
production of a $W$ boson in association with two jets (\ewk{}+\qcd \Wjj) is measured.
The measurement of the \ewk{}+\qcd \Wjj process constitutes an important addition
to benchmark theoretical calculations of \qcd \Wjj production, which are
necessary for accurate MC sample production.
The same fiducial phase space is used, also considering as signal the
summed contributions from the electron and muon decay modes of the $W$ boson,
while its $\tau$-lepton decay mode is treated as a background.

Along with the samples described in Section~\ref{sec:samples}, two alternative
MC generators are employed for the simulation of \qcd \Wjj events, namely
\Sherpa[2.2.11] and \MGNLO FxFx~\cite{Alwall:2014hca}, as summarised in
Table~\ref{tab:app:samples_and_generators}.
The \qcd \Wjj sample from \Sherpa[2.2.11] has MEs calculated at NLO for events
with up to two partons in the final state, and at LO for up to three additional partons.
The \AtlasMC{Comix} and \OPENLOOPS libraries
were used to produce the parton predictions, while the CKKW procedure
was used to merge final-state topologies. The \NNPDF[3.0nnlo]
PDF set~\cite{Ball:2014uwa} was used
for the ME calculation, and the default parameters of \Sherpa for
the parton showering, hadronisation, and underlying-event activity.
The \MGNLO FxFx sample has NLO accuracy in perturbative \qcd for MEs with up
to three partons, with the FxFx prescription~\cite{Frederix:2012ps} allowing different
jet multiplicities to be merged. The events were generated using the
\NNPDF[3.1nnlo] PDF set~\cite{Bertone_2018},
and the parton showering and subsequent hadronisation
were performed using \PYTHIA[8] with the A14 tune and the \NNPDF[2.3lo] PDF set.

\begin{table}[h]
\begin{center}
\caption{Summary of the additional MC generators used for the
simulation of \qcd \Wjj events.
For each sample, the parton showering algorithm and the tune used
are specified.
The generated jet multiplicity at the highest accuracy in
perturbative \qcd for the ME computation is provided, as
well as the PDF used.
}
\label{tab:app:samples_and_generators}
\scalebox{0.85}{
\begin{tabular}{lllll}
\hline\hline
Process & Simulation (+hadronisation) & ME accuracy & PDF & Tune \\ \hline\hline

\qcd \wjets
& \Sherpa[2.2.11]
& 0,1,2j@NLO or 3,4,5j@LO
& \NNPDF[3.0nnlo]
& Default \\

\qcd \wjets
& \MGNLOPY[8]
& 0,1,2,3j@NLO
& \NNPDF[3.1nnlo]
& A14 \\ \hline\hline

\end{tabular}
}
\end{center}
\end{table}

Measurements of the combined \ewk{}+\qcd \Wjj differential cross-sections
are performed in the SR as functions of $m_{jj}$, $\pt^{jj}$, $|\Delta y_{jj}|$, $\pt^{j1}$, $\pt^{\ell}$, and $\Delta\phi_{jj}$. The inclusive spectra are
obtained by subtracting, bin-by-bin, the multijet and other background
contributions from the observed data. Resulting signal distributions are then
unfolded to particle level using the IBU method described in
Section~\ref{sec:unfolding} to yield the \Wjj measurements.

To perform the unfolding procedure, the response matrix is built using the
simulations of the \ewk and \qcd \Wjj events.
The \ewk \Wjj simulation from \POWPY[8] is utilised as it provides
the best accuracy available in pertubative \qcd.
The unfolded measurements are evaluated separately by using the \qcd \Wjj
predictions from \POWPY[8] and \Sherpa[2.2.11], since those predictions
differ the most. The arithmetic average of the two estimates is
then taken as the nominal unfolded measurement.
In contrast to Section~\ref{sec:bkg-determination:lh}, no likelihood-fit is
used to constrain the \qcd \Wjj process since it is part of the measured signal.

The measured \ewk{}+\qcd \Wjj differential cross-sections have small statistical uncertainties, well below 6\% across all observables, and the overall precision is limited by systematic effects.
The same sources of experimental uncertainties as in the \ewk-only
measurement are considered.
The combined contribution from the integrated luminosity, lepton reconstruction, pile-up, MC statistics, multijet estimate, and modelling of the simulated background constitutes less than 4\% over most of the measured phase space, although it increases to approximately 7\% at central $\Delta\phi_{jj}$ values, and 10\% at high $\pt^{\ell}$. The leading experimental uncertainty arises from the jet energy scale. Jet-related effects contribute almost 12\% across the observables, with an increase to approximately 14\% at low $m_{jj}$, low $|\Delta y_{jj}|$, and central $\Delta\phi_{jj}$ values.
The theoretical uncertainty in the modelling of the detector response used
in the unfolding procedure is also relatively large. It is estimated as half of the difference
between the unfolded measurements using the two aforementioned \qcd \Wjj generators,
and is then added in quadrature to the unfolding bias originating from
distribution shape differences between the data and the
simulated \ewk{}+\qcd \Wjj predictions to estimate the total
uncertainty in the unfolding method. In the tails of the $m_{jj}$ and
$|\Delta y_{jj}|$ distributions, this uncertainty reaches 18\% and
28\%, respectively.

\begin{figure}[h]
\centering
\subfloat[]{
\includegraphics[width=0.5\textwidth]{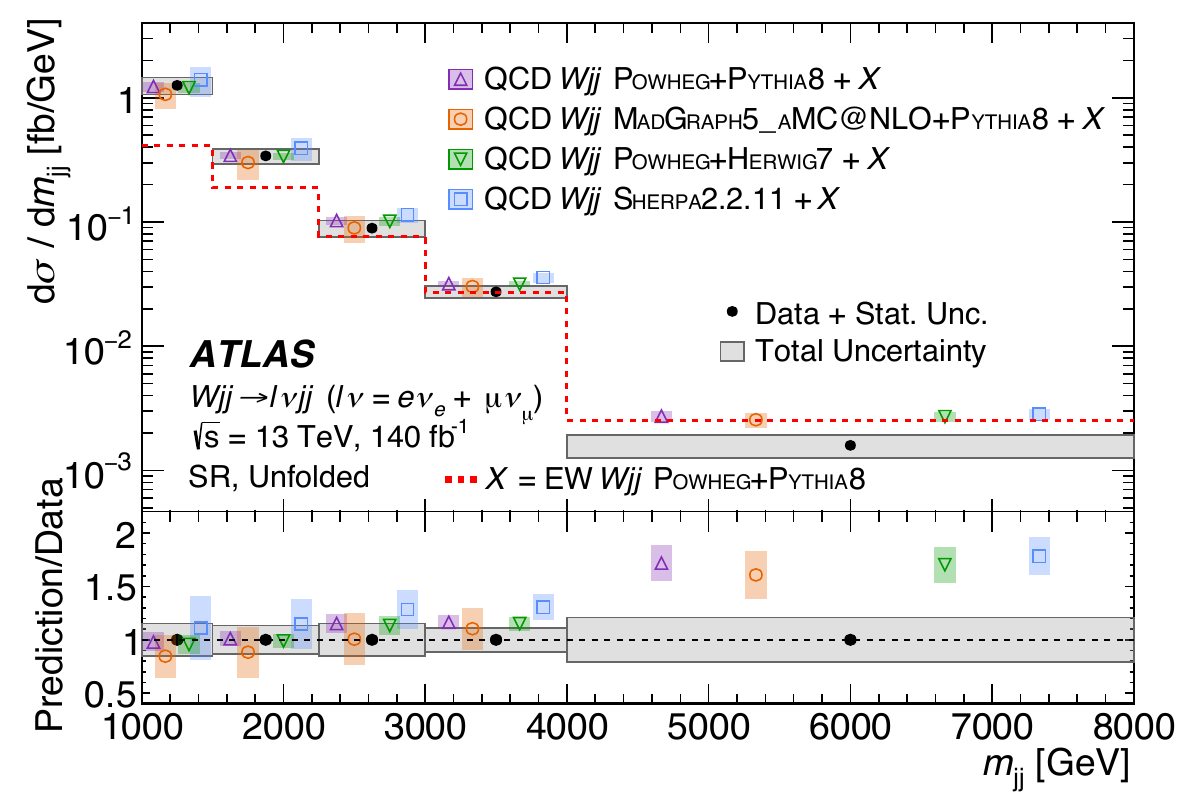}
}
\subfloat[]{
\includegraphics[width=0.5\textwidth]{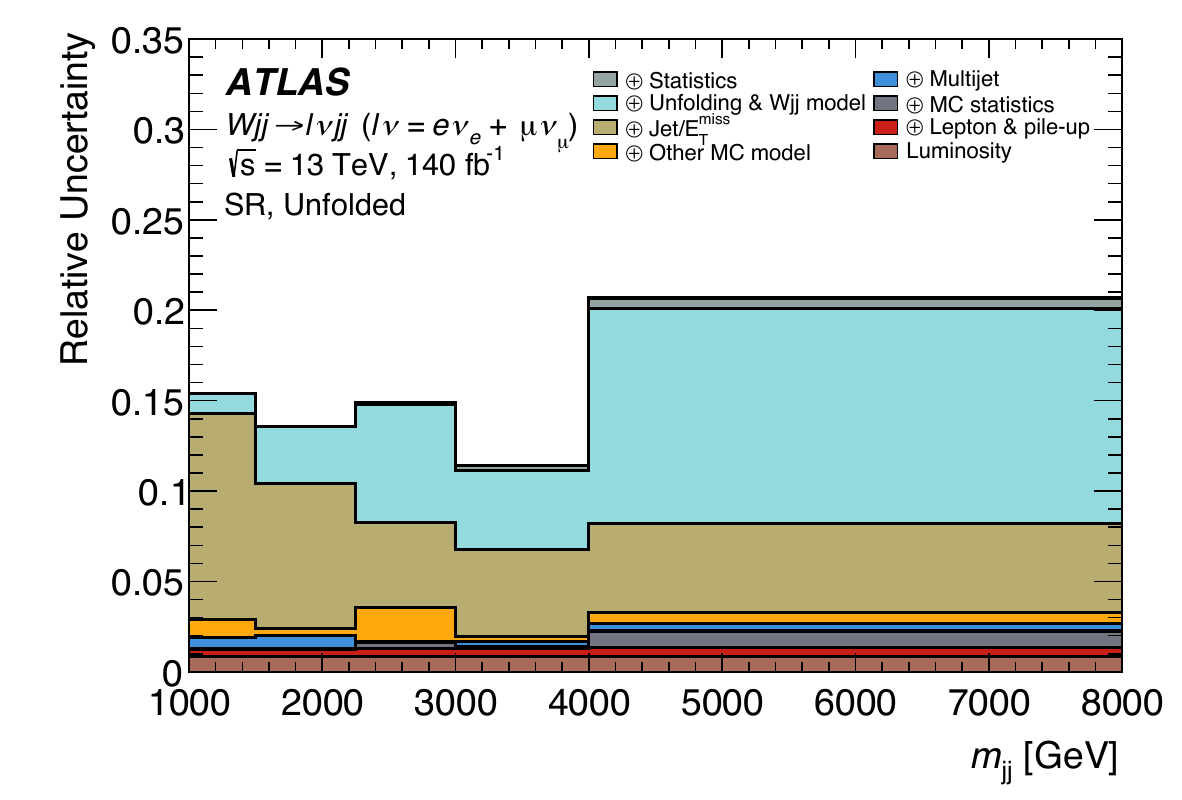}
}
\caption{(a) Unfolded differential cross-section for combined \ewk{}+\qcd \Wjj production and (b) associated statistical, experimental, and theoretical uncertainties
as a function of $m_{jj}$.
Measurements are compared with the \ewk \Wjj prediction from \POWPY[8] summed with \qcd \Wjj predictions from \Sherpa[2.2.11], \POWPY[8], \POWHER[7.2.1], and \MGNLO FxFx+\PYTHIA[8] including their theoretical uncertainties. The underlying \ewk \Wjj prediction is shown. All provided SM expectations are NLO in QCD accuracy. Uncertainties on the measurement are grouped into categories and summed in quadrature ($\oplus$) to yield the total uncertainty.
}
\label{fig:results:unf_incl_mjj}
\end{figure}

\begin{figure}[h]
\centering
\subfloat[]{
\includegraphics[width=0.5\textwidth]{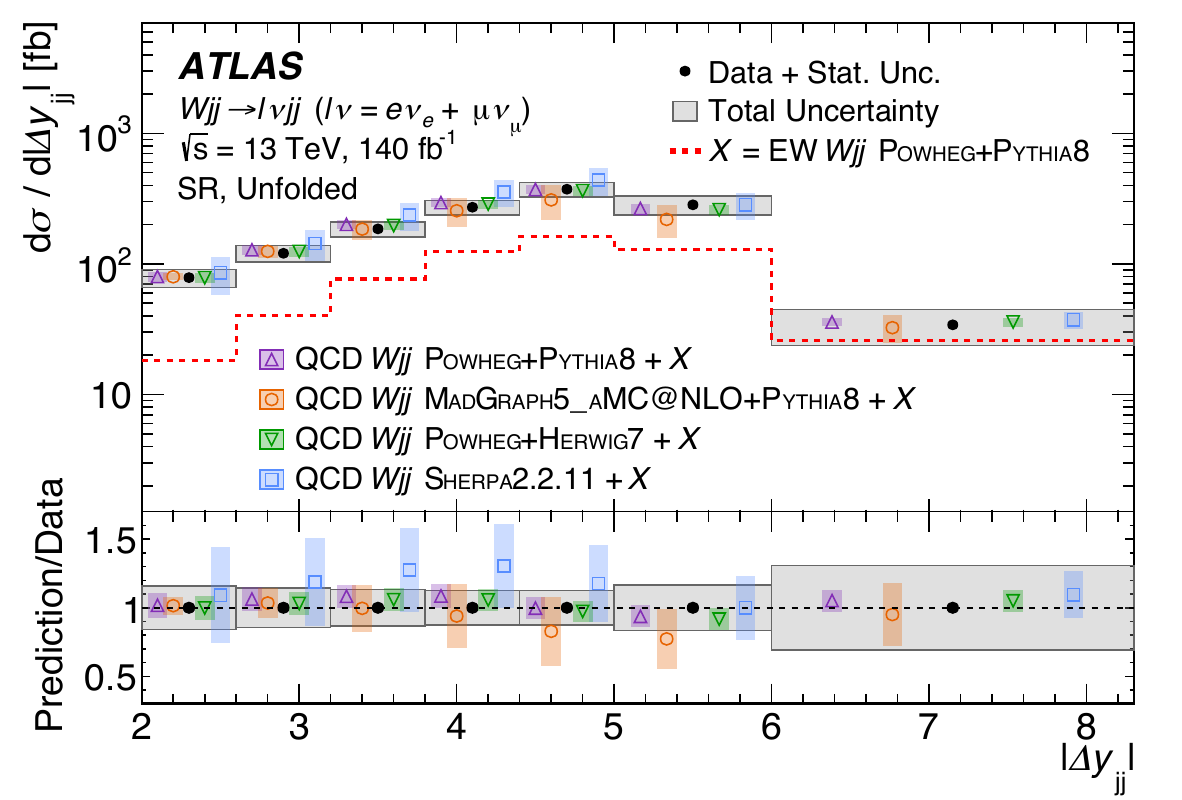}
}
\subfloat[]{
\includegraphics[width=0.5\textwidth]{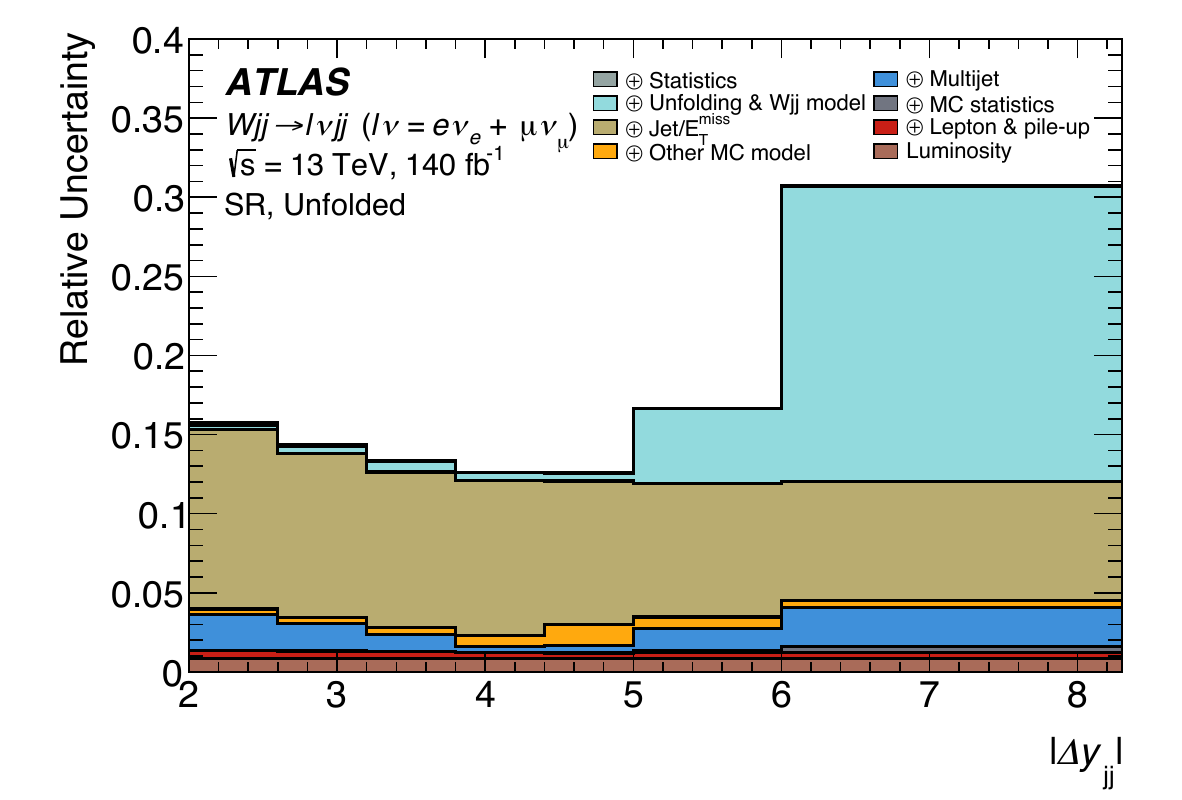}
}
\caption{(a) Unfolded differential cross-section for combined \ewk{}+\qcd \Wjj production and (b) associated statistical, experimental, and theoretical uncertainties
as a function of $|\Delta y_{jj}|$.
Measurements are compared with the \ewk \Wjj prediction from \POWPY[8] summed with \qcd \Wjj predictions from \Sherpa[2.2.11], \POWPY[8], \POWHER[7.2.1], and \MGNLO FxFx+\PYTHIA[8] including their theoretical uncertainties. The underlying \ewk \Wjj prediction is shown. All provided SM expectations are NLO in QCD accuracy. Uncertainties on the measurement are grouped into categories and summed in quadrature ($\oplus$) to yield the total uncertainty.
}
\label{fig:results:unf_incl_dyjj}
\end{figure}

\begin{figure}[h]
\centering
\subfloat[]{
\includegraphics[width=0.5\textwidth]{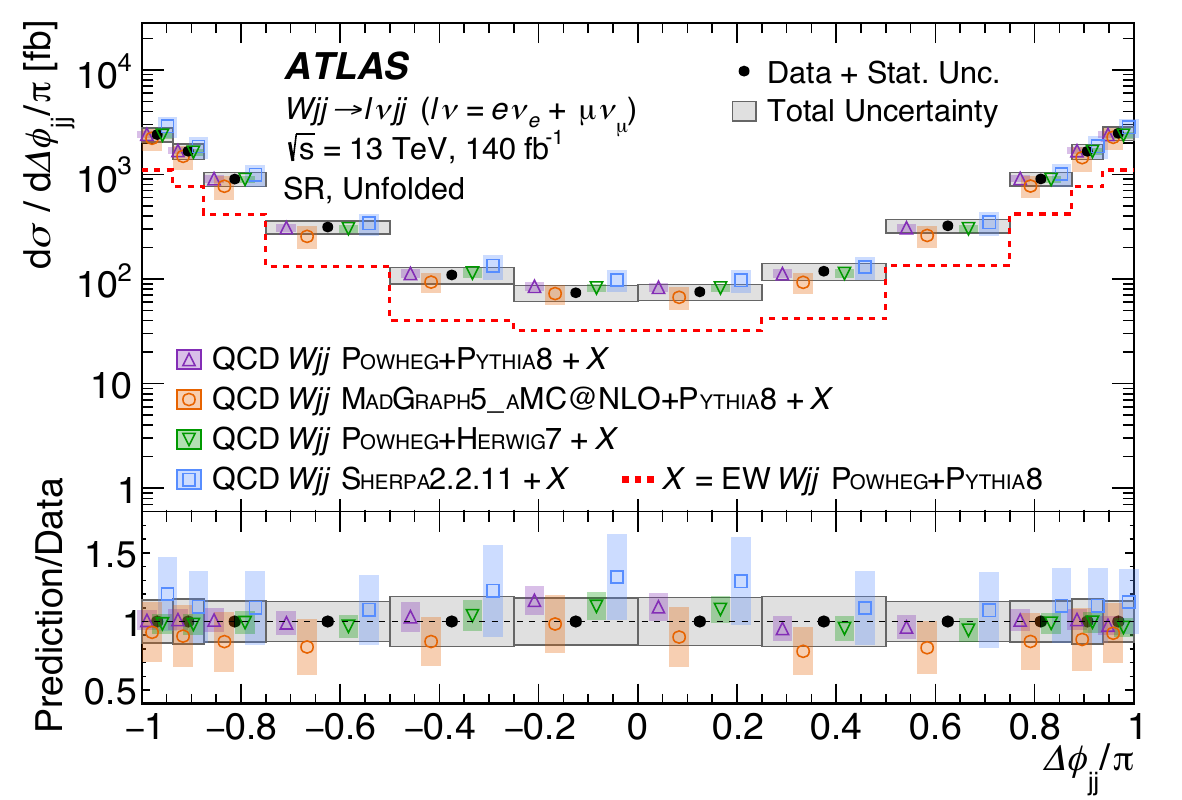}
}
\subfloat[]{
\includegraphics[width=0.5\textwidth]{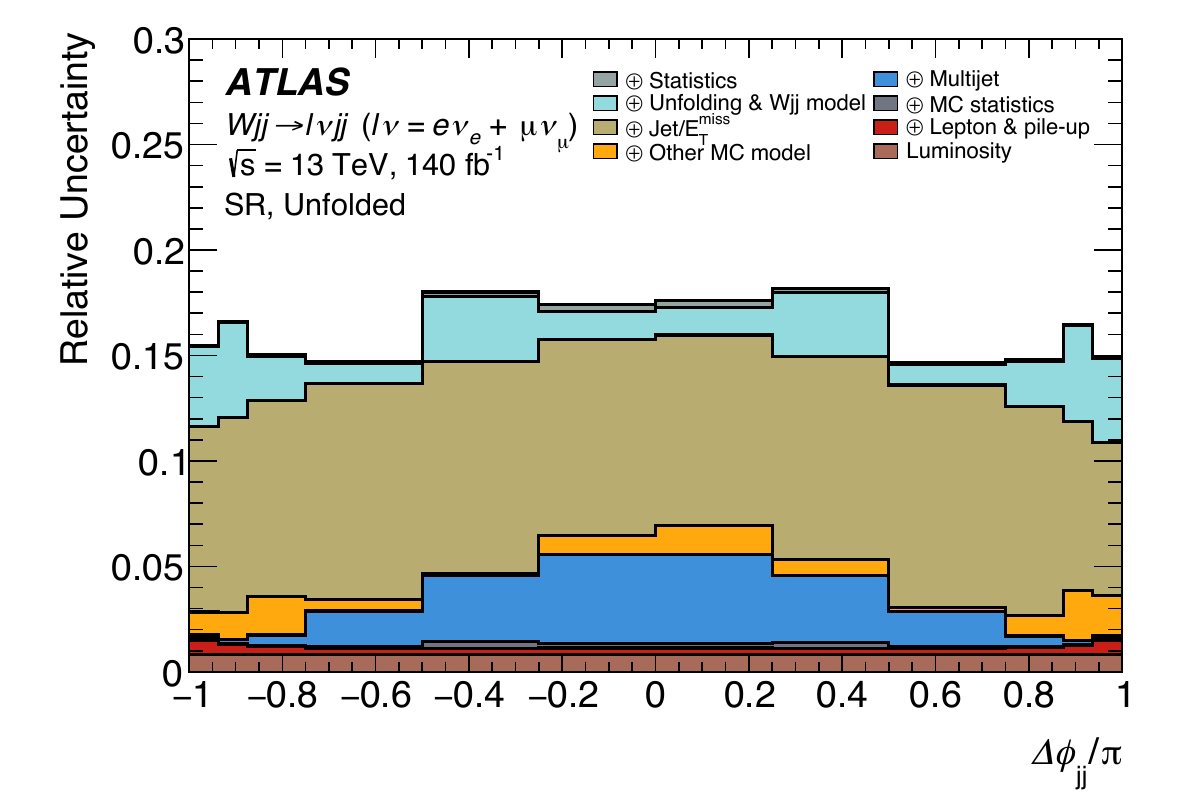}
}
\caption{(a) Unfolded differential cross-section for combined \ewk{}+\qcd \Wjj production and (b) associated statistical, experimental, and theoretical uncertainties
as a function of $\Delta\phi_{jj}$.
Measurements are compared with the \ewk \Wjj prediction from \POWPY[8] summed with \qcd \Wjj predictions from \Sherpa[2.2.11], \POWPY[8], \POWHER[7.2.1], and \MGNLO FxFx+\PYTHIA[8] including their theoretical uncertainties. The underlying \ewk \Wjj prediction is shown. All provided SM expectations are NLO in QCD accuracy. Uncertainties on the measurement are grouped into categories and summed in quadrature ($\oplus$) to yield the total uncertainty.
}
\label{fig:results:unf_incl_dphijj}
\end{figure}

Results are shown in Figures~\ref{fig:results:unf_incl_mjj}--\ref{fig:results:unf_incl_dphijj}, together with a full breakdown of the uncertainties.
Since the focus of this combined \ewk{}+\qcd \Wjj measurement is the modelling of
\qcd production, the \qcd \Wjj simulations from \POWPY[8],
\POWHER[7.2.1], \MGNLO FxFx~\cite{Alwall:2014hca} and
\Sherpa[2.2.11] are employed for comparison with the measurement at
particle level, together with the nominal \ewk{} \Wjj simulation from
\POWPY[8].
Across most of the phase space, the \ewk{}+\qcd \POWPY[8] description is in good agreement with the data. However, at large $m_{jj}$ the \ewk{}+\qcd \Wjj predictions are in tension with the data. This region of phase space is dominated by \ewk \Wjj production, and the observed behaviour is consistent with the results presented in Section~\ref{sec:results}. In other regions,
the nominal prediction of \MGNLO FxFx+\PYTHIA[8] tends to underestimate the data, while that of \Sherpa[2.2.11] is observed to overestimate it in almost all cases. Both \MGNLO FxFx+\PYTHIA[8] and \Sherpa{2.2.11} exhibit sizeable effects from renormalisation and factorisation scale variations, which result in large theory uncertainty bands.
These are understood to be due to higher parton multiplicities being
computed at LO and to the specific SR investigated, which impacts
the effects of \qcd scale variations in multi-leg generators.
Overall, the measurements are consistent with the SM within uncertainties.

The cross-section for combined \ewk{}+\qcd \Wjj production in the particle-level fiducial volume, obtained by integrating the $m_{jj}$ distribution, is measured to be:
\begin{equation*}
\sigma_{EW+QCD} = 1055 \pm 4 \, (\text{stat.}) \pm 148 \, (\text{exp.}) \pm 18 \, (\text{theory}) \; \text{fb}.
\end{equation*}
Consistent values are obtained by integrating the other measured observables, with agreement within 1\%.
As with the \ewk-only measurement, the statistical and systematic bin-to-bin correlations from the \ewk{}+\qcd \Wjj measurement are accounted for in this computation.
The measured cross-section is in good agreement with the predictions of
$984 \pm 59 \; \text{fb}$, $962 \pm 58 \; \text{fb}$, $857 \pm 143 \; \text{fb}$, and $1115 \pm 200 \; \text{fb}$,
obtained using the \POWPY[8], \POWHER[7.2.1], \MGNLO FxFx, and \Sherpa[2.2.11] \qcd \Wjj predictions, respectively, each combined with the \POWPY[8] \ewk \Wjj prediction. The sources of uncertainty in the predictions are the same as those reported for the \ewk \Wjj predictions in Section~\ref{sec:results}.

\begin{figure}[h]
\centering
\subfloat[]{
\includegraphics[width=0.5\textwidth]{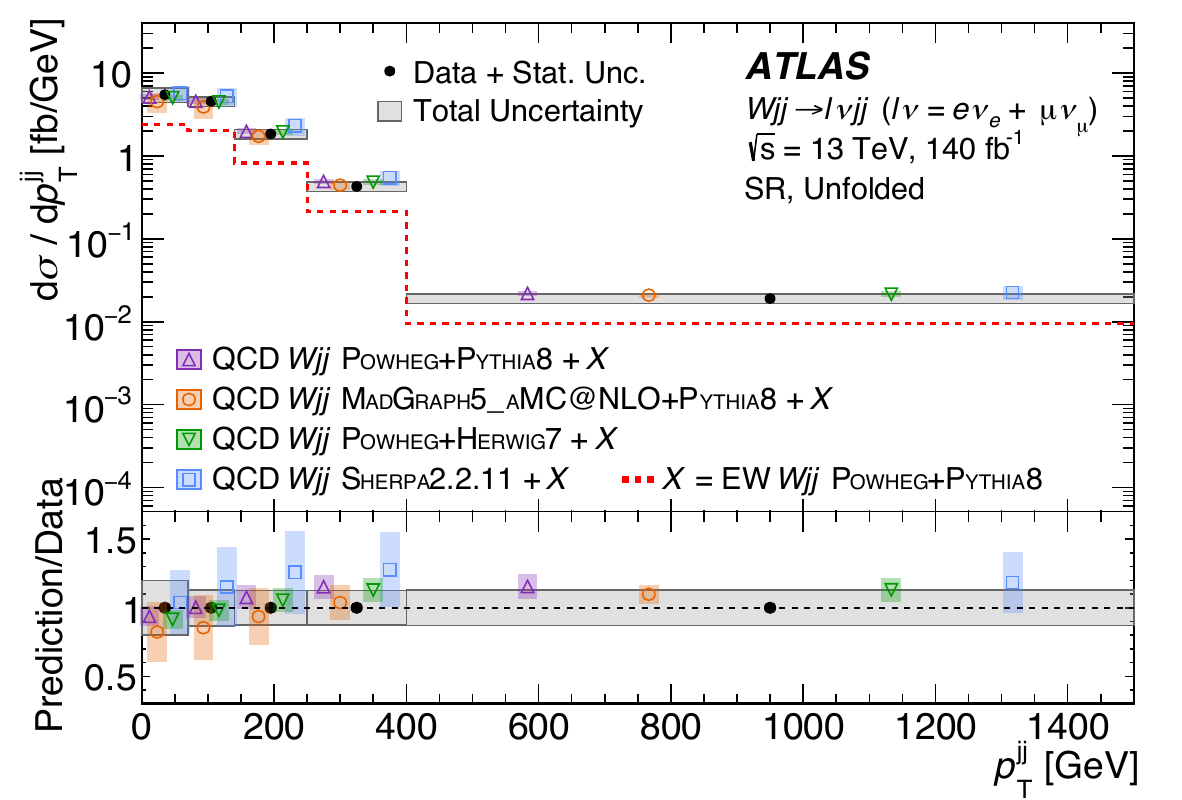}
}
\subfloat[]{
\includegraphics[width=0.5\textwidth]{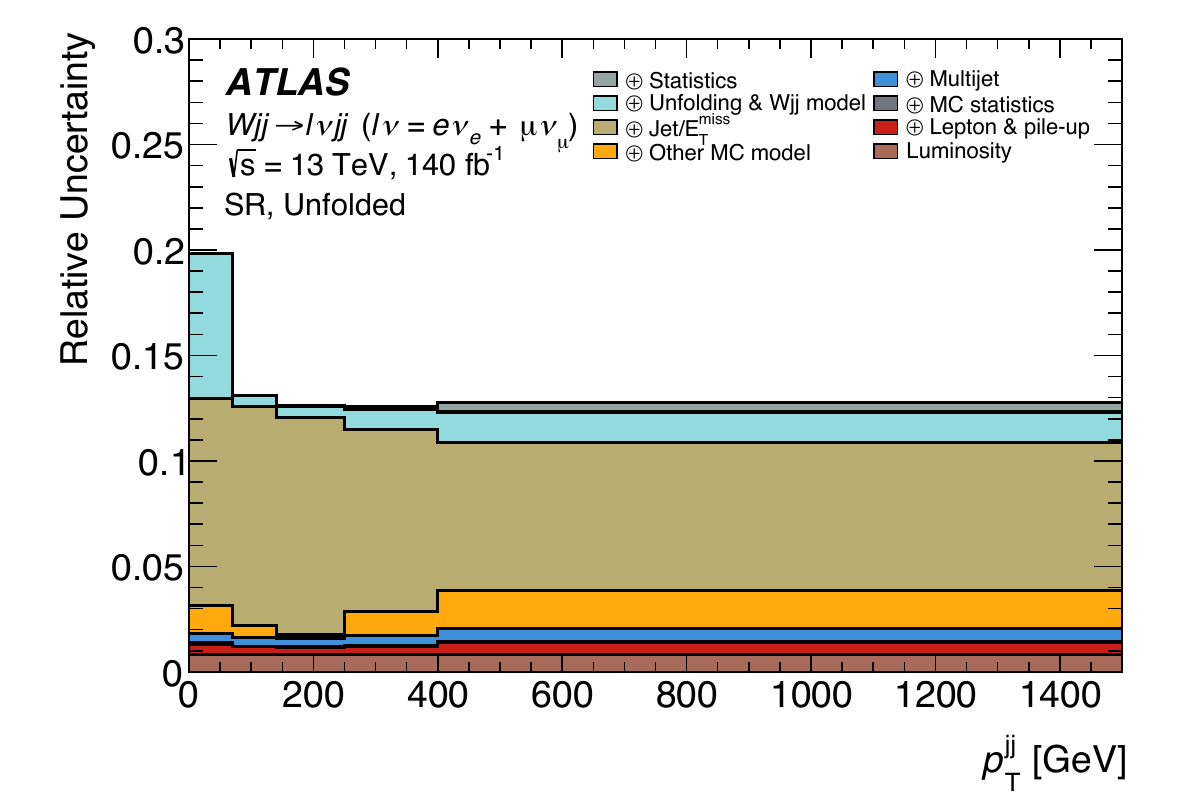}
}
\caption{(a) Unfolded differential cross-section for combined \ewk{}+\qcd \Wjj production and (b) associated statistical, experimental, and theoretical uncertainties
as a function of $\pt^{jj}$.
Measurements are compared with the \ewk \Wjj prediction from \POWPY[8] summed with \qcd \Wjj predictions from \Sherpa[2.2.11], \POWPY[8], \POWHER[7.2.1], and \MGNLO FxFx+\PYTHIA[8] including their theoretical uncertainties. The underlying \ewk \Wjj prediction is shown. All provided SM expectations are NLO in QCD accuracy. Uncertainties on the measurement are grouped into categories and summed in quadrature ($\oplus$) to yield the total uncertainty.
}
\end{figure}

\begin{figure}[h]
\centering
\subfloat[]{
\includegraphics[width=0.5\textwidth]{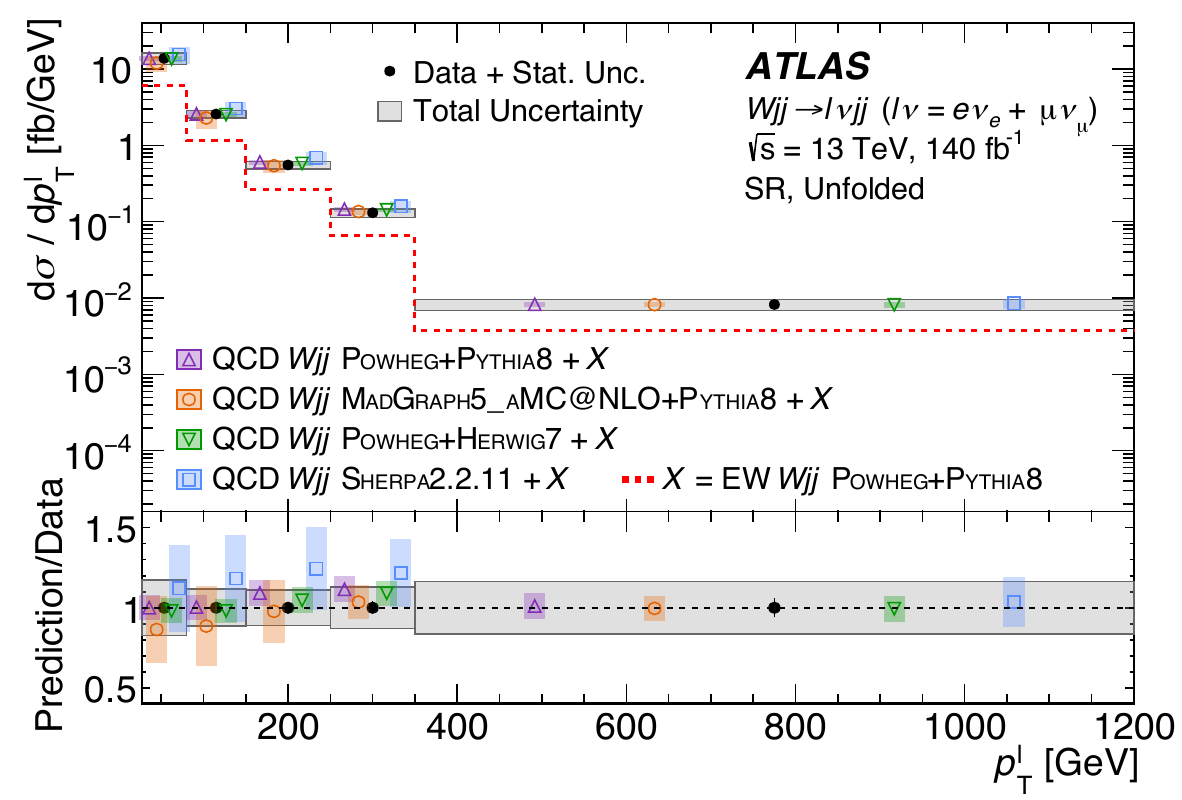}
}
\subfloat[]{
\includegraphics[width=0.5\textwidth]{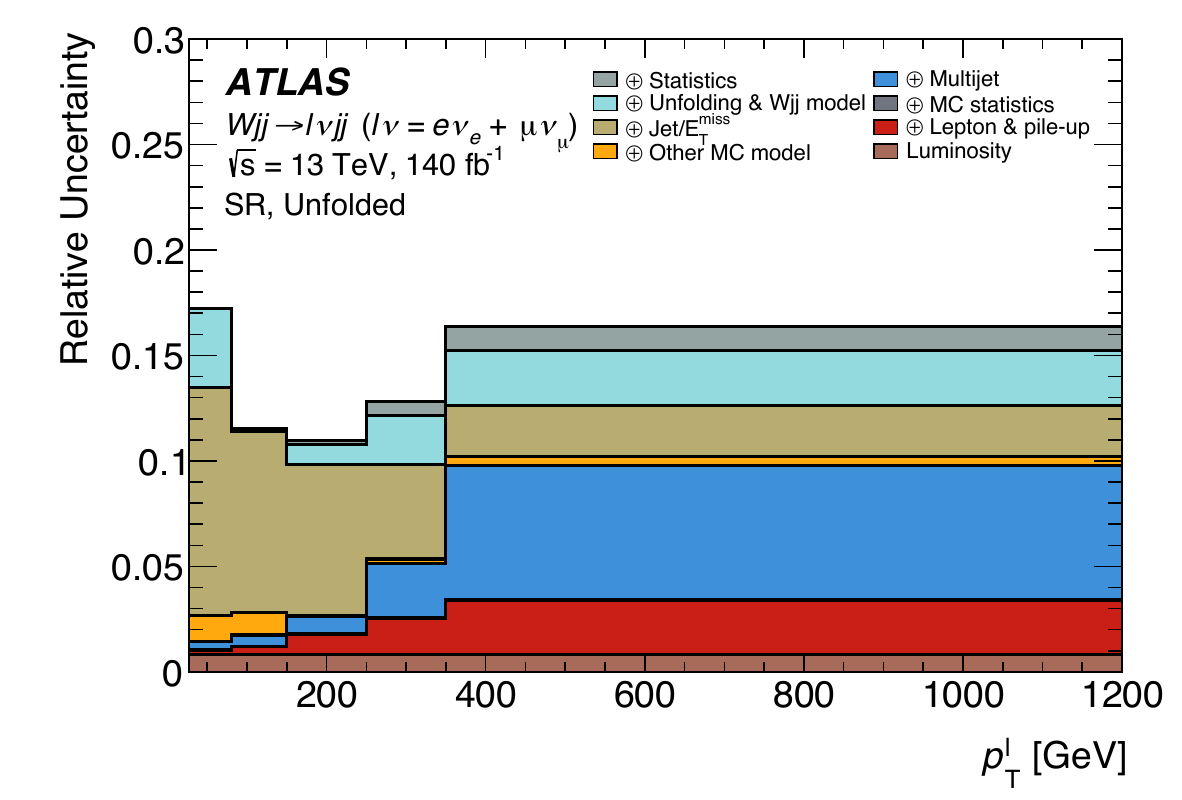}
}
\caption{(a) Unfolded differential cross-section for combined \ewk{}+\qcd \Wjj production and (b) associated statistical, experimental, and theoretical uncertainties
as a function of $\pt^{\ell}$.
Measurements are compared with the \ewk \Wjj prediction from \POWPY[8] summed with \qcd \Wjj predictions from \Sherpa[2.2.11], \POWPY[8], \POWHER[7.2.1], and \MGNLO FxFx+\PYTHIA[8] including their theoretical uncertainties. The underlying \ewk \Wjj prediction is shown. All provided SM expectations are NLO in QCD accuracy. Uncertainties on the measurement are grouped into categories and summed in quadrature ($\oplus$) to yield the total uncertainty.
}
\end{figure}

\begin{figure}[h]
\centering
\subfloat[]{
\includegraphics[width=0.5\textwidth]{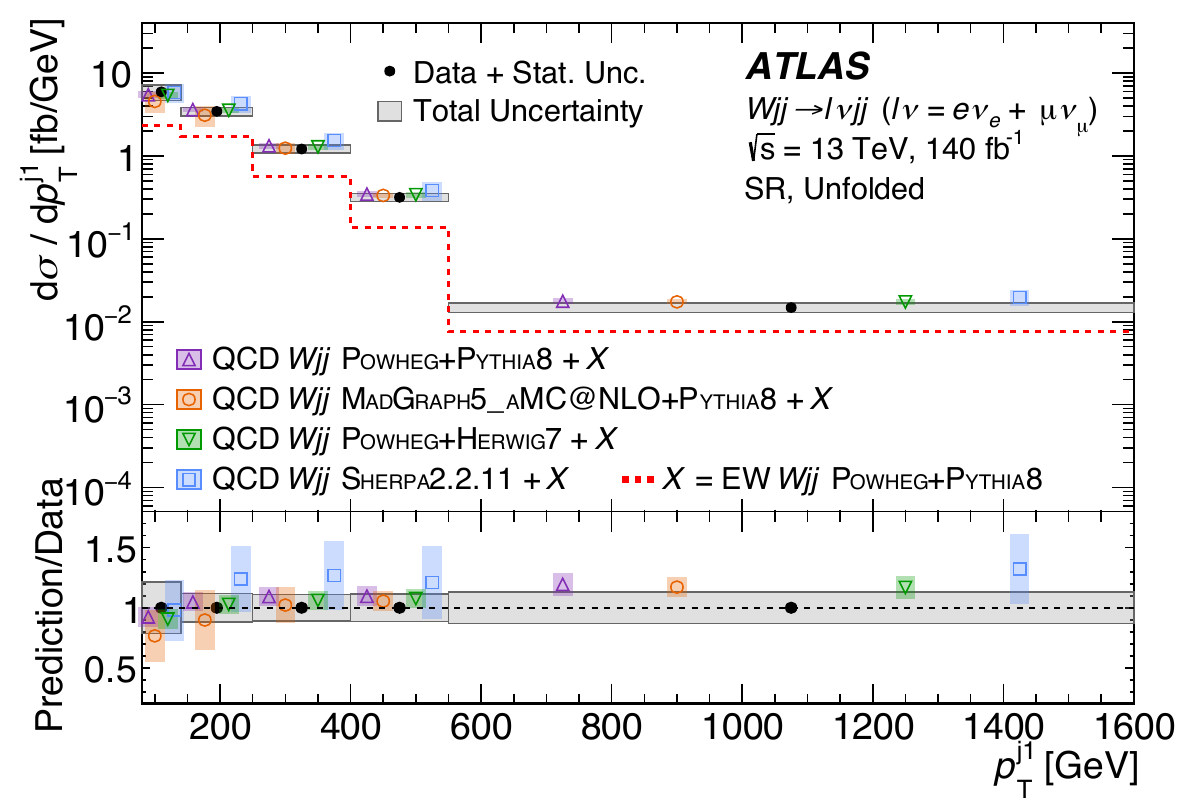}
}
\subfloat[]{
\includegraphics[width=0.5\textwidth]{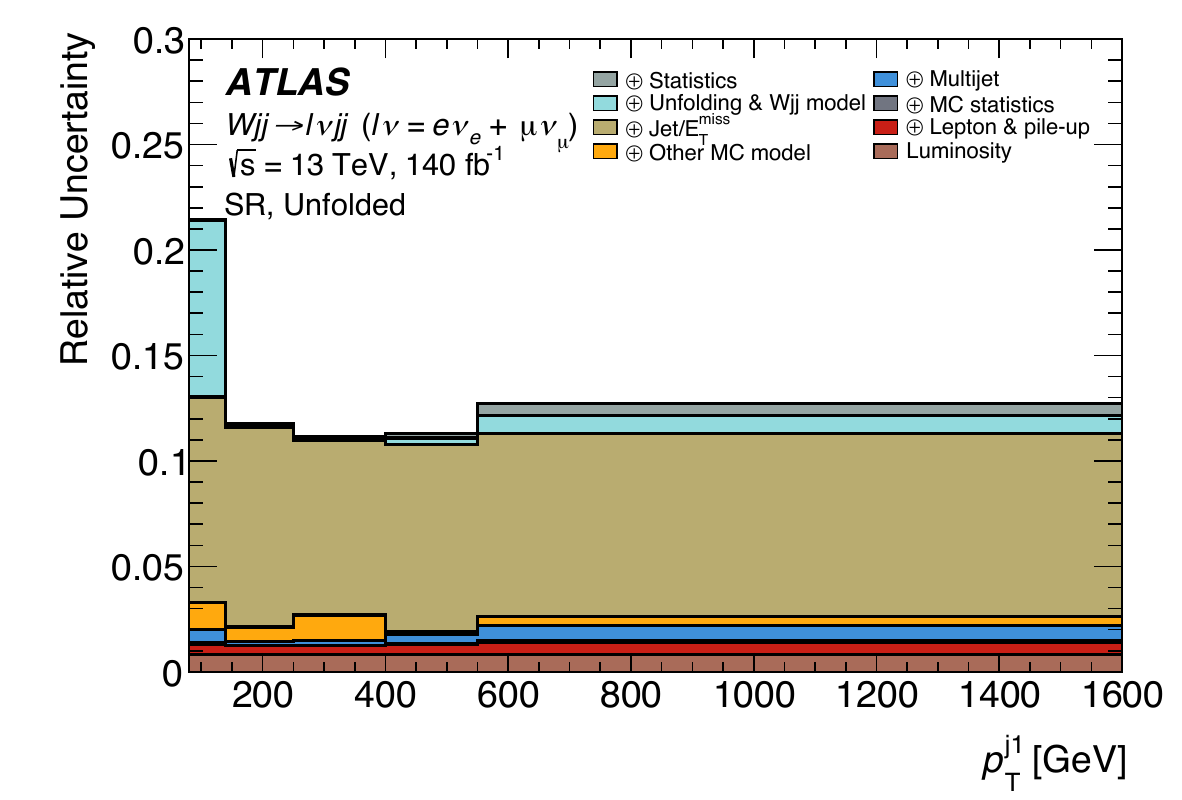}
}
\caption{(a) Unfolded differential cross-section for combined \ewk{}+\qcd \Wjj production and (b) associated statistical, experimental, and theoretical uncertainties
as a function of $\pt^{j1}$.
Measurements are compared with the \ewk \Wjj prediction from \POWPY[8] summed with \qcd \Wjj predictions from \Sherpa[2.2.11], \POWPY[8], \POWHER[7.2.1], and \MGNLO FxFx+\PYTHIA[8] including their theoretical uncertainties. The underlying \ewk \Wjj prediction is shown. All provided SM expectations are NLO in QCD accuracy. Uncertainties on the measurement are grouped into categories and summed in quadrature ($\oplus$) to yield the total uncertainty.
}
\end{figure}

%
\FloatBarrier
\section*{Acknowledgements}
%

%

%
%

%
%

We thank CERN for the very successful operation of the LHC and its injectors, as well as the support staff at
CERN and at our institutions worldwide without whom ATLAS could not be operated efficiently.

The crucial computing support from all WLCG partners is acknowledged gratefully, in particular from CERN, the ATLAS Tier-1 facilities at TRIUMF/SFU (Canada), NDGF (Denmark, Norway, Sweden), CC-IN2P3 (France), KIT/GridKA (Germany), INFN-CNAF (Italy), NL-T1 (Netherlands), PIC (Spain), RAL (UK) and BNL (USA), the Tier-2 facilities worldwide and large non-WLCG resource providers. Major contributors of computing resources are listed in Ref.~\cite{ATL-SOFT-PUB-2026-001}.

We gratefully acknowledge the support of ANPCyT, Argentina; YerPhI, Armenia; ARC, Australia; BMWFW and FWF, Austria; ANAS, Azerbaijan; CNPq and FAPESP, Brazil; NSERC, NRC and CFI, Canada; CERN; ANID, Chile; CAS, MOST, NSFC and FRFCU, China; Minciencias, Colombia; MEYS CR, Czech Republic; DNRF and DNSRC, Denmark; IN2P3-CNRS and CEA-DRF/IRFU, France; SRNSFG, Georgia; BMFTR, HGF and MPG, Germany; GSRI, Greece; RGC and Hong Kong SAR, China; ICHEP and Academy of Sciences and Humanities, Israel; INFN, Italy; MEXT and JSPS, Japan; CNRST, Morocco; NWO, Netherlands; RCN, Norway; MNiSW, Poland; FCT, Portugal; MNE/IFA, Romania; MSTDI, Serbia; MSSR, Slovakia; ARIS and MVZI, Slovenia; DSI/NRF, South Africa; MICIU/AEI, Spain; SRC and Wallenberg Foundation, Sweden; SERI, SNSF and Cantons of Bern and Geneva, Switzerland; NSTC, Taipei; TENMAK, T\"urkiye; STFC/UKRI, United Kingdom; DOE and NSF, United States of America.

Individual groups and members have received support from BCKDF, CANARIE, CRC and DRAC, Canada; CERN-CZ, FORTE and PRIMUS, Czech Republic; COST, ERC, ERDF, Horizon 2020 and Marie Sk{\l}odowska-Curie Actions, European Union; Investissements d'Avenir Labex, Investissements d'Avenir Idex and ANR, France; DFG and AvH Foundation, Germany; Herakleitos, Thales and Aristeia programmes co-financed by EU-ESF and the Greek NSRF, Greece; BSF-NSF and MINERVA, Israel; NCN and NAWA, Poland; La Caixa Banking Foundation, CERCA and AGAUR programs from Generalitat de Catalunya and PROMETEO and GenT Programmes Generalitat Valenciana, Spain; G\"{o}ran Gustafssons Stiftelse, Sweden; The Royal Society and Leverhulme Trust, United Kingdom; Eric and Wendy Schmidt Fund for Strategic Innovation, United States of America.

In addition, individual members wish to acknowledge support from Chile: Agencia Nacional de Investigaci\'on y Desarrollo (ANID FONDECYT reg. 1230987, FONDECYT 1230812, FONDECYT 1240864, Fondecyt 3240661, Fondecyt Regular 1240721); China: Fundamental Research Funds for the Central Universities (010-63263105), Chinese Ministry of Science and Technology (MOST-2023YFA1605700, MOST-2023YFA1609300), National Natural Science Foundation of China (NSFC 12275265, NSFC-W2543005); Czech Republic: Czech Science Foundation (GACR - 24-11373S), Ministry of Education Youth and Sports (ERC-CZ-LL2327, FORTE CZ.02.01.01/00/22\_008/0004632), PRIMUS Research Programme (PRIMUS/21/SCI/017); EU: H2020 European Research Council (ERC - 101002463); European Union: European Research Council (BARD No. 101116429, ERC - 101219398, ERC - 948254, ERC 101089007), European Regional Development Fund (HE COFUND GA No.101081355, ERDF), Marie Sklodowska-Curie Actions (GAP-101168829); France: Agence Nationale de la Recherche (ANR-21-CE31-0013, ANR-22-EDIR-0002, ANR-24-CE31-0504-01); Germany: Deutsche Forschungsgemeinschaft (DFG - 469666862); China: Research Grants Council (GRF); Italy: Istituto Nazionale di Fisica Nucleare (LHC-MIUR - 28003/2025), Ministero dell'Università e della Ricerca (NextGenEU 153D23001490006 M4C2.1.1, NextGenEU I53D23000820006 M4C2.1.1, NextGenEU I53D23001490006 M4C2.1.1, SOE2024\_0000023); Japan: Japan Society for the Promotion of Science (JSPS KAKENHI  JP25H0063, JSPS KAKENHI JP22H01227, JSPS KAKENHI JP22H04944, JSPS KAKENHI JP22KK0227, JSPS KAKENHI JP24K23939, JSPS KAKENHI JP24KK0251, JSPS KAKENHI JP25H00650, JSPS KAKENHI JP25H01291, JSPS KAKENHI JP25K01011, JSPS KAKENHI JP25K01023); Poland: Polish National Science Centre (NCN 2021/42/E/ST2/00350, NCN OPUS 2023/51/B/ST2/02507, NCN OPUS nr 2022/47/B/ST2/03059, NCN UMO-2019/34/E/ST2/00393, UMO-2022/47/O/ST2/00148, UMO-2023/49/B/ST2/04085, UMO-2023/51/B/ST2/00920, UMO-2024/53/N/ST2/00869); Spain: Agència de Gestió d'Ajuts Universitaris i de Recerca. (AGAUR - 2023 BP 00141), Ministry of Science and Innovation (RYC2019-028510-I, RYC2020-030254-I, RYC2021-031273-I, RYC2022-038164-I), Ministerio de Ciencia, Innovación y Universidades/Agencia Estatal de Investigaci\'on (EU NextGenerationEU (PRTR-C17.I1), PID2022-142604OB-C22); Sweden: Carl Trygger Foundation (Carl Trygger Foundation CTS 22:2312), Swedish Research Council (Swedish Research Council 2023-04654, VR 2021-03651, VR 2022-03845, VR 2022-04683, VR 2023-03403, VR 2024-05451, VR 2025-05940), Knut and Alice Wallenberg Foundation (KAW 2023.0366); Switzerland: Swiss National Science Foundation (SNSF - PCEFP2\_194658); United Kingdom: The Binks Trust, Royal Society (NIF-R1-231091); United States of America: U.S. Department of Energy (ECA DE-AC02-76SF00515), John Templeton Foundation (John Templeton Foundation 63206), Neubauer Family Foundation.

%
%


%
%
%
\printbibliography
\clearpage
 
\begin{flushleft}
\hypersetup{urlcolor=black}
{\Large The ATLAS Collaboration}

\bigskip

\AtlasOrcid[0000-0002-6665-4934]{G.~Aad}$^\textrm{\scriptsize 102}$,
\AtlasOrcid[0000-0001-7616-1554]{E.~Aakvaag}$^\textrm{\scriptsize 17}$,
\AtlasOrcid[0000-0002-5888-2734]{B.~Abbott}$^\textrm{\scriptsize 121}$,
\AtlasOrcid[0000-0002-0287-5869]{S.~Abdelhameed}$^\textrm{\scriptsize 83b}$,
\AtlasOrcid[0000-0002-1002-1652]{K.~Abeling}$^\textrm{\scriptsize 54}$,
\AtlasOrcid[0000-0001-5763-2760]{N.J.~Abicht}$^\textrm{\scriptsize 48}$,
\AtlasOrcid[0000-0002-8496-9294]{S.H.~Abidi}$^\textrm{\scriptsize 30}$,
\AtlasOrcid[0009-0003-6578-220X]{M.~Aboelela}$^\textrm{\scriptsize 44}$,
\AtlasOrcid[0000-0002-9987-2292]{A.~Aboulhorma}$^\textrm{\scriptsize 36e}$,
\AtlasOrcid[0000-0001-5329-6640]{H.~Abramowicz}$^\textrm{\scriptsize 154}$
% ...
\bigskip
\\

% ...
$^{17}$Department for Physics and Technology, University of Bergen, Bergen; Norway.\\
% ...
$^{30}$Physics Department, Brookhaven National Laboratory, Upton NY; United States of America.\\
% ...
$^{36}$$^{(a)}$Facult\'e des Sciences Ain Chock, Universit\'e Hassan II de Casablanca;$^{(b)}$Facult\'{e} des Sciences, Universit\'{e} Ibn-Tofail, K\'{e}nitra;$^{(c)}$Facult\'e des Sciences Semlalia, Universit\'e Cadi Ayyad, LPHEA-Marrakech;$^{(d)}$LPMR, Facult\'e des Sciences, Universit\'e Mohamed Premier, Oujda;$^{(e)}$Facult\'e des sciences, Universit\'e Mohammed V, Rabat;$^{(f)}$Institute of Applied Physics, Mohammed VI Polytechnic University, Ben Guerir; Morocco.\\
% ...
$^{44}$Physics Department, Southern Methodist University, Dallas TX; United States of America.\\
% ...
$^{48}$Fakult\"{a}t Physik, Technische Universit{\"a}t Dortmund, Dortmund; Germany.\\
% ...
$^{54}$II. Physikalisches Institut, Georg-August-Universit\"{a}t G\"ottingen, G\"ottingen; Germany.\\
% ...
$^{83}$$^{(a)}$Khalifa University of Science and Technology, Abu Dhabi;$^{(b)}$New York University Abu Dhabi, Abu Dhabi;$^{(c)}$United Arab Emirates University, Al Ain;$^{(d)}$University of Sharjah, Sharjah; United Arab Emirates.\\
% ...
$^{102}$CPPM, Aix-Marseille Universit\'e, CNRS/IN2P3, Marseille; France.\\
% ...
$^{121}$Homer L. Dodge Department of Physics and Astronomy, University of Oklahoma, Norman OK; United States of America.\\
% ...
$^{154}$Raymond and Beverly Sackler School of Physics and Astronomy, Tel Aviv University, Tel Aviv; Israel.\\
% ...
$^{a}$ Also at Affiliated with an institute formerly covered by a cooperation agreement with CERN.\\
$^{b}$ Also at An-Najah National University, Nablus; Palestine.\\
$^{c}$ Also at Borough of Manhattan Community College, City University of New York, New York NY; United States of America.\\
$^{d}$ Also at Center for Interdisciplinary Research and Innovation (CIRI-AUTH), Thessaloniki; Greece.\\
$^{e}$ Also at Centre of Physics of the Universities of Minho and Porto (CF-UM-UP); Portugal.\\
$^{f}$ Also at CERN, Geneva; Switzerland.\\
% ...

\end{flushleft}


%

%
%

%

%
%
%
%

\end{document}